\documentclass[aj,twocolumn]{aastex62}

\newcommand{\starAteff}{\ensuremath{5875_{-190}^{+100}}}
\newcommand{\starAlogg}{\ensuremath{4.560_{-0.046}^{+0.046}}}
\newcommand{\starAfeh}{\ensuremath{-0.106_{-0.070}^{+0.079}}}
\newcommand{\starAvsini}{\ensuremath{11.5 \pm 1.0}}
\newcommand{\starAvmac}{\ensuremath{7.9 \pm 1.0}}

\newcommand{\starAmass}{\ensuremath{1.036_{-0.009}^{+0.013}}}
\newcommand{\starAradius}{\ensuremath{0.881_{-0.047}^{+0.038}}}
\newcommand{\starAlum}{\ensuremath{0.831_{-0.072}^{+0.105}}}
\newcommand{\starAirot}{\ensuremath{78_{-14}^{+7}\,(>43\,3\sigma)}}

\newcommand{\starAperiod}{\ensuremath{4.937770_{-0.000029}^{+0.000028}}}
\newcommand{\starAperiodshort}{\ensuremath{4.94}}
\newcommand{\starATc}{\ensuremath{2458357.30548_{-0.0015}^{+0.0013}}}
\newcommand{\starATdur}{\ensuremath{0.0930_{-0.0020}^{+0.0026}}}
\newcommand{\starAars}{\ensuremath{14.02_{-0.61}^{+0.82}}}
\newcommand{\starArprs}{\ensuremath{0.02858_{-0.0010}^{+0.0011}}}
\newcommand{\starAb}{\ensuremath{0.607_{-0.070}^{+0.047}}}
\newcommand{\starAinc}{\ensuremath{87.52_{-0.31}^{+0.40}}}

\newcommand{\starAKrv}{\ensuremath{<120}}
\newcommand{\starArvjitter}{\ensuremath{81_{-22}^{+34}}}
\newcommand{\starAsystemic}{\ensuremath{-2.059_{-0.030}^{+0.026}}}

\newcommand{\starAplrad}{\ensuremath{2.74_{-0.18}^{+0.18}}}
\newcommand{\starAplmass}{\ensuremath{<1.0}}
\newcommand{\starAa}{\ensuremath{0.05741_{-0.00017}^{+0.00023}}}
\newcommand{\starAdist}{\ensuremath{99.52_{-0.43}^{+0.44}}}
\newcommand{\starAage}{\ensuremath{40-320}}

\newcommand{\starAGPQ}{\ensuremath{-0.55_{-0.40}^{+0.57}}}
\newcommand{\starAGPS}{\ensuremath{-4.25_{-0.56}^{+0.56}}}
\newcommand{\starAGPw}{\ensuremath{-1.26_{-0.14}^{+0.14}}}

\newcommand{\starBteff}{\ensuremath{4928_{-85}^{+125}}}
\newcommand{\starBlogg}{\ensuremath{4.314_{-0.028}^{+0.028}}}
\newcommand{\starBfeh}{\ensuremath{-0.221_{-0.076}^{+0.131}}}
\newcommand{\starBvsini}{\ensuremath{14.3 \pm 0.5}}
\newcommand{\starBvmac}{\ensuremath{9.8 \pm 3.4}}

\newcommand{\starBmass}{\ensuremath{0.788_{-0.031}^{+0.037}}}
\newcommand{\starBradius}{\ensuremath{1.022_{-0.020}^{+0.018}}}
\newcommand{\starBlum}{\ensuremath{0.428_{-0.079}^{+0.166}}}
\newcommand{\starBirot}{\ensuremath{76_{-11}^{+9}\, (>45\,3\sigma)}}

\newcommand{\starBage}{\ensuremath{20-160}}
\newcommand{\starBdist}{\ensuremath{152.76_{-0.71}^{+0.71}}}

\newcommand{\starBGPQ}{\ensuremath{-1.10_{-0.31}^{+0.43}}}
\newcommand{\starBGPS}{\ensuremath{-2.37_{-0.50}^{+0.48}}}
\newcommand{\starBGPw}{\ensuremath{-1.19_{-0.11}^{+0.11}}}

\newcommand{\starBrvjitter}{\ensuremath{86_{-29}^{+56}}}
\newcommand{\starBsystemic}{\ensuremath{25.321_{-0.032}^{+0.034}}}

\newcommand{\starBperioda}{\ensuremath{4.324190_{-0.000030}^{+0.000030}}}
\newcommand{\starBperiodashort}{\ensuremath{4.32}}
\newcommand{\starBTca}{\ensuremath{2458441.5762_{-0.0021}^{+0.0021}}}
\newcommand{\starBTdura}{\ensuremath{0.1418_{-0.0018}^{+0.0025}}}
\newcommand{\starBarsa}{\ensuremath{10.12_{-0.18}^{+0.13}}}
\newcommand{\starBrprsa}{\ensuremath{0.04322_{-0.0016}^{+0.0015}}}
\newcommand{\starBba}{\ensuremath{0.05_{-0.04}^{+0.05}}}
\newcommand{\starBinca}{\ensuremath{89.97_{-0.51}^{+0.34}}}

\newcommand{\starBKrva}{\ensuremath{<370}}

\newcommand{\starBplrada}{\ensuremath{4.81_{-0.20}^{+0.20}}}
\newcommand{\starBplmassa}{\ensuremath{<2.6}}
\newcommand{\starBaa}{\ensuremath{0.04796_{-0.00065}^{+0.00073}}}

\newcommand{\starBperiodb}{\ensuremath{10.156430_{-0.000079}^{+0.000069}}}
\newcommand{\starBperiodbshort}{\ensuremath{10.16}}
\newcommand{\starBTcb}{\ensuremath{2458447.0563_{-0.0023}^{+0.0023}}}
\newcommand{\starBTdurb}{\ensuremath{0.07994_{-0.00088}^{+0.00082}}}
\newcommand{\starBarsb}{\ensuremath{17.88_{-0.32}^{+0.22}}}
\newcommand{\starBrprsb}{\ensuremath{0.05202_{-0.0013}^{+0.0012}}}
\newcommand{\starBbb}{\ensuremath{0.149_{-0.093}^{+0.119}}}
\newcommand{\starBincb}{\ensuremath{89.54_{-0.42}^{+0.68}}}

\newcommand{\starBKrvb}{\ensuremath{<284}}

\newcommand{\starBplradb}{\ensuremath{5.79_{-0.18}^{+0.19}}}
\newcommand{\starBplmassb}{\ensuremath{<2.5}}
\newcommand{\starBab}{\ensuremath{0.0847_{-0.0011}^{+0.0012}}}

\usepackage{placeins}
\usepackage{comment}

\newcommand{\starA}{TOI-251}
\newcommand{\starB}{TOI-942}

\newcommand{\ctbd}[1]{}

\newcommand{\lc}{light curve}

\newcommand{\Lc}{Light curve}

\newcommand{\kms}{\ensuremath{\rm km\,s^{-1}}}
\newcommand{\ms}{\ensuremath{\rm m\,s^{-1}}}

\newcommand{\teff}{\ensuremath{T_{\rm eff}}}

\newcommand{\vsini}{\ensuremath{v \sin{I_\star}}}
\newcommand{\feh}{\ensuremath{\rm [Fe/H]}}

\newcommand{\rsun}{\ensuremath{R_\sun}}
\newcommand{\msun}{\ensuremath{M_\sun}}
\newcommand{\lsun}{\ensuremath{L_\sun}}

\newcommand{\rstar}{\ensuremath{R_\star}}
\newcommand{\mstar}{\ensuremath{M_\star}}
\newcommand{\lstar}{\ensuremath{L_\star}}

\newcommand{\teffstar}{\ensuremath{T_{\rm eff\star}}}

\newcommand{\loggstar}{\ensuremath{\log{g_{\star}}}}

\newcommand{\rpl}{\ensuremath{R_{p}}}
\newcommand{\mpl}{\ensuremath{M_{p}}}

\newcommand{\arstar}{\ensuremath{a/\rstar}}

\graphicspath{{./}{figures/}}

\received{\today}
\submitjournal{AJ}

\shorttitle{Planets around young field stars}
\shortauthors{Zhou et al.}

\begin{document}

\title{Two young planetary systems around field stars with ages between $20-320$\,Myr from \emph{TESS}}

\correspondingauthor{George~Zhou}
\email{george.zhou@cfa.harvard.edu}

\author[0000-0002-4891-3517]{George Zhou}  
\affiliation{Center for Astrophysics \textbar{} Harvard \& Smithsonian, 60 Garden St., Cambridge, MA 02138, USA.}
\affiliation{Hubble Fellow}

\author[0000-0002-8964-8377]{Samuel N. Quinn}  
\affiliation{Center for Astrophysics \textbar{} Harvard \& Smithsonian, 60 Garden St., Cambridge, MA 02138, USA.}

\author{Jonathan Irwin}  
\affiliation{Center for Astrophysics \textbar{} Harvard \& Smithsonian, 60 Garden St., Cambridge, MA 02138, USA.}

\author[0000-0003-0918-7484]{Chelsea X. Huang}  
\affiliation{Department of Physics, and Kavli Institute for Astrophysics and Space Research, Massachusetts Institute of Technology, Cambridge, MA 02139, USA.}

\author[0000-0001-6588-9574]{Karen A.~Collins}  
\affiliation{Center for Astrophysics \textbar{} Harvard \& Smithsonian, 60 Garden St., Cambridge, MA 02138, USA.}

\author{Luke G.~Bouma}  
\affiliation{Department of Astrophysical Sciences, Princeton University, NJ 08544, USA.}

\author{Lamisha Khan}  
\affiliation{Cambridge Rindge and Latin High School}
\affiliation{Center for Astrophysics \textbar{} Harvard \& Smithsonian, 60 Garden St., Cambridge, MA 02138, USA.}

\author{Anaka Landrigan}  
\affiliation{Cambridge Rindge and Latin High School}
\affiliation{Center for Astrophysics \textbar{} Harvard \& Smithsonian, 60 Garden St., Cambridge, MA 02138, USA.}

\author[0000-0001-7246-5438]{Andrew M. Vanderburg}  
\affiliation{Department of Astronomy, The University of Texas at Austin, Austin, TX 78712, USA.}

\author[0000-0001-8812-0565]{Joseph E.~Rodriguez}  
\affiliation{Center for Astrophysics \textbar{} Harvard \& Smithsonian, 60 Garden St., Cambridge, MA 02138, USA.}

\author[0000-0001-9911-7388]{David W.~Latham}  
\affiliation{Center for Astrophysics \textbar{} Harvard \& Smithsonian, 60 Garden St., Cambridge, MA 02138, USA.}

\author{Guillermo Torres}  
\affiliation{Center for Astrophysics \textbar{} Harvard \& Smithsonian, 60 Garden St., Cambridge, MA 02138, USA.}

\author{Stephanie T. Douglas}  
\affiliation{Center for Astrophysics \textbar{} Harvard \& Smithsonian, 60 Garden St., Cambridge, MA 02138, USA.}

\author{Allyson Bieryla}  
\affiliation{Center for Astrophysics \textbar{} Harvard \& Smithsonian, 60 Garden St., Cambridge, MA 02138, USA.}

\author[0000-0002-9789-5474]{Gilbert A. Esquerdo}  
\affiliation{Center for Astrophysics \textbar{} Harvard \& Smithsonian, 60 Garden St., Cambridge, MA 02138, USA.}

\author{Perry Berlind}  
\affiliation{Center for Astrophysics \textbar{} Harvard \& Smithsonian, 60 Garden St., Cambridge, MA 02138, USA.}

\author{Michael L. Calkins}  
\affiliation{Center for Astrophysics \textbar{} Harvard \& Smithsonian, 60 Garden St., Cambridge, MA 02138, USA.}

\author[0000-0003-1605-5666]{Lars A. Buchhave}  
\affiliation{DTU Space, National Space Institute, Technical University of Denmark, Elektrovej 328, DK-2800 Kgs. Lyngby, Denmark}

\author{David Charbonneau}  
\affiliation{Center for Astrophysics \textbar{} Harvard \& Smithsonian, 60 Garden St., Cambridge, MA 02138, USA.}

\author[0000-0003-2781-3207]{Kevin I.\ Collins}  
\affiliation{George Mason University, 4400 University Drive, Fairfax, VA, 22030 USA}

\author[0000-0003-0497-2651]{John F.\ Kielkopf}   
\affiliation{Department of Physics and Astronomy, University of Louisville, Louisville, KY 40292, USA}

\author[0000-0002-4625-7333]{Eric L. N. Jensen}   
\affiliation{Dept.\ of Physics \& Astronomy, Swarthmore College, Swarthmore PA 19081, USA}"

\author[0000-0001-5603-6895]{Thiam-Guan Tan}  
\affiliation{Perth Exoplanet Survey Telescope, Perth, Western Australia}

\author{Rhodes Hart}  
\affiliation{Centre for Astrophysics, University of Southern Queensland, Toowoomba, QLD, 4350, Australia}

\author{Brad Carter}  
\affiliation{Centre for Astrophysics, University of Southern Queensland, Toowoomba, QLD, 4350, Australia}

\author[0000-0003-2163-1437]{Christopher Stockdale}  
\affiliation{Hazelwood Observatory, Victoria, Australia}

\author{Carl Ziegler} 
\affiliation{Dunlap Institute for Astronomy and Astrophysics, University of Toronto, 50 St. George Street, Toronto, Ontario M5S 3H4, Canada}


\author{Nicholas Law} 
\affiliation{Department of Physics and Astronomy, The University of North Carolina at Chapel Hill, Chapel Hill, NC 27599-3255, USA}

\author[0000-0003-3654-1602]{Andrew W. Mann} 
\affiliation{Department of Physics and Astronomy, The University of North Carolina at Chapel Hill, Chapel Hill, NC 27599-3255, USA}

\author[0000-0002-2532-2853]{Steve~B.~Howell}  
\affiliation{NASA Ames Research Center, Moffett Field, CA 94035, USA}

\author[0000-0001-7233-7508]{Rachel A.~Matson} 
\affiliation{NASA Ames Research Center, Moffett Field, CA 94035, USA}
\affiliation{U.S. Naval Observatory, Washington, DC 20392 USA}

\author[0000-0003-1038-9702]{Nicholas J. Scott} 
\affiliation{NASA Ames Research Center, Moffett Field, CA 94035, USA}

\author{Elise Furlan} 
\affiliation{NASA Exoplanet Science Institute, Caltech/IPAC, Mail Code 100-22, 1200 E. California Blvd., Pasadena, CA 91125, USA}

\author{Russel J. White} 
\affiliation{IPAC, Mail Code 314-6, Caltech, 1200 E. California Boulevard, Pasadena, CA 91125, USA}

\author{Coel Hellier}  
\affiliation{Astrophysics Group, Keele University, Staffordshire, ST5 5BG, UK}

\author{David R. Anderson} 
\affiliation{Astrophysics Group, Keele University, Staffordshire, ST5 5BG, UK}
\affiliation{Department of Physics, University of Warwick, Gibbet Hill Road, Coventry CV4 7AL, UK}

\author{Richard G. West}
\affiliation{Department of Physics, University of Warwick, Gibbet Hill Road, Coventry CV4 7AL, UK}


\author{George Ricker} 
\affiliation{Department of Physics, and Kavli Institute for Astrophysics and Space Research, Massachusetts Institute of Technology, Cambridge, MA 02139, USA.}

\author{Roland Vanderspek} 
\affiliation{Department of Physics, and Kavli Institute for Astrophysics and Space Research, Massachusetts Institute of Technology, Cambridge, MA 02139, USA.}

\author[0000-0002-6892-6948]{Sara Seager} 
\affiliation{Department of Physics, and Kavli Institute for Astrophysics and Space Research, Massachusetts Institute of Technology, Cambridge, MA 02139, USA.}
\affiliation{Department of Earth, Atmospheric and Planetary Sciences, Massachusetts Institute of Technology, Cambridge, MA 02139, USA}
\affiliation{Department of Aeronautics and Astronautics, MIT, 77 Massachusetts Avenue, Cambridge, MA 02139, USA}

\author{Jon M. Jenkins} 
\affiliation{NASA Ames Research Center, Moffett Field, CA 94035, USA}

\author[0000-0002-4265-047X]{Joshua N.~Winn} 
\affiliation{Department of Astrophysical Sciences, Princeton University, NJ 08544, USA.}

\author[0000-0002-4510-2268]{Ismael~Mireles} 
\affiliation{Department of Physics and Kavli Institute for Astrophysics and Space Research, Massachusetts Institute of Technology, Cambridge, MA 02139, USA}

\author[0000-0002-4829-7101]{Pamela~Rowden}  
\affiliation{School of Physical Sciences, The Open University, Milton Keynes MK7 6AA, UK}

\author[0000-0003-4755-584X]{Daniel A. Yahalomi}  
\affiliation{Center for Astrophysics \textbar{} Harvard \& Smithsonian, 60 Garden St., Cambridge, MA 02138, USA.}


\author[0000-0002-5402-9613]{Bill~Wohler} 
\affiliation{NASA Ames Research Center, Moffett Field, CA, 94035, USA}
\affiliation{SETI Institute, Mountain View, CA 94043, USA}

\author[0000-0002-9314-960X]{Clara. E. Brasseur} 
\affiliation{Space Telescope Science Institute, USA}

\author{Tansu Daylan}  
\affiliation{Department of Physics, and Kavli Institute for Astrophysics and Space Research, Massachusetts Institute of Technology, Cambridge, MA 02139, USA.}


\author[0000-0001-8020-7121]{Knicole D. Col\'{o}n} 
\affiliation{NASA Goddard Space Flight Center, Exoplanets and Stellar Astrophysics Laboratory (Code 667), Greenbelt, MD 20771, USA}                                              



\begin{abstract}

Planets around young stars trace the early evolution of planetary systems. We report the discovery and validation of two planetary systems with ages $\lesssim 300$\,Myr from observations by the Transiting Exoplanet Survey Satellite. \starA{} is a \starAage{}\,Myr old G star hosting a $\starAplrad{}\,R_\oplus$ mini-Neptune with a \starAperiodshort{}\,day period. \starB{} is a \starBage{}\,Myr old K star hosting a system of inflated Neptune-sized planets, with \starB{}b orbiting with a period of \starBperiodashort{}\,days, with a radius of $\starBplrada{}\,R_\oplus$, and \starB{}c orbiting in a period of \starBperiodbshort{}\,days with a radius of $\starBplradb{}\,R_\oplus$. Though we cannot place either host star into a known stellar association or cluster, we can estimate their ages via their photometric and spectroscopic properties. Both stars exhibit significant photometric variability due to spot modulation, with measured rotation periods of $\sim 3.5$\,days. These stars also exhibit significant chromospheric activity, with age estimates from the chromospheric calcium emission lines and X-ray fluxes matching that estimated from gyrochronology. Both stars also exhibit significant lithium absorption, similar in equivalent width to well-characterized young cluster members. \emph{TESS} has the potential to deliver a population of young planet-bearing field stars, contributing significantly to tracing the properties of planets as a function of their age. 

\end{abstract}



\section{Introduction}
\label{sec:introduction}

The first few hundred million years of planet evolution sculpts the population of exoplanetary systems we observe today. The properties of young planetary systems help unravel the factors that shape the present-day population: in-situ formation, migration, and photo-evaporation. Probing the timescales of these mechanisms motivates us to search for planets within the first few hundred million years of their birth. 

Recent searches for young planets have yielded a handful of discoveries with targeted radial velocity surveys and space-based transit monitoring. Dozens of planets have been identified in the 600-800 Myr old Praesepe and Hyades clusters \citep{2012ApJ...756L..33Q,2014ApJ...787...27Q,2016AJ....152..223O,2016ApJ...818...46M,2018AJ....155...10C,2017AJ....153...64M,2018AJ....155..115L,2018AJ....155....4M,2018AJ....156...46V,2018AJ....156..195R,2019MNRAS.484....8L}. These discoveries enabled estimates of planet occurrence rates in million-year old cluster environments \citep[e.g.][]{2017AJ....154..224R}. 

To sample the earlier stages of planet evolution, transit searches have focused on members of known young stellar associations, finding planets with ages ranging from $\sim 10-150$\,Myr. Observations from the \emph{K2} mission revealed planets within the Upper Scorpius moving group \citep{2016AJ....152...61M,2016Natur.534..658D} and Taurus-Auriga star-forming region \citep{2019AJ....158...79D,2019ApJ...885L..12D}.

With the library of all sky photometry made available by the Transiting Exoplanet Survey Satellite \citep[TESS,][]{2016SPIE.9904E..2BR}, planets around bright, young stars have been identified \citep[e.g.][Plavchan et al. 2020, Newton et al. in-prep]{2019ApJ...880L..17N,2020arXiv200500013R,2020arXiv200500047M}. These are the first planets around bright young stars that are suitable for in-depth characterizations, enabling the first obliquity measurements of newly formed planets \citep{2020ApJ...892L..21Z,2020AJ....159..112M}, as well as atmospheric studies in the near future. 

However, star forming clusters begin to disperse within the first hundred million years of formation \citep{2019ARA&A..57..227K}. Left behind are relatively young stars that can no longer be traced back to their source of origin. These stars can still be identified by their signatures of youth. \citet{2008ApJ...687.1264M} summarizes the activity signatures that can be used to provide approximate ages of Sun-like stars. The rapid rotation of these young stars lead to enhanced chromospheric activity and X-ray emission. Their current rotation rates can help guide the age estimates via gyrochronology. The abundance  of lithium in the atmospheres of main-sequence Sun-like stars is also a useful youth indicator. Recent efforts to trace young field stars have made use of all-sky spectroscopic surveys to catalog large numbers of chromospherically active \citep[e.g.][]{2013ApJ...776..127Z} and lithium bearing \citep[e.g.][]{2019MNRAS.484.4591Z} stars. 

These photometric and spectroscopic signatures of youth, however, are often detrimental to our ability to identify and characterize the planets these stars harbor. The photometric variations due to star spots and stellar rotation make transit planet searches more difficult. The same spot activity induces radial velocity variations on the $\sim 100\,\ms$ level, masking the planetary orbits. Despite these difficulties, \citet{2013ApJ...775...54S} identified Kepler-63\,b as a giant planet around a $\sim 300$\,Myr young field star, and made use of the spot activity to infer its orbital obliquity via transit spot-crossing events. The small planets around K2-233 \citep{2018AJ....155..222D} and EPIC 247267267 \citep{2018AJ....156..302D} are additional examples of young field stars with ages of 100-700\,Myr discovered by the \emph{K2} mission. 

We report the discovery of a mini-Neptune around \starA{}, which we determine to have an age of \starAage{}\,Myr based on its rotational, spectroscopic, and X-ray properties. We also report the two inflated Neptunes around \starB{}, with an age of \starBage{}\,Myr from its rotation, spectroscopic, and X-ray age indicators. These planets were validated by a campaign of ground-based photometric and spectroscopic observations.   In particular, the ability to predict future transit times degraded substantially over the year between the \emph{TESS} discovery and subsequent follow-up efforts. Our ground based photometric follow-up campaign demonstrates the effectiveness of small telescopes in recovering the shallow 1\,mmag transits that these small planets exhibit. These young planets around field stars open an untapped population that can help us construct the properties of planetary systems over time. 

\section{Candidate identification and follow-up observations}
\label{sec:obs}

\subsection{Identification of planet candidates by \emph{TESS}}
\label{sec:tess}

\begin{figure*}
    \centering
    \includegraphics[width=0.8\textwidth]{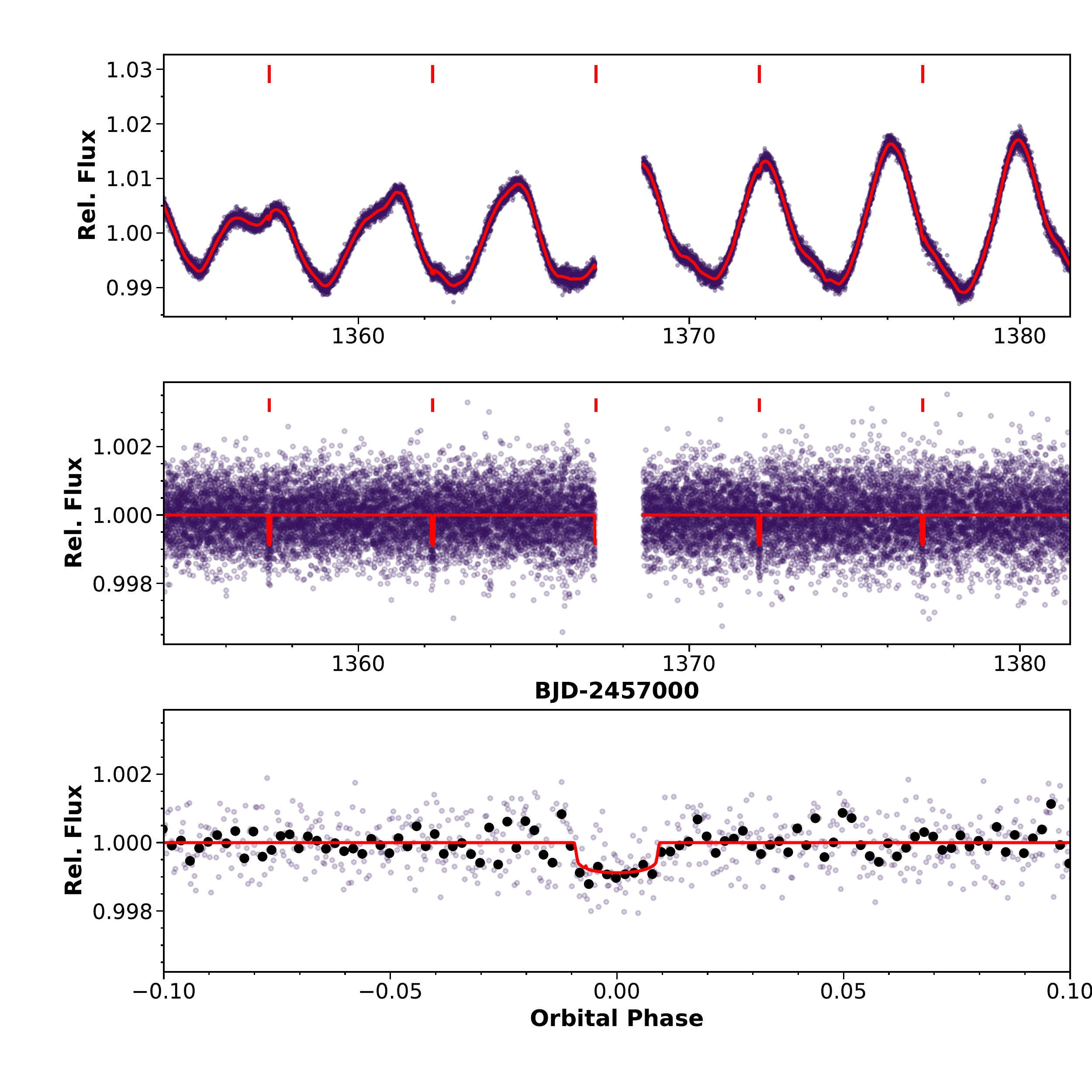} 
    \caption{The full \emph{TESS} light curve of \starA{} during Sector 2 of short cadence observations. The \textbf{top panel} shows the SPOC Simple Aperture Photometry fluxes. The red ticks above the light curve mark the times of transit for \starA{}b. The \textbf{middle panel} shows the light curve after removal of the stellar activity signal as modeled via our global analysis. The \textbf{bottom panel} shows the phased light curve around the transit of \starA{}b. The black points show the binned light curve at phase intervals of 0.002. The best fit models are shown in red in each panel.}
    \label{fig:toi251_tess}
\end{figure*}

\begin{figure*}
    \centering
    \includegraphics[width=0.7\textwidth]{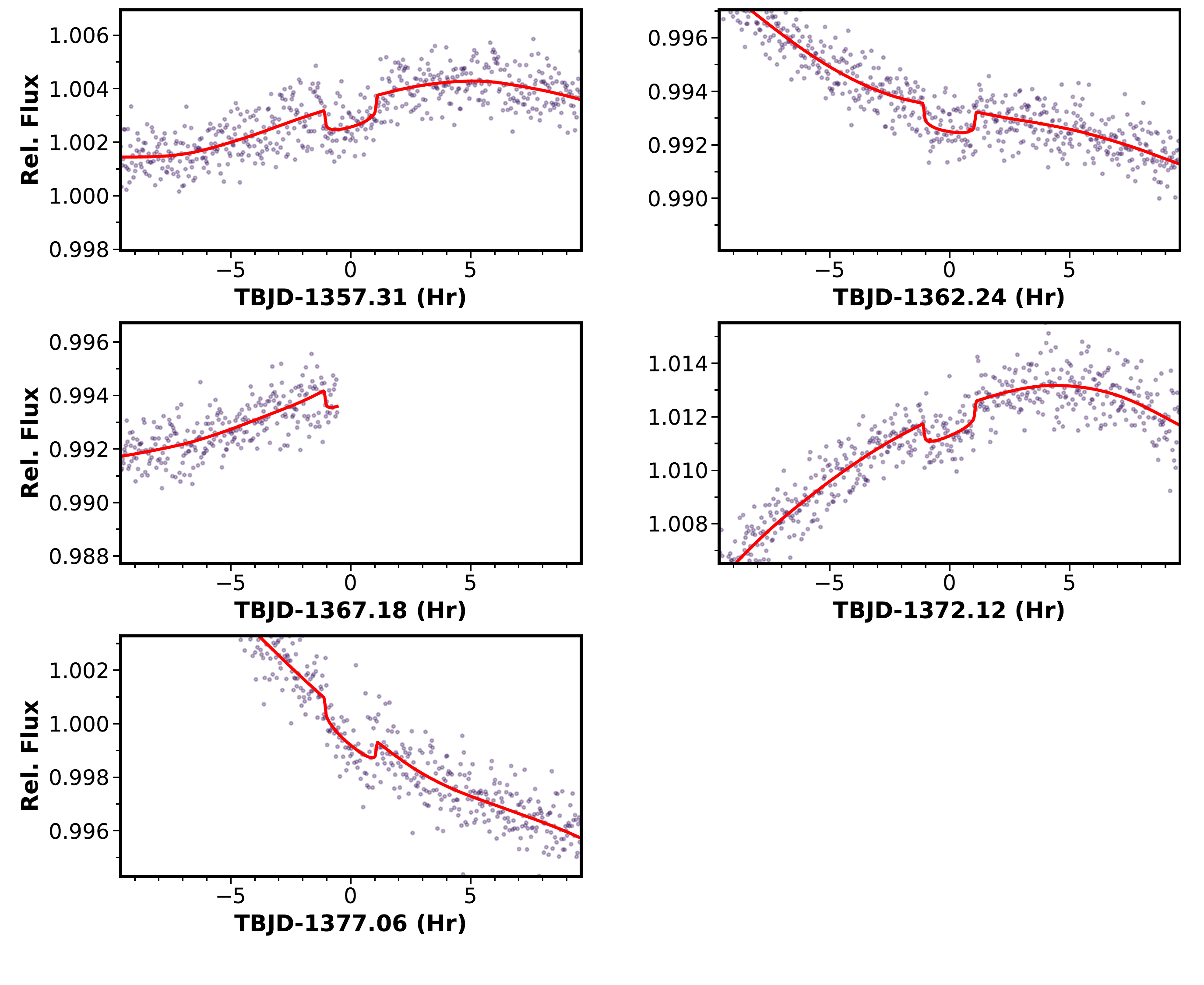} 
    \caption{Individual transits of \starA{}b during the \emph{TESS} observations. Each panel shows an individual transit, with the mid-transit epoch labelled on the X-axis. The best fit model is overlaid in red, showing the transit model and the stellar activity model.}
    \label{fig:toi251_tess_transits}
\end{figure*}

TIC 224225541 received \emph{TESS} 2 minute cadenced target pixel stamp observations, as part of the TESS Candidate Target List \citep[CTL][]{2018AJ....156..102S} during Sector 2, between 2018 August 08 and September 09, with the target star being located on Camera 2, CCD 4 of the \emph{TESS} array. Light curves and transit identification were made possible via the Science Processing Observation Center \citep[SPOC,][]{2016SPIE.9913E..3EJ}. The transits of the mini-Neptune were detected as per \citet{Twicken:DVdiagnostics2018} and \citet{Li:DVmodelFit2019}, with a multiple event statistic of 10.9, and were released to the community as \starA{}b, with an orbital period of \starAperiodshort{} days. We make use of the Simple Aperture Photometry made available for this star for further analysis \citep{twicken:PA2010SPIE,morris2020}. The \emph{TESS} light curve of \starA{} is shown in Figure~\ref{fig:toi251_tess}, and individual transits of \starA{}b in Figure~\ref{fig:toi251_tess_transits}. 

TIC 146520535 was observed during Sector 5 of the primary mission, between 2018 November 15 and December 11, by the Camera 2 CCD 2 Full Frame Images at 30 minute cadence. The MIT quicklook pipeline \citep{2019AAS...23320908H} identified a set of transit events at 4.331 day with signal to pink noise ratio of 11.7, and released as \starB{}b. 
In subsequent visual examinations of the light curves, two single transits of slightly deeper depth than \starB{}b were identified in the light curves spaced \starBperiodbshort{} days apart. 
Further analysis showed that these two single transits are identical in depth and duration, and are due to a second planet candidate \starB{}c. 

\begin{figure*}
    \centering
    \includegraphics[width=0.8\textwidth]{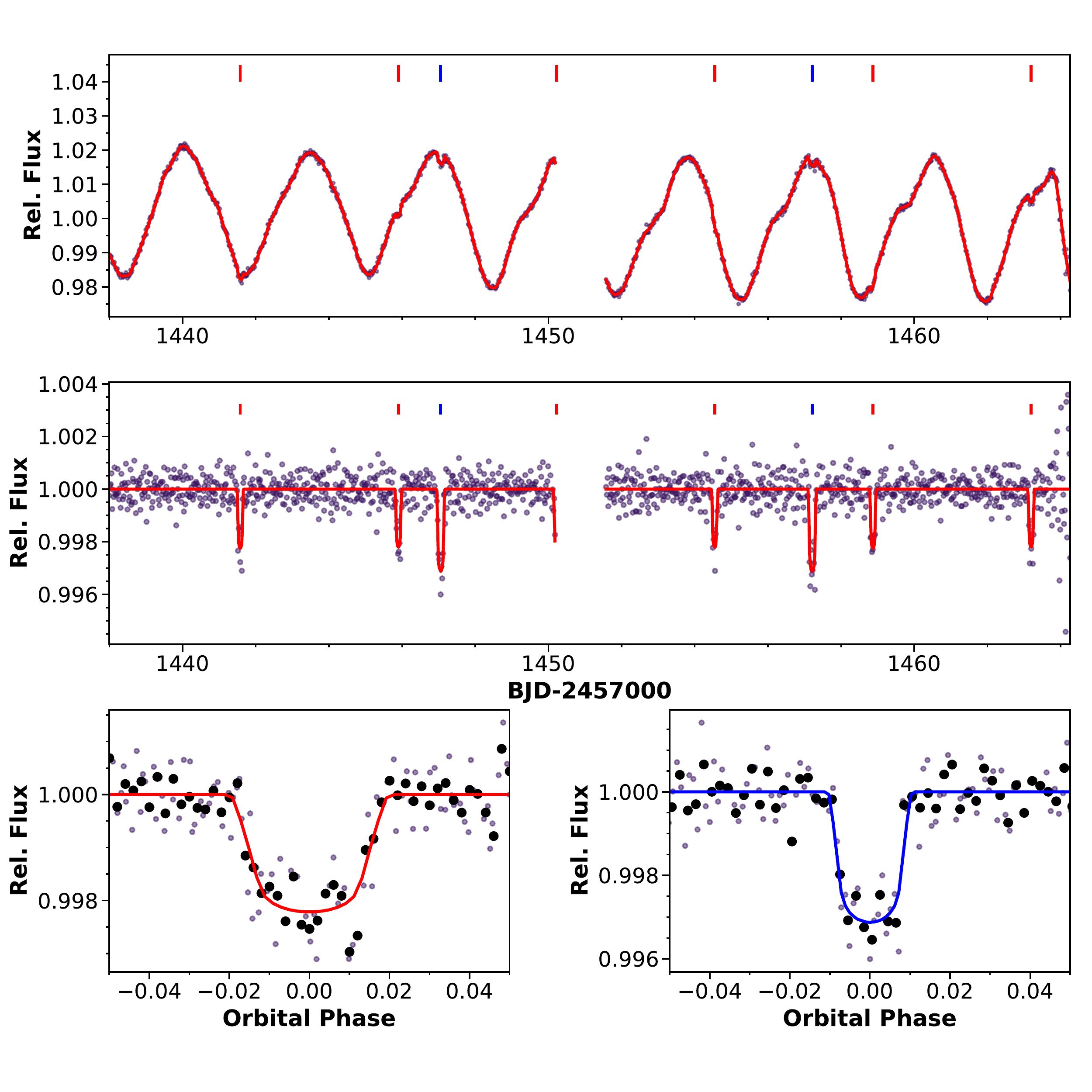} 
    \caption{The \emph{TESS} light curve of \starB{} from 30-minute cadence Full Frame Images during Sector 5 of observations. The panels are arranged as per Figure~\ref{fig:toi251_tess}. The red ticks above the light curve mark the times of transit for \starB{}b, the blue ticks mark the transit times for \starB{}c. The phase folded transit of \starB{}b is shown on the bottom left, and \starB{}c on the bottom right. The black points show the binned light curve at phase intervals of 0.002. }
    \label{fig:toi942_tess}
\end{figure*}

\begin{figure*}
    \centering
    \includegraphics[width=0.7\textwidth]{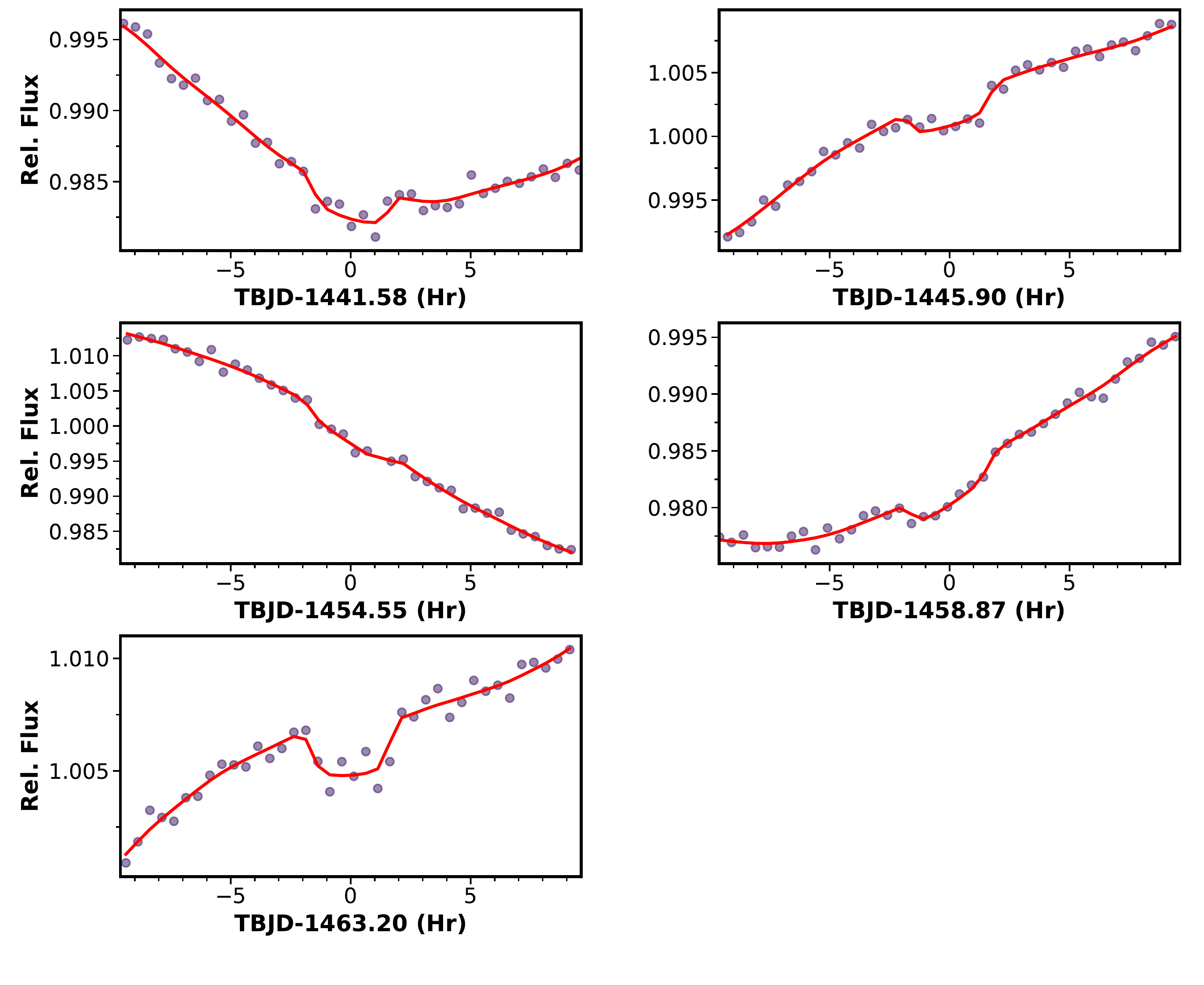} 
    \caption{Individual transits of \starB{}b during the \emph{TESS} observations. Each panel shows an individual transit, with hours from the mid-transit epoch labeled. The best fit model is overlaid in red, showing the transit model and the stellar activity model. Note that the vertical scales differ from transit to transit due to the sharply varying stellar activity at each transit event. }
    \label{fig:toi942_tess_transits1}
\end{figure*}

\begin{figure*}
    \centering
    \includegraphics[width=0.7\textwidth]{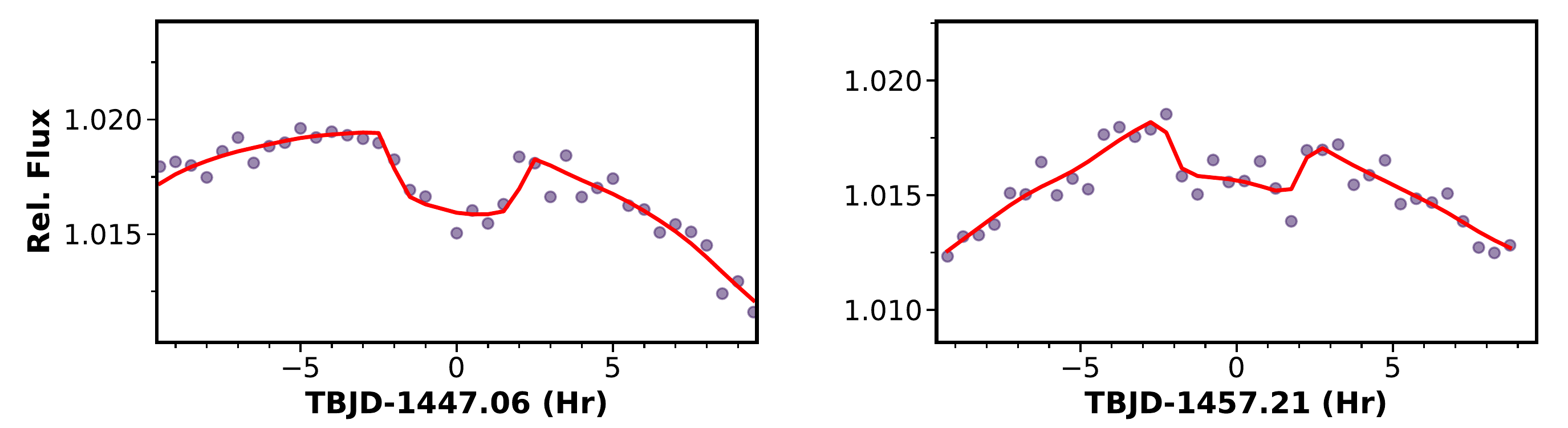} 
    \caption{Individual transits of \starB{}c during the \emph{TESS} observations. Each panel shows an individual transit. The best fit model is overlaid in red, showing the transit model and the stellar activity model.}
    \label{fig:toi942_tess_transits2}
\end{figure*}

We then extracted the light curves for \starB{} from the public Full Frame Images (FFIs) made available on the MAST archive using the \emph{lightkurve} package \citep{lightkurve}. A $10\times10$ pixel FFI cutout was extracted around the target star using the TESScut function \citep{2019ascl.soft05007B}. The photometric aperture was defined to encompass the brightest $68\%$ of pixels, within a maximum radius of 3 pixels from the target star coordinates. Surrounding pixels without nearby stars are used for background subtraction. The \emph{TESS} light curve of \starB{} is shown in Figure~\ref{fig:toi942_tess}, individual transits of \starB{}b and c are shown in Figure~\ref{fig:toi942_tess_transits1} and \ref{fig:toi942_tess_transits2} respectively. 

\subsection{Spectroscopic follow-up}
\label{sec:spec}

\subsubsection{SMARTS 1.5\,m / CHIRON}
\label{sec:chiron}

We obtained a series of spectroscopic observations with the CHIRON facility to characterize the host star properties and constrain the masses of the planets in each system. CHIRON is a high resolution spectrograph on the 1.5\,m SMARTS telescope, located at Cerro Tololo Inter-American Observatory (CTIO), Chile \citep{2013PASP..125.1336T}. CHIRON is fed through an image slicer via a fiber, yielding a spectral resolving power of $\lambda / \Delta \lambda \equiv R = 80,000$ over the wavelength region 4100 to $8700$\,\AA{}. 

\begin{figure}
    \centering
    \includegraphics[width=0.4\textwidth]{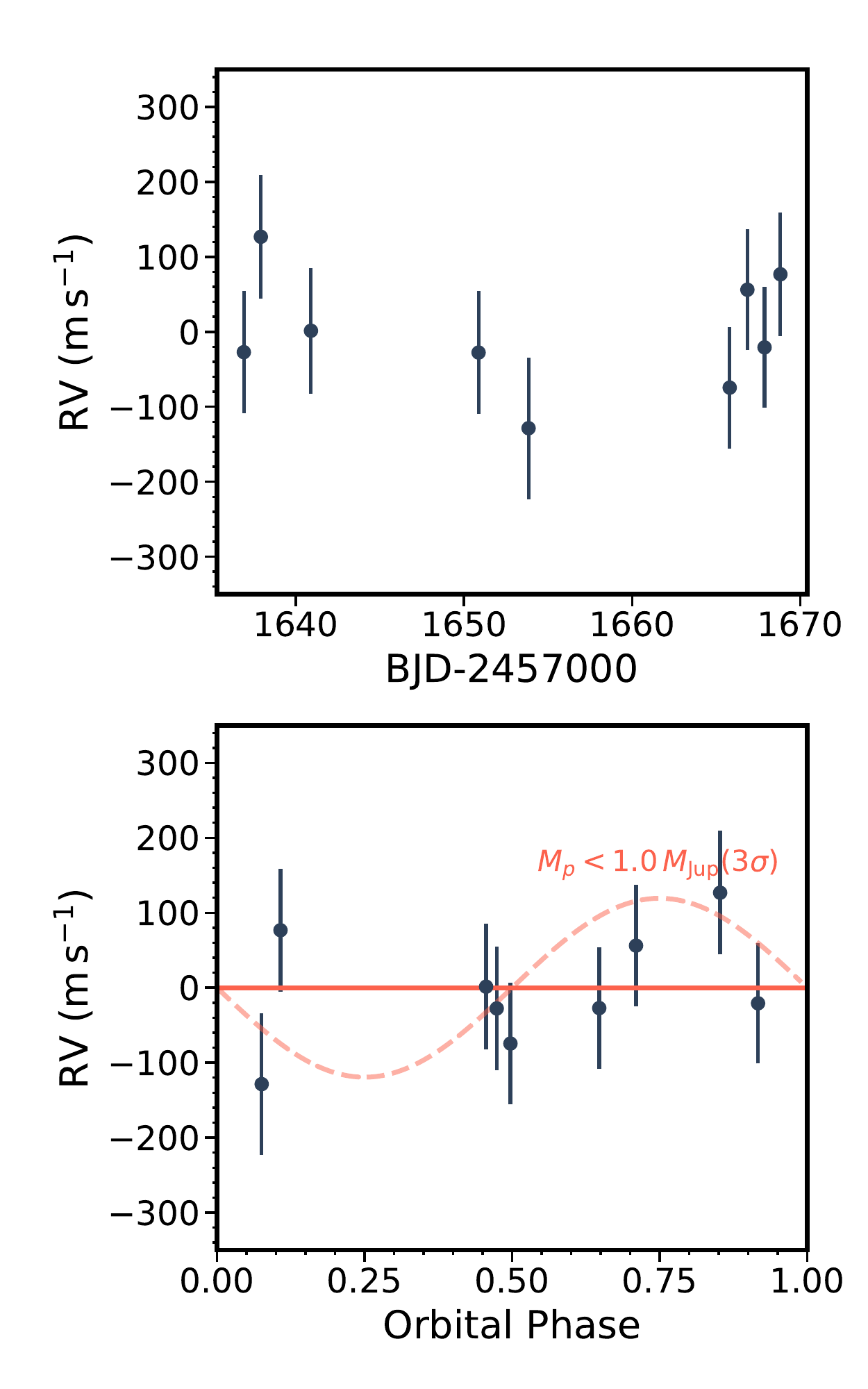} 
    \caption{Radial velocities of \starA{} obtained with CHIRON. \textbf{Top panel} shows the radial velocities as a function of their observation. The plotted uncertainties are the quadrature addition of the per-point measurement uncertainty and stellar jitter. \textbf{Bottom panel} shows the velocities phase folded to the period of \starA{}b, with the transit occurring at phase 0. The dashed radial velocity model show the 3$\sigma$ mass upper limit that we can place on \starA{}b based on these velocities. }
    \label{fig:toi251_rv}
\end{figure}

\begin{figure}
    \centering
    \includegraphics[width=0.4\textwidth]{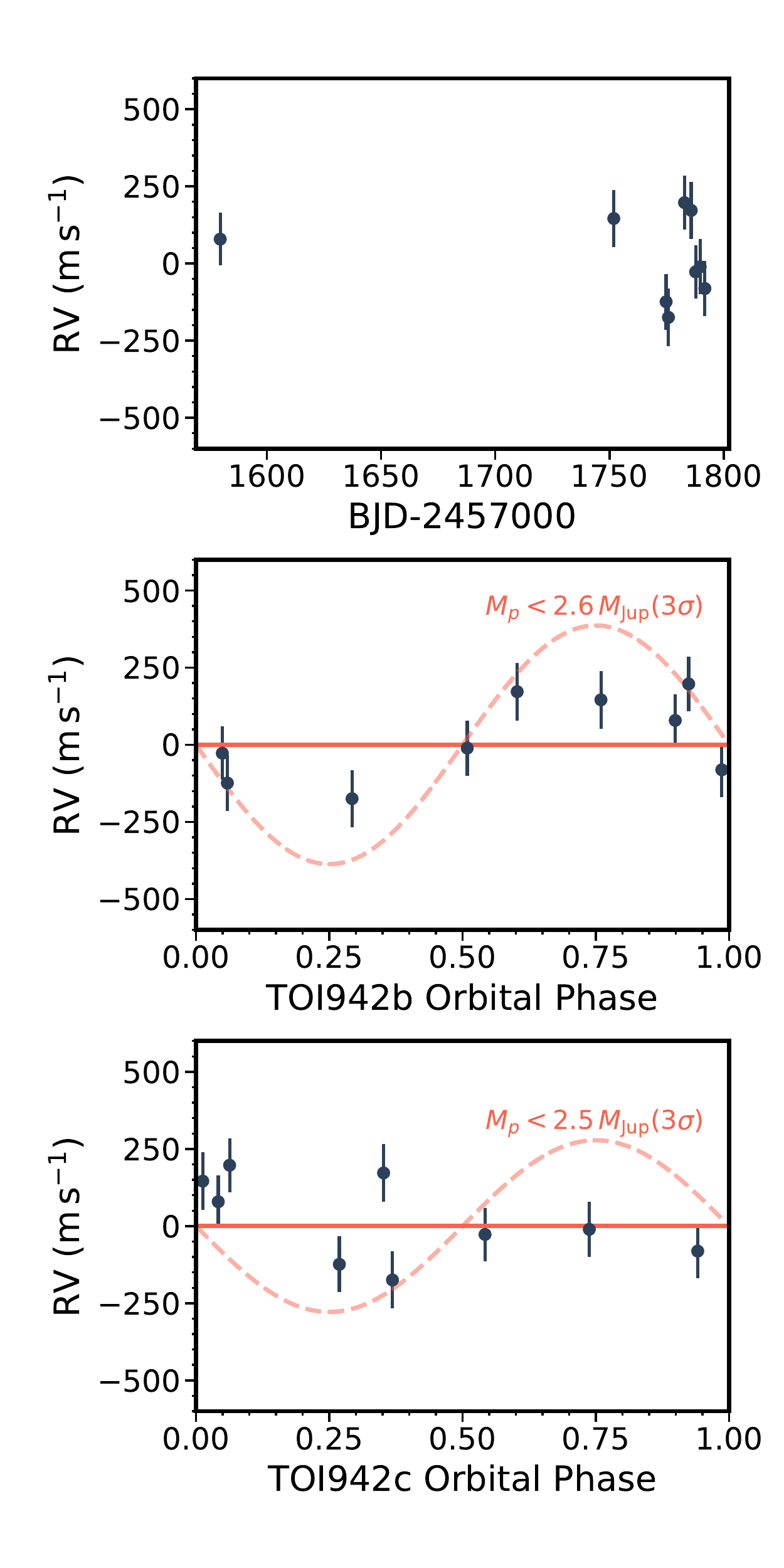} 
    \caption{Radial velocities of \starB{} obtained with CHIRON. \textbf{Top panel} shows the radial velocities as a function of their observation. The plotted uncertainties are the quadrature addition of the per-point measurement uncertainty and stellar jitter. \textbf{Bottom panels} show the velocities phased at the ephemerides of \starB{}b and \starB{}c, with the corresponding 3$\sigma$ upper limit for each planet marked by the dashed curve. Note that while the velocities of \starB{}b appear to phase well with its transit ephemeris, this should be attributed to the stellar activity signal being at a similar timescale to the orbital period. }
    \label{fig:toi942_rv}
\end{figure}

Radial velocities were extracted from CHIRON spectra by fitting the line profiles derived from each spectra. The line profiles are measured via a least-squares deconvolution of the observed spectra against synthetic templates \citep[following][]{1997MNRAS.291..658D}, and are listed in Table~\ref{tab:rv_toi251} and Table~\ref{tab:rv_toi942}, and plotted in Figures~\ref{fig:toi251_rv} and \ref{fig:toi942_rv}. 

\begin{figure}
    \centering
    \textbf{TOI-251}\\
    \includegraphics[width=0.4\textwidth]{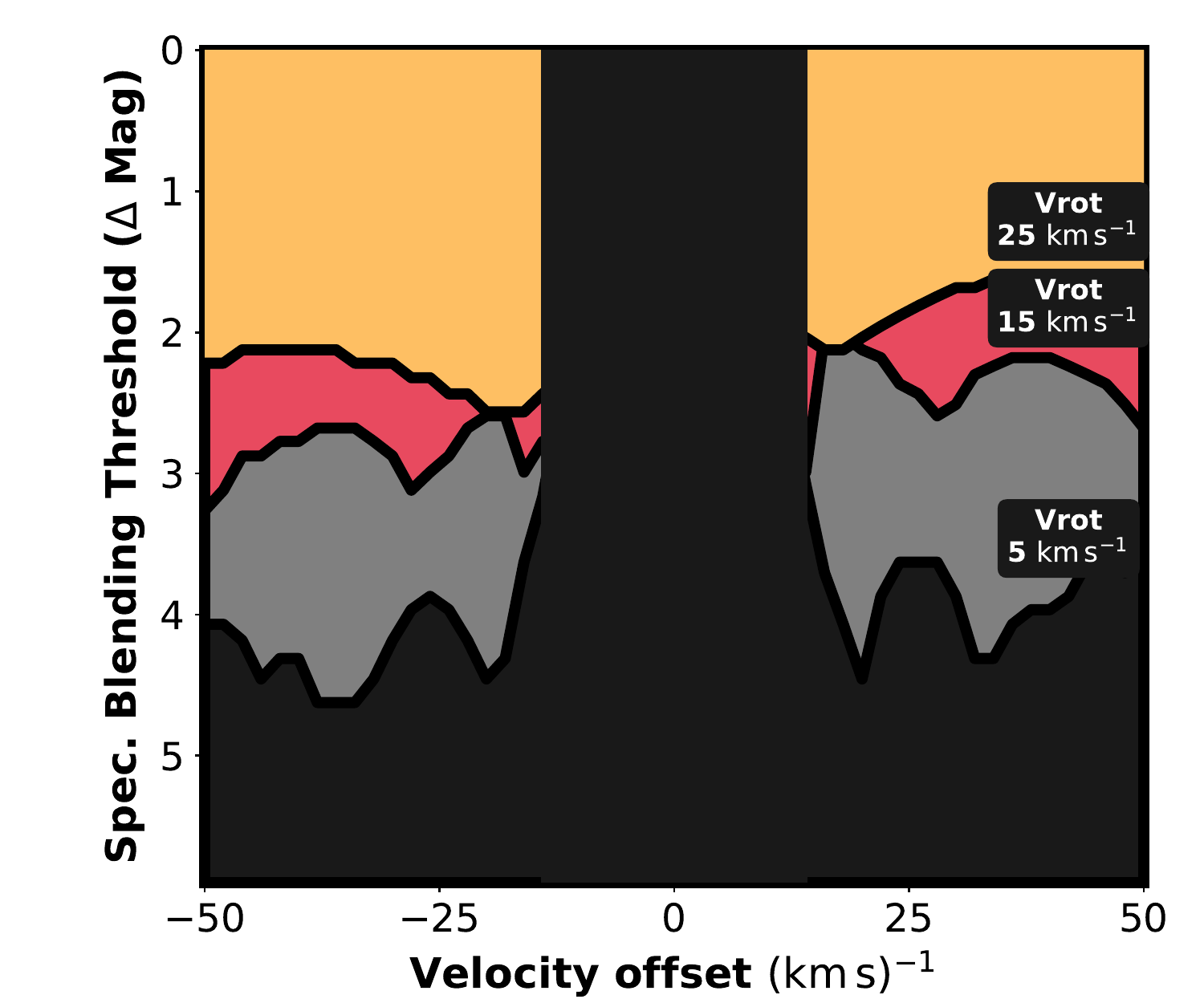} 
    \textbf{TOI-942}\\
        \includegraphics[width=0.4\textwidth]{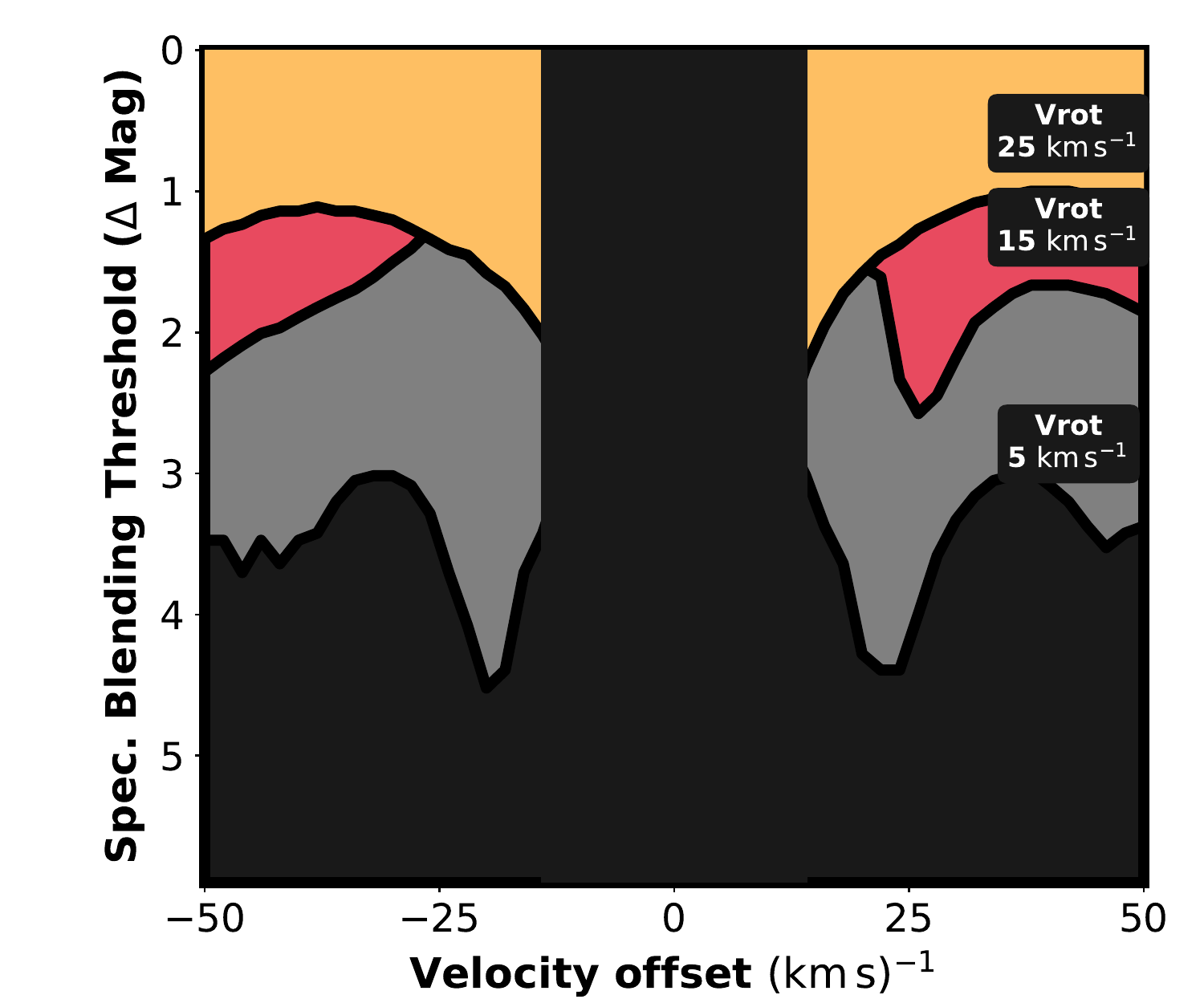} 
    \caption{Detection threshold for any blended spectroscopic companions to the target stars via CHIRON observations. Colored regions show the excluded parameter space for any blended stars, as a function of their magnitude difference to the target star $\Delta \mathrm{mag}$, and their velocity separation. Our ability to detect such companions depend on their rotational line broadening. The simulation shows the detection thresholds for a blended star of $v_\mathrm{rot}$ = 5, 15, and 25 $\mathrm{km\,s}^{-1}$, with the detectability progressively worsening for more rapidly rotating companions. The black regions show where no companions can be detected. For low velocity separations, there is degeneracy between any blended companion and the target star, and as such we are not sensitive to such scenarios. As the broadening profiles have structured noise features, some velocities are systematically more sensitive to secondary companions than other regions. }
    \label{fig:sb2_blend}
\end{figure}

In addition to providing radial velocity and stellar atmosphere parameter measurements, the CHIRON spectra also allow us to search for additional blended spectral companions that may be indicative of other astrophysical false positive scenarios for these systems. We perform a signal injection and recovery exercise to determine the detection thresholds for any additional spectroscopic stellar companion that may be blended in the spectrum. The detectability of the blended source is determined by its flux ratio and velocity separation to the target star, and its rotational broadening. As such, we performed $\sim 10,000$ iterations of the injection, with different combinations of these factors, to the averaged spectroscopic line profile for each target star. The derived detection thresholds are shown in Figure~\ref{fig:sb2_blend}. We can rule out non-associated any stellar companions with $\Delta \mathrm{mag} < 4$ for \starA{}, assuming they exhibit minimal rotational broadening and substantial velocity separation between the target star and the blended companion. Similarly, we can rule out blended, non-associated, slowly rotating stellar companions with $\Delta \mathrm{mag} < 3.5$ for \starB{}. Our ability to detect such companions degrade significantly if they are rapidly rotating, or if they have similar systemic velocities to the target star. 

We also made use of the CHIRON observations to measure the projected rotational velocity $\vsini$ of the host stars. Following \citet{2018AJ....156...93Z}, we model the line profiles derived from the CHIRON spectra via a convolution of kernels representing the rotation, radial-tangential macroturbulence, and instrument broadening terms. The rotational and radial-tangential macroturbulence kernels follow the prescription in \citet{2005oasp.book.....G}, while the instrument broadening kernel is represented by a Gaussian function of width that of the instrument resolution. For \starA{}, we measured a projected rotational broadening velocity of $\vsini = \starAvsini{}\,\kms$, and macroturbulent velocity of $v_\mathrm{mac} = \starAvmac{}\,\kms$. For \starB{}, we measure $\vsini = \starBvsini{}\,\kms$, and $v_\mathrm{mac} = \starBvmac{}\,\kms$. We note that for slowly rotating stars, there is significant degeneracy between various line broadening parameters. These degeneracies can systematically impact future observations, such as transit spectroscopic obliquity observations, or estimates of the line-of-sight inclination of the system. 

\begin{table}
    \centering
    \caption{Radial Velocities for \starA{}}
    \label{tab:rv_toi251}
    \begin{tabular}{cccl}
    \hline\hline
    BJD-TDB & RV $(\kms)$ & RV Error $(\kms)$ & Inst \\
    \hline
2458472.59427 & -2.323 & 0.031 & TRES\vspace{2mm}\\
2458636.91068 & -2.087 & 0.013 & CHIRON \\ 
2458637.92204 & -1.933 & 0.019 & CHIRON \\ 
2458640.90185 & -2.058 & 0.024 & CHIRON \\ 
2458650.86649 & -2.087 & 0.018 & CHIRON \\ 
2458653.83789 & -2.188 & 0.050 & CHIRON \\ 
2458665.79446 & -2.134 & 0.013 & CHIRON \\ 
2458666.84512 & -2.004 & 0.010 & CHIRON \\ 
2458667.86580 & -2.080 & 0.006 & CHIRON \\ 
2458668.80856 & -1.983 & 0.019 & CHIRON \\
\hline
    \end{tabular}
\end{table}

\begin{table}
    \centering
    \caption{Radial Velocities for \starB{}}
    \label{tab:rv_toi942}
    \begin{tabular}{cccl}
    \hline\hline
    BJD-TDB & RV $(\kms)$ & RV Error $(\kms)$ & Inst \\
    \hline
2458579.51343 & 25.400 & 0.024 & CHIRON \\ 
2458751.87828 & 25.467 & 0.043 & CHIRON \\ 
2458774.79336 & 25.197 & 0.038 & CHIRON \\ 
2458775.80474 & 25.147 & 0.043 & CHIRON \\ 
2458782.85938 & 25.518 & 0.031 & CHIRON \\ 
2458785.79240 & 25.493 & 0.044 & CHIRON \\ 
2458787.72494 & 25.294 & 0.026 & CHIRON \\ 
2458789.71153 & 25.310 & 0.036 & CHIRON \\ 
2458791.77641 & 25.240 & 0.033 & CHIRON \vspace{2mm}\\ 
2458775.91595 & 25.181 & 0.267 & TRES \\ 
2458786.92907 & 25.024 & 0.074 & TRES \\
\hline
    \end{tabular}
\end{table}

\subsubsection{FLWO 1.5\,m / TRES}
\label{sec:tres}

We also obtained observations of \starA{} and \starB{} with the Tillinghast Reflect Echelle Spectrograph \citep[TRES,][]{Furesz:2008} on the 1.5\,m reflector at the Fred Lawrence Whipple Observatory (FLWO), Arizona, US. TRES is a fiber-fed echelle spectrograph with a resolution of $R\sim 44000$ over the spectral region of $3850-9100\,\AA$. The observing strategy and reduction procedure are outlined in \citet{2012Natur.486..375B}. 

One observation was obtained for \starA{}, and two for \starB{}. These spectra were used to measure the stellar atmospheric properties of the host stars via the Stellar Parameter Classification pipeline \citep{2010ApJ...720.1118B}, subsequently used as priors in our global analyses (Section~\ref{sec:modelling}). We find \starA{} to be a Sun-like stellar surface effective temperature of $\teff = 5833 \pm 143$\,K, surface gravity of $\log g = 4.49 \pm 0.23$\,dex, and metallicity of $\mathrm{[m/H]} = -0.16 \pm 0.13$\,dex. \starB{} is an early K star with $\teff = 5187 \pm 52$\,K, $\log g = 4.59 \pm 0.10$\,dex, and $\mathrm{[m/H]}=0.07 \pm 0.08$\,dex.

\subsection{Photometric follow-up}
\label{sec:photfu}

\begin{deluxetable*}{lllrrr}
\tablewidth{0pc}
\tabletypesize{\scriptsize}
\tablecaption{
    Summary of photometric observations
    \label{tab:photobs}
}
\tablehead{
    &
    &
    &
    &
    &\\
    \multicolumn{1}{c}{Target}          &
    \multicolumn{1}{c}{Facility}          &
    \multicolumn{1}{c}{Date(s)}             &
    \multicolumn{1}{c}{Number of Images}\tablenotemark{a}      &
    \multicolumn{1}{c}{Cadence (s)}\tablenotemark{b}         &
    \multicolumn{1}{c}{Filter}            
}
\startdata
\starA{} & WASP & 2006-05-15 -- 2014-12-06 & 132002 & 37 & WASP \\
&& (Over 8 observing campaigns) &&&\\
\starA{} & \emph{TESS} & 2018-08-23 -- 2018-09-20 & 18316 & 120 & $TESS$ \\
\starA{} & LCO-SSO & 2019-07-23 & 148 & 127 & $Y$\\
\starA{} & MKO & 2019-07-28 & 168 & 89 & $i'$\\
\starA{} & MEarth & 2019-08-22 & 1749 & 12 & MEarth band \\
[1.5ex]
\starB{} & WASP & 2006-09-15 -- 2014-12-09  & 87361 & 30 & WASP \\
&& (Over 10 observing campaigns) &&&\\
\starB{} & \emph{TESS} & 2018-11-15 -- 2018-12-11 & 1188 & 1799 & $TESS$ \\
\starB{} & MEarth & 2019-11-05 & 1680 & 10 & MEarth band \\
\starB{} & MEarth & 2019-11-13 & 1791 & 10 & MEarth band \\
\starB{} & MEarth & 2019-11-15 & 1814 & 11 & MEarth band \\
\starB{} & MEarth & 2019-11-17 & 2346 & 6 & MEarth band \\
\starB{} & MEarth & 2019-11-25 & 1974 & 10 & MEarth band \\
\starB{} & MEarth & 2019-11-30 & 2049 & 11 & MEarth band \\
\starB{} & MEarth & 2019-12-05 & 2488 & 10 & MEarth band \\
\starB{} & MEarth & 2019-12-26 & 2223 & 5 & MEarth band \\
\enddata 
\tablenotetext{a}{
  Outlying exposures have been discarded.
}
\tablenotetext{b}{
  Median time difference between points in the \lc. Uniform sampling was not possible due to visibility, weather, pauses.
}
\end{deluxetable*}

We obtained a series of ground-based photometric follow-up observations to confirm that the transits are on-target, eliminate false-positive scenarios, and refine the ephemeris and orbital parameters. 

Full and partial transits of \starA{}b, \starB{}b, and \starB{}c were obtained by an array of ground-based observatories. In particular, with only two transits in the \emph{TESS} observations, the ephemeris uncertainty for \starB{}c was as large as 5 hours, whilst the uncertainties for the shorter period planets \starA{}b and \starB{}b were 2 hours. The first attempts at recovering the transits of both planets failed due to these large uncertainties, but were still useful in refining the transit predictions. Subsequent attempts fortuitously captured partial transits, leading to the recovery of the ephemeris.  

\subsubsection{Las Cumbres Observatory}

 Las Cumbres Observatory Global Telescope (LCOGT) is a global network of small robotic telescopes \citep{2013PASP..125.1031B}. A full transit of \starA{}b was observed by the 1\,m LCOGT telescope located at Siding Spring Observatory, Australia. The observations are scheduled via the {\tt TESS Transit Finder}, which is a customized version of the {\tt Tapir} software package \citep{Jensen:2013}. The observations were obtained with the Sinistro camera, in the $Y$ band on the night of 2019-07-23. The LCOGT images were calibrated by the standard LCOGT BANZAI pipeline \citep{McCully:2018} and the photometric data were extracted using the {\tt AstroImageJ} ({\tt AIJ}) software package \citep{Collins:2017}, showing a likely detection of the 1\,mmag transit event. Nearby stars were cleared for signs of eclipsing binaries, showing that the transit event is on target to within seeing limits. The observation is shown in Figure~\ref{fig:toi251_fu}.
 
\subsubsection{Mt. Kent Observatory}

A full transit of \starA{}b was observed with the University of Louisville Research Telescope at Mt. Kent Observatory, Queensland, Australia. The University of Louisville Research Telescope is a PlaneWave Instruments CDK700 0.7\,m telescope equipped with a 4\,K$\times$4\,K detector. The observation on 2019-07-28 was obtained in the $i'$ band, spanning four hours, with an average exposure time of $\sim 90\,s$. Photometry was extracted using AstroImageJ \citep{2017AJ....153...77C}, showing a successful detection of the transit event. The nearby stars were again checked for signatures of eclipsing binaries, with none detected.

\subsubsection{MEarth}

The MEarth instruments are described in detail by
\citet{2015csss...18..767I} and for brevity we do not repeat those
details here.

For the majority of the MEarth observations listed in Table~\ref{tab:photobs}, we
adopted a standard observational strategy used for bright TOIs where
all but one of the available telescopes at a given observing site are
operated defocused to obtain photometry of the target star, and one
telescope observes in focus with the target star saturated to obtain
photometry of any nearby contaminating stars not properly resolved in
the defocused observations.  Once these stars had been fully ruled out
as the source of the transits, we simply used all telescopes in
defocused mode to obtain a slight improvement in sampling.  This was
done for the observation of \starA{} and for the last two observations
of \starB{} (2019-12-05 and 2019-12-26) only.

With the exception of the observation of \starA{} on 2019-11-17 which
was gathered from both MEarth-North and MEarth-South simultaneously
(using a total of 15 telescopes, 8 at MEarth-North and 7 at
MEarth-South), all other observations were gathered from MEarth-South
using 7 telescopes.  For \starB{} exposure times were 60s on all
telescopes and the defocused telescopes used a half flux diameter (HFD)
of 12 pixels.  For \starB{} we used a defocus setting of 6 pixels HFD
and an exposure time of 60s for the defocus telescopes and 30s for the
single in-focus telescope at MEarth-South, and 8 pixels HFD with all
telescopes using 60s exposure times at MEarth-North.  These
instruments use a different model of CCD so have different pixel
scales (0.84 arcsec/sec at MEarth-South and 0.76 arcsec/sec at
MEarth-North) and require slightly different observational setups to
achieve the same saturation limit.

Data were gathered continuously subject to twilight, zenith distance,
and weather constraints.  Telescope 7 at MEarth-South used in the
defocus ensemble had a shutter stuck in the open position for all
observations but this does not appear to affect the light curves
despite visible smearing during readout.  All data were reduced
following standard procedures detailed in
\citet{2007MNRAS.375.1449I,2012AJ....144..145B}.  Photometric aperture
radii for extraction of the defocus time series of the target star
were 24 pixels for \starA{}, 9.9 pixels for \starB{} at MEarth-South,
and 12.7 pixels for \starA{} at MEarth-North.

\begin{figure}
    \centering
    \includegraphics[width=0.4\textwidth]{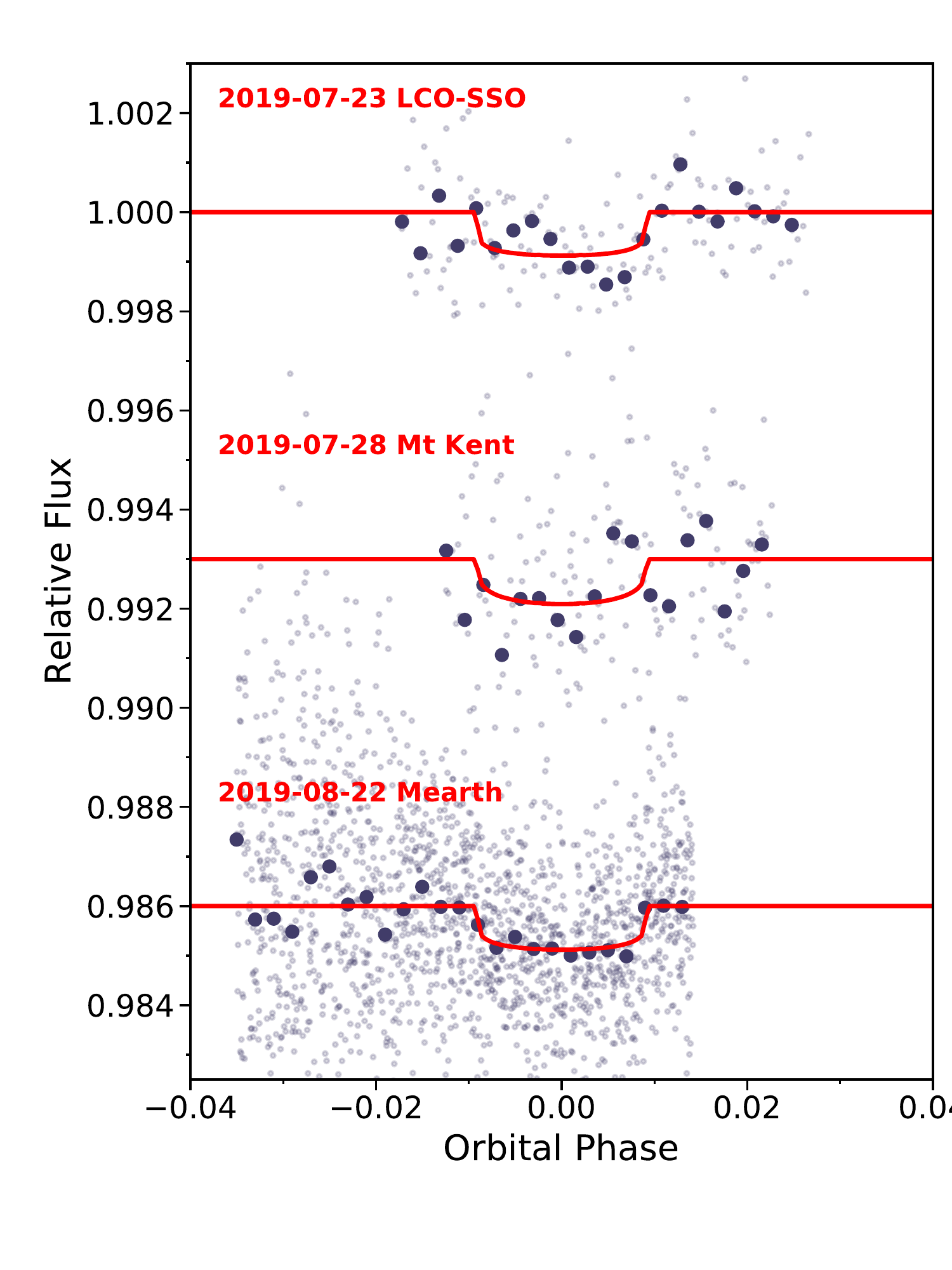} 
    \caption{Ground based follow-up light curves of \starA{}b. The light curves from each facility are shown with an arbitrary vertical offset applied. The solid points mark the binned fluxes at phase intervals of 0.002. The observatory, date, and filter of each observation are labelled. Observations by the MEarth observatory were taken in the MEarth band.}
    \label{fig:toi251_fu}
\end{figure}

\begin{figure*}
    \centering
    \begin{tabular}{cc}
            \includegraphics[width=0.4\textwidth]{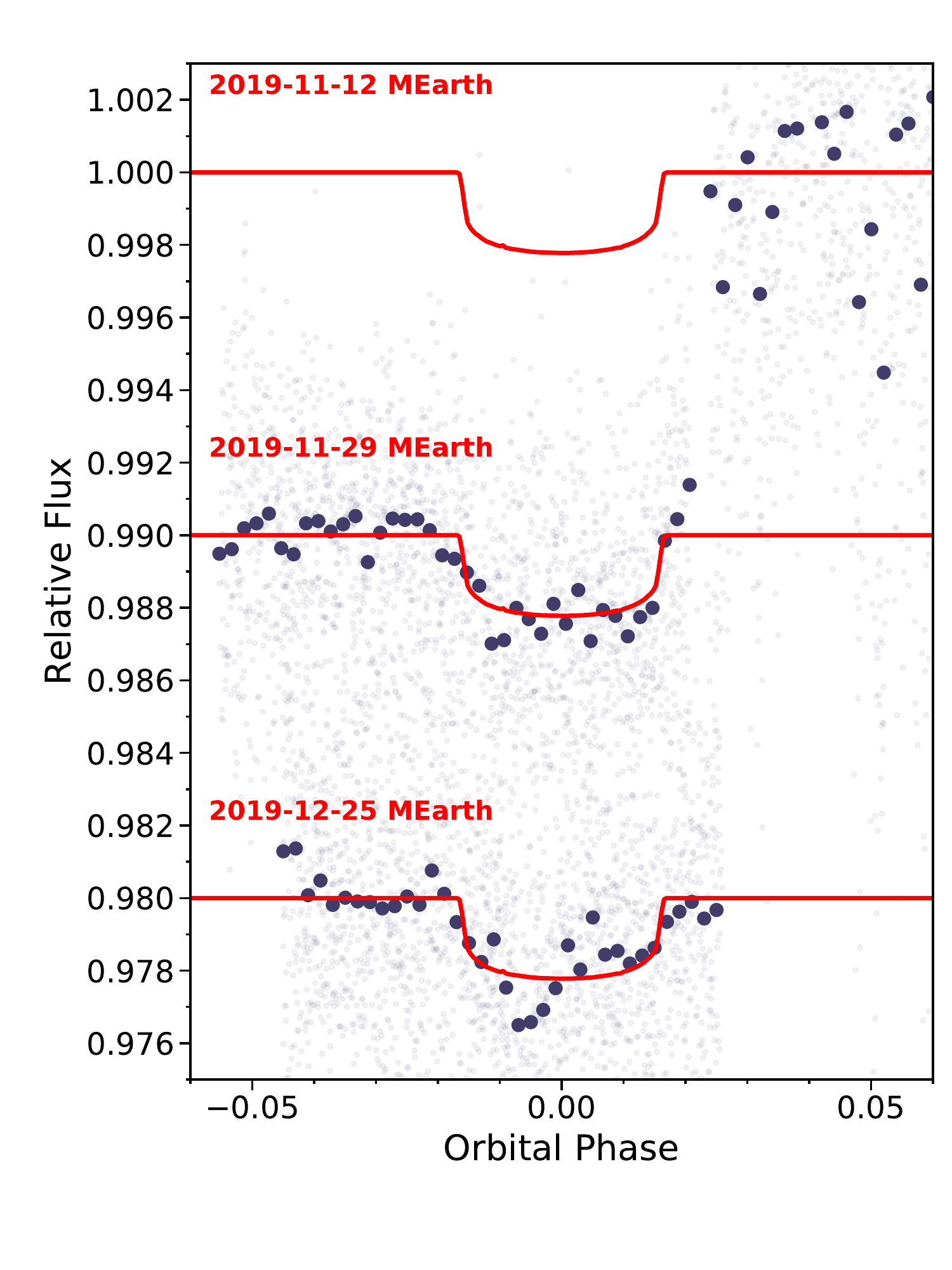} &
            \includegraphics[width=0.4\textwidth]{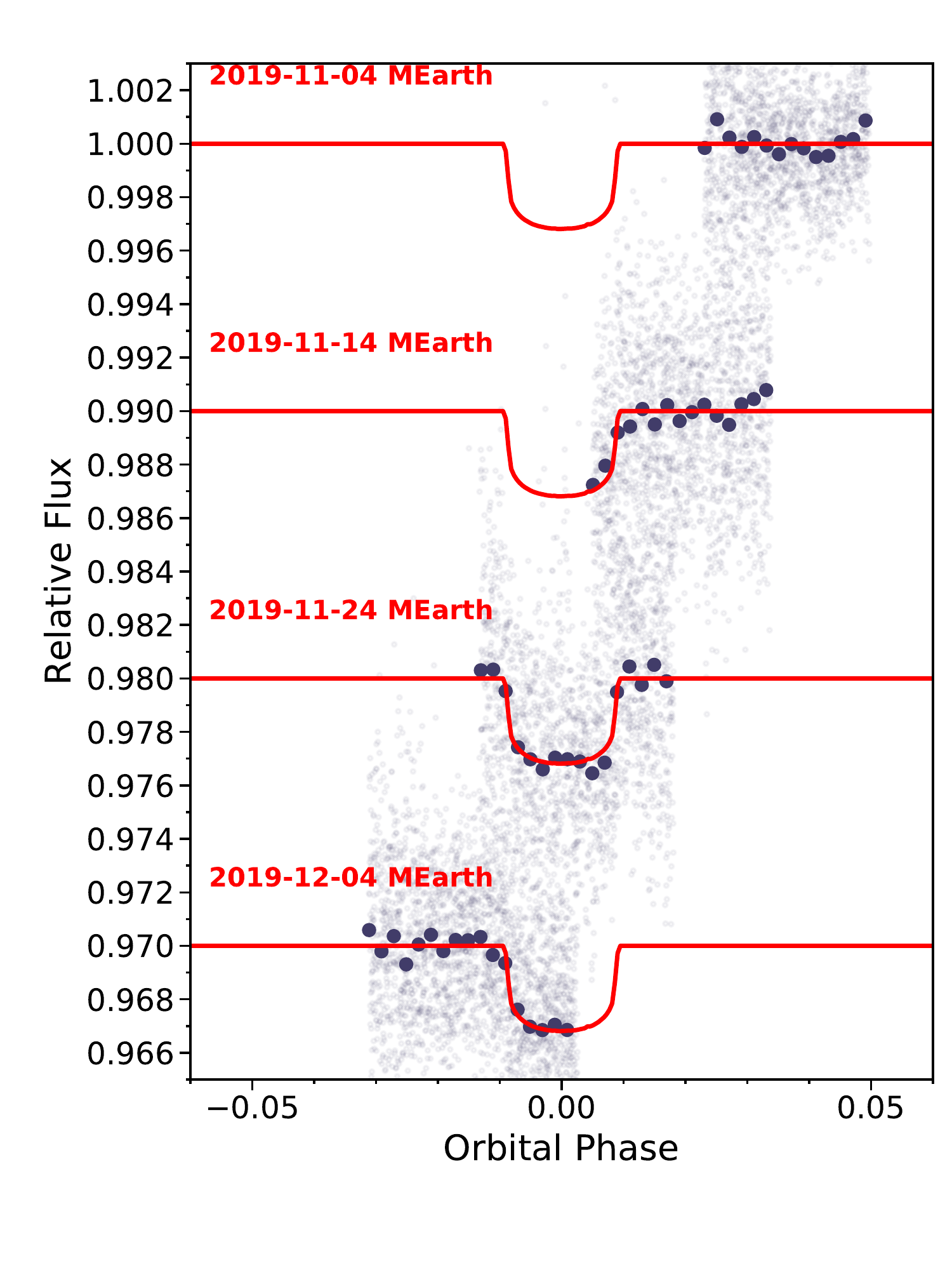} \\
    \end{tabular}
    \caption{Ground based follow-up light curve of \starB{}b \textbf{Left} and \starB{}c \textbf{Right} are shown. The light curves are gathered by the MEarth observatory. }
    \label{fig:toi942_fu}
\end{figure*}

\subsubsection{WASP archival observations}

\starA{} and \starB{} also received multi-season monitoring by ground-based transit surveys. The Wide Angle Search for Planets (WASP) Consortium \citep{2006PASP..118.1407P} observed both target stars with the Southern SuperWASP facility, located at the Sutherland Station of the South African Astronomical Observatory. Each SuperWASP station consists of arrays of eight commonly mounted 200\,mm f/1.8 Canon telephoto lenses, equipped with a $2\mathrm{K} \times 2\mathrm{K}$ detector yielding a field of view of $7.8^\circ \times 7.8^\circ$ per camera. 

\starA{} and \starB{} received eight years of WASP observations, spanning 2006 to 2014. The $\sim 2\%$ spot modulation signal is clear in the WASP datasets for both host stars. We made use of these observations in Section~\ref{sec:age} to confirm that the rotation period we measure from \emph{TESS} is robust and consistent with that seen over the significantly longer timescales of the WASP observations. 

\subsection{High spatial resolution imaging}
\label{sec:AO}

We obtained a series of high spatial resolution imagery of the target stars to check for blended nearby stellar companions. Such companions can be the source of false positive signals due to stellar eclipsing binaries, or dilute the planetary transit signal leading to systematically smaller planet radius measurements. 

\begin{figure*}
    \centering
    \begin{tabular}{cc}
        \includegraphics[width=0.4\linewidth]{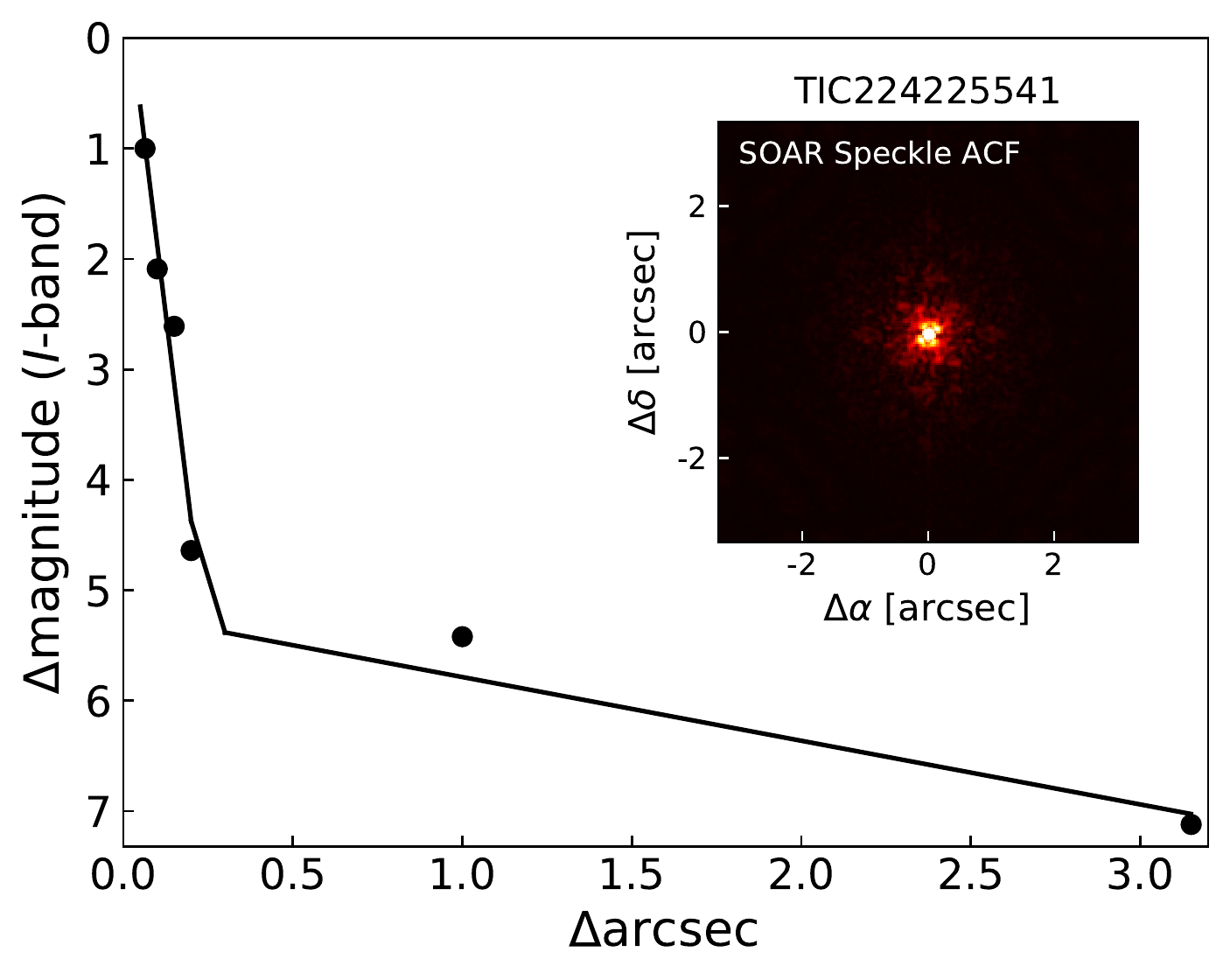} & 
        \includegraphics[width=0.4\linewidth]{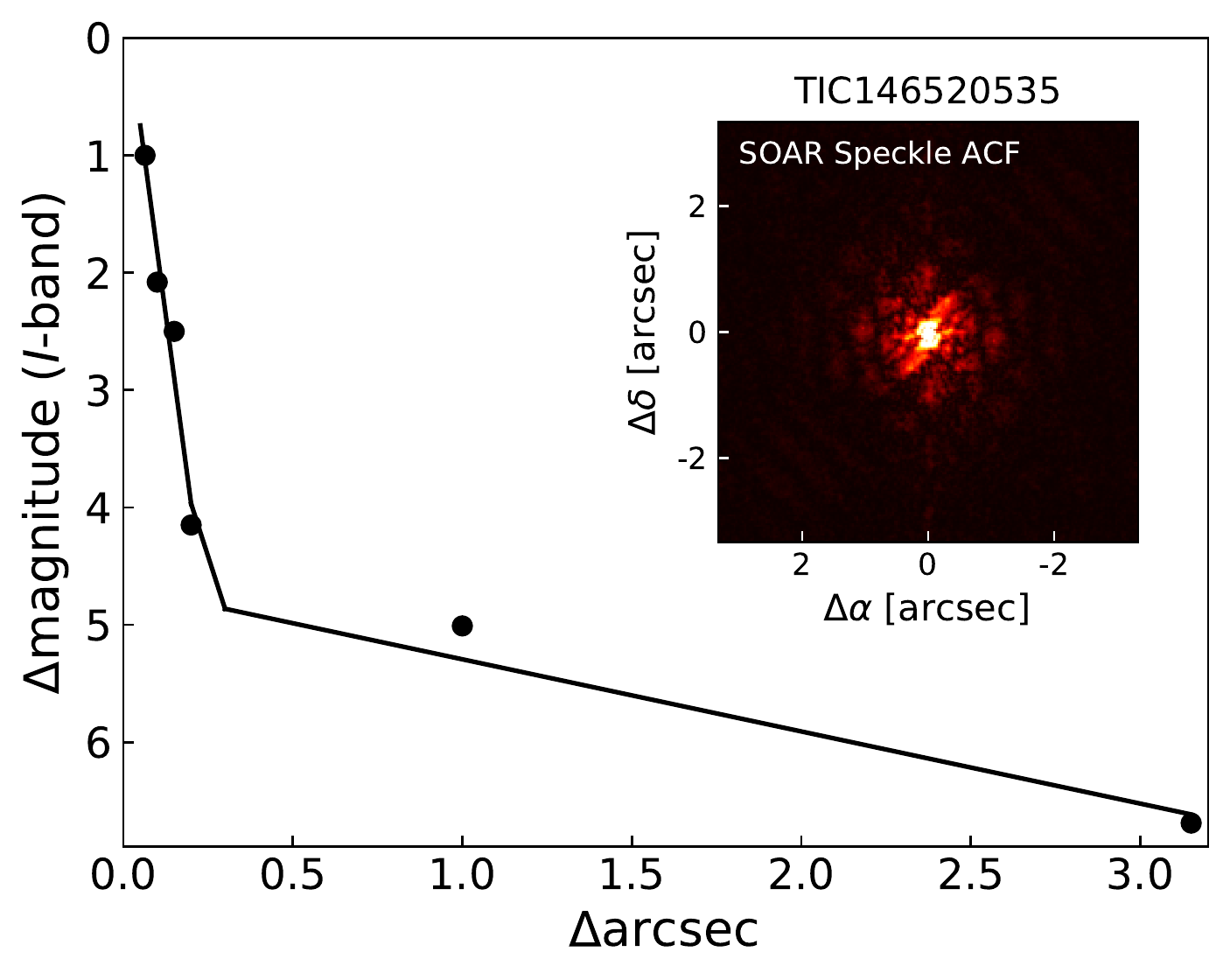}\\
    \end{tabular}
    \caption{SOAR speckle observations of \starA{} (left) and \starB{} (right). The $5\sigma$ detection sensitivity to companions are shown in each figure. The inserts show the auto correlation functions (ACF) for each observation. No stellar companions were detected near either target star.}
    \label{fig:hr_imaging}
\end{figure*}

Observations of \starA{} and \starB{} were obtained as part of the Southern Astrophysical Research (SOAR) \emph{TESS} survey \citep{2020AJ....159...19Z}. Speckle imaging observations were obtained with the Andor iXon-888 camera on the 4.1\,m SOAR telescope. Observations of \starA{} were obtained on 2019 May 18, and \starB{} on 2019 November 19. Each target star observation involved 400 frames of $200\times200$ binned pixels about the target, with a pixel scale of $0\farcs01575$, obtained over the course of 11\,s. The observations were reduced as per \citet{2018PASP..130c5002T}. The speckle auto cross correlation functions from these observations are shown in Figure~\ref{fig:hr_imaging}. No stellar companions were detected for either target star. 

\begin{figure}
    \centering
        \includegraphics[width=1\linewidth]{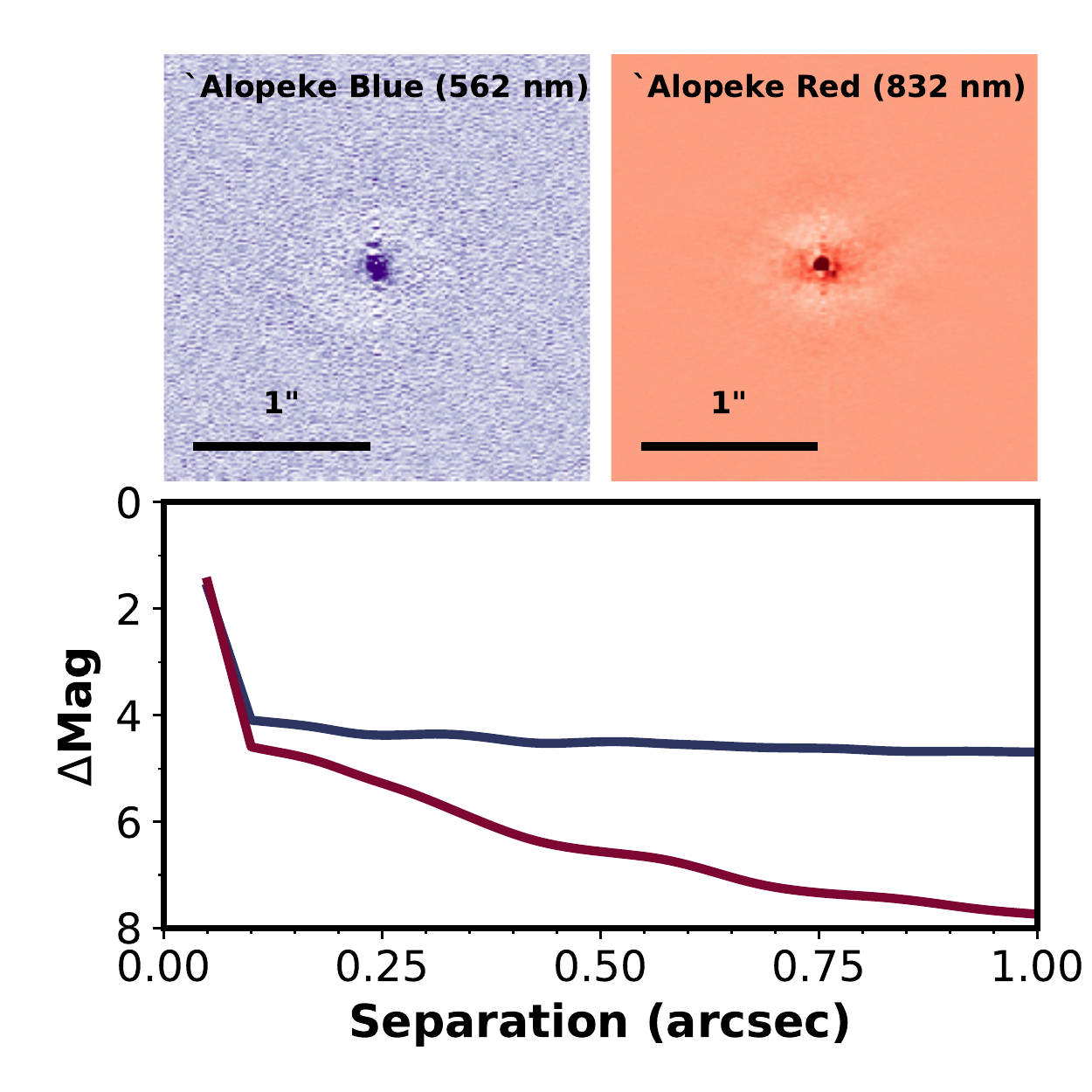}\\
    \caption{Gemini-N `Alopeke speckle images of \starB{}, in the blue (top left) and red (top right) channels. (Lower panel) The corresponding limiting magnitudes for companions as a function of angular distance to the target star.  }
    \label{fig:alopeke_imaging}
\end{figure}

We also obtained speckle observations of \starB{} with $'$Alopeke at the 8\,m Gemini-North observatory, located on Mauna Kea, Hawaii. The observation, obtained on 2019 October 14, incorporates a 1-minute integration involving 1000 60\,ms exposures of $256\times256$ pixel subarrays about the target star. These observations have resulting spatial resolutions of $0.016\arcsec$ FWHM in the blue, and $0.025\arcsec$ in the red, yielding an inner working angle of $\sim$3 AU at the distance to \starB{}. The analysis and detection limits were derived as per \citet{2011AJ....142...19H} and \citet{2016ApJ...829L...2H}. The speckle images and limits on companions are shown in Figure~\ref{fig:alopeke_imaging}. Note that the observations were obtained at relatively high airmass, and as such the blue reconstructed image is adversely affected. No stellar companions are detected in either channel. 

\section{Elimination of false positive scenarios}
\label{sec:fp}

A number of astrophysical false positive scenarios can imitate the transit signals of planetary systems. Eclipsing binaries in grazing geometries, or nearby faint eclipsing binaries can exhibit transits of depths similar to that of transiting super-Earths and Neptunes in extreme scenarios.

The possibility that our target stars are actually grazing eclipsing binaries is extremely unlikely given their well-resolved box-shaped transits. Nevertheless we obtained a series of radial velocity observations for both target stars. Radial velocity variation at the $>1\,\kms$ level can be indicative of the companion being of stellar mass, or the target star being spectroscopically blended with a background eclipsing binary. 

We obtained nine spectra of each target star, and derived radial velocities from each via a least-squares deconvolution analysis (Section~\ref{sec:chiron}). As expected for young stars, these velocities exhibited significant astrophysical jitter beyond their velocity uncertainties.  The mean uncertainty of the velocities for \starA{} is $17\,\ms$, while the velocity scatter is significantly higher at $73\,\ms$. Similarly, the mean uncertainty in the velocities of \starB{} is $35\,\ms$, but the velocity scatter is at $127\,\ms$. 

Nevertheless, the lack of a detectable radial velocity orbit to within $1\,\kms$ for both target stars rules out the possibility that they are orbited by stellar companions. In section~\ref{sec:modelling}, we derive upper limits to the masses of any orbiting companions around our target stars. The velocities of each system are modeled assuming circular orbits for each planet; the stellar activity is accounted for via a jitter term. We make no attempts at correcting for the stellar jitter via decorrelation against stellar activity indicators, nor make use of Gaussian processes to model possible rotational signals in the velocities. Through this simple exercise, we find $3\sigma$ mass upper limits of $\starAplmass{}\,M_\mathrm{Jup}$ for \starA{}b, $\starBplmassa{}\,M_\mathrm{Jup}$ for \starB{}b, and  $\starBplmassb{}\,M_\mathrm{Jup}$ for \starB{}c. Future examination making use of stellar activity markers can further refine the masses of these planets. 

The $22\arcsec\,\mathrm{pixel}^{-1}$ plate scale of \emph{TESS} often makes it difficult to distinguish the true source of transit signals in crowded fields. Fainter eclipsing binaries whose depths are diluted by the flux from the target star can often be misinterpreted as planet candidates. We obtained ground-based follow-up photometric confirmation of all planet candidates around \starA{} (Figure~\ref{fig:toi251_fu}) and \starB{} (Figure~\ref{fig:toi942_fu}). The on-target detection of the transits shows that the transit signal originates from the target stars to within the spatial resolution of $\sim 1\arcsec$ of our ground-based follow-up facilities.

From the high spatial resolution speckle images presented in Section~\ref{sec:AO}, we can also eliminate the presence of any stellar companions with $\Delta M < 4$ within $\sim 0.2\arcsec$ of our target stars. No \emph{Gaia} stars are catalogued within $2\arcsec$ of these target stars. Similarly, no slowly rotating spectroscopic blended companions were detected from our CHIRON observations (Section~\ref{sec:chiron}) at $\Delta M < 4$.

We can estimate the probability that a faint eclipsing binary lying unresolved by our high resolution speckle images is causing the transit signals. The transit shape, and the ratio between the ingress and totality timescales, can inform us about the brightness of any diluted background eclipsing binary that may be causing the transit signal \citep{2003ApJ...585.1038S}. We follow \citet{2019ApJ...881L..19V} and estimate the brightest possible background eclipsing binary that may be inducing our transit signal. The maximum difference in magnitude is given by $\Delta m_\mathrm{TESS} \leq 2.5 \log_{10} (T_{12}^2 / T_{13}^2 \delta)$, where $T_{12}$ is the duration of ingress, $T_{13}$ is the duration between first and third contact, and $\delta$ is the depth of the transit. 

We find that for \starA{}b, the transit can only be caused by a background eclipsing binary with $\Delta m _{\rm TESS} \leq 3.40$ and 3.0$\sigma$ significance.
From our diffraction limited observations of \starA{}, we excluded stellar blends within 0.2$''$ to a brightness contrast of $\approx 3$ magnitudes. For a randomly chosen star in a direction near \starA{}, the density of $2.0 \leq \Delta m _{\rm TESS} \leq 3.4$  stars within the ground-based exclusion radius of 0.2$''$ is  $<3 \times 10^{-5}$. Though TOI-251 is not randomly chosen, it is far from the galactic plane, where background eclipsing binaries are expected to be roughly two orders of magnitude less common than true planets \citep[][ Figure 30]{2015ApJ...809...77S}. While not formally impossible, the combined transit shape, lack of companions detectable by diffraction limited imaging, lack of secondary spectroscopic lines, and line of sight of TOI-251 lead background eclipsing binary scenarios to be highly disfavored.

For \starB{}, a similar calculation yields that a background eclipsing binary no fainter than 2.68 magnitudes is required to cause the transits of either one of its two planets. If it were a randomly chosen star, then the chance of there being another background star within $0.2$'' is $<6\times10^{-5}$ at $\Delta m_\mathrm{TESS} < 2.68$. The chance that two such background eclipsing binaries exist inducing our multi-planet transiting signal is $<4\times10^{-9}$. The low probability of such an occurrence, the multi-planet nature of the system, and the lack of any companions from diffraction limited imaging and spectroscopy, lead to eclipsing binary scenarios being disfavored.

\section{Estimating the age of young field stars}
\label{sec:age}

It is notoriously difficult to estimate the ages of field stars. However, young Sun-like stars exhibit rotational and activity induced photometric and spectroscopic behavior that make it possible to approximate their ages. These activity and rotation period properties can typically be calibrated via well-characterized clusters \citep{2008ApJ...687.1264M} to provide age relationships that can be used for dating purposes. 

Sun-like stars spin down over their main-sequence lifetimes through mass loss via stellar-wind processes. Whilst the Sun has a present-day rotation period of 24\,days, similar stars in the 120\,Myr old Pleiades cluster have rotation periods of $\sim 4$\,days. The rapid rotation and consequentially stronger dynamo of young stars excites strong spot and chromospheric activity. Observable chromospheric activity indicators, such as emission in the core of the calcium II lines, as well as X-ray emission, can be used as proxies for rotation and age of young Sun-like stars. The element lithium is destroyed in the cores of Sun-like stars through proton collisions. Through convective mixing processes, this leads to a rapid depletion of lithium in the stellar atmosphere within the first $\sim 500$\,Myr of their lives. The strength of the lithium 6708\,\AA{} absorption feature has traditionally being used as a youth indicator for Sun-like stars. 

We made use of these photometric and spectroscopic properties to estimate the ages of \starA{} and \starB{}. Though these estimates are imprecise given the field nature of our target stars, they generally agree sufficiently for us to place meaningful age constraints on these target stars. 

\subsection{Lithium depletion}

The strength of atmospheric lithium absorption in the spectra of convective envelope stars is also a commonly adopted age indicator. We make use of the Li $6708\,\AA$ line as another age estimator for our targets. The Li line equivalent widths are estimated by fitting three Gaussian profiles, accounting for the Li doublet at $6707.76\,\AA$ and $6707.91\,\AA$, and the nearby Fe I line at $6707.43\,\AA$ that is often blended with the Li features. Each line is assumed to be of equal width, and the two Li lines are also assumed to be of equal height. Using the TRES observations we obtained for both target stars, we measured equivalent widths of the Li $6708\AA$ line to be $0.134 \pm 0.017\,\AA$ for \starA{}, and $0.257\pm0.054 \,\AA$ for \starB{}.

To compare the lithium absorption feature of our target stars against stars in well characterized clusters and associations, we obtained a series of spectra of members of the IC 2602, IC 2391, Pleiades, and Praesepe groups. Spectra and measurements of Pleiades and Praespe members come from long-term radial-velocity surveys for planets in open clusters using the TRES spectrograph. Two such surveys, in Praesepe and the Hyades, are described in \citet{2012ApJ...756L..33Q} and \citet{2014ApJ...787...27Q}. With the goal of measuring low-amplitude RV variation, the spectra are typically high SNR, and therefore support precise measurement of Li equivalent widths. Li measurements for IC 2602 and IC 2391 members are adapted from \citet{1997A&A...323...86R} and \citet{2001A&A...372..862R}. 

\begin{figure*}
    \centering
    \begin{tabular}{cc}
    \includegraphics[width=0.4\textwidth]{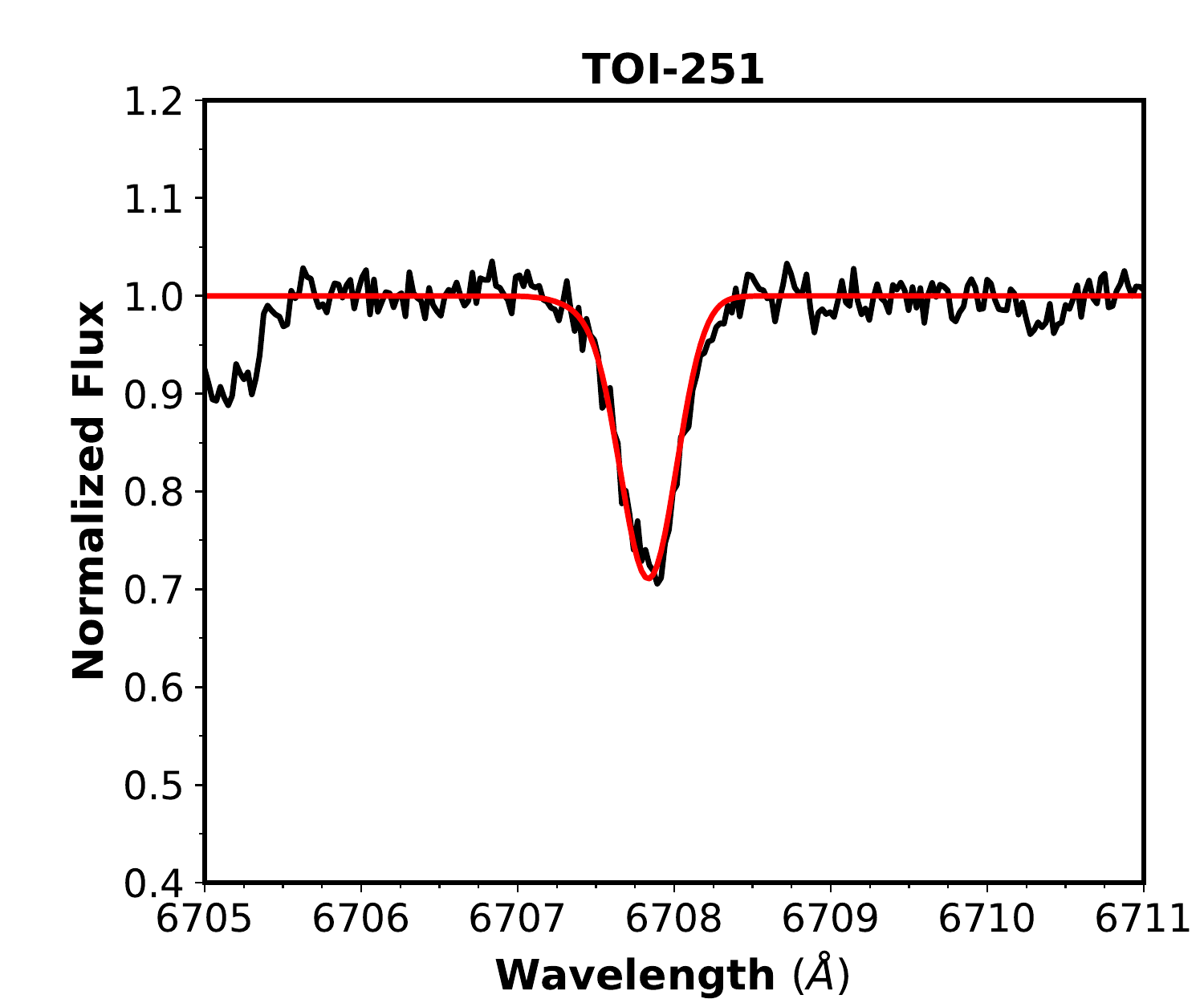} &
    \includegraphics[width=0.4\textwidth]{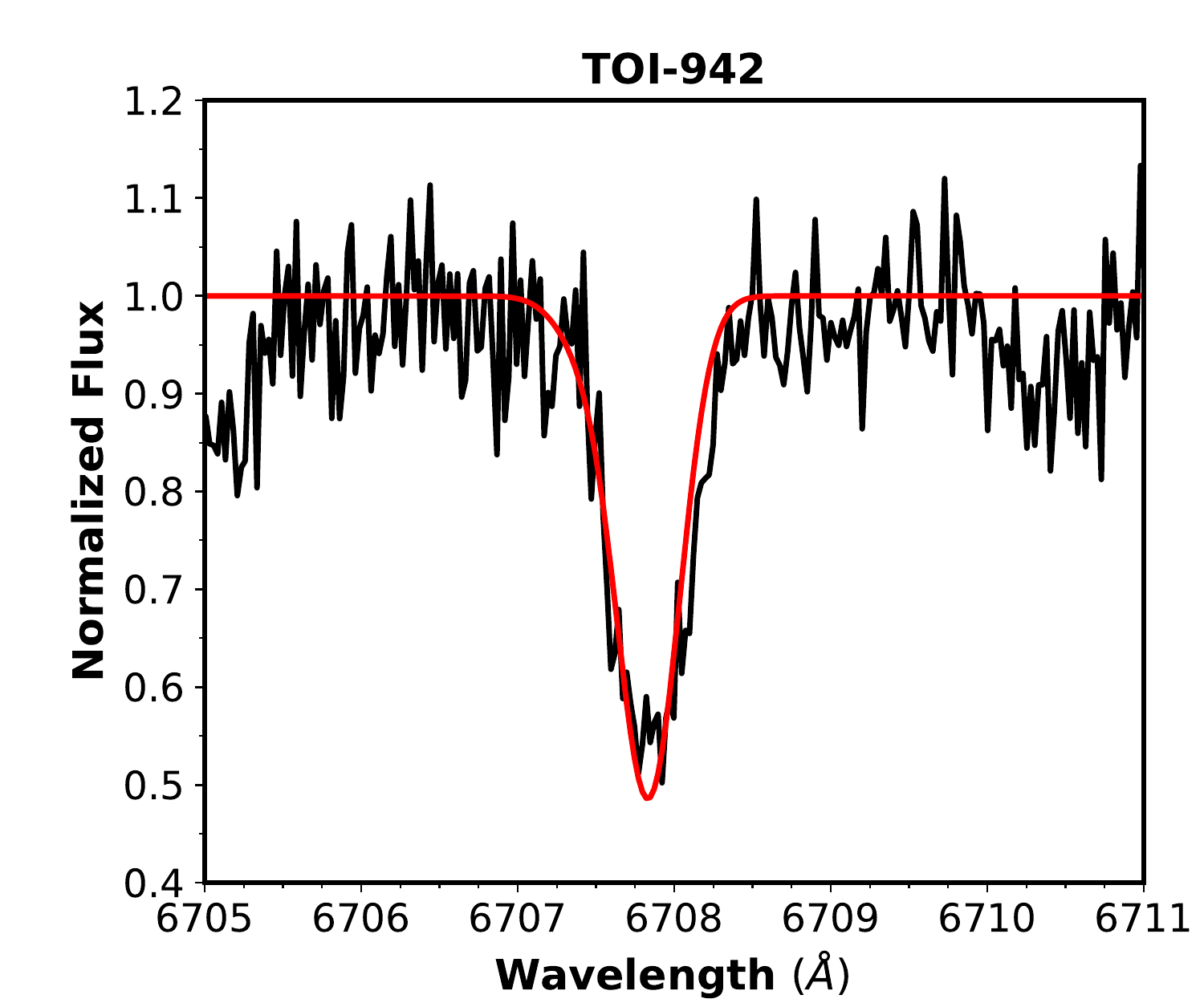} \\
    \end{tabular}
    \includegraphics[width=0.6\textwidth]{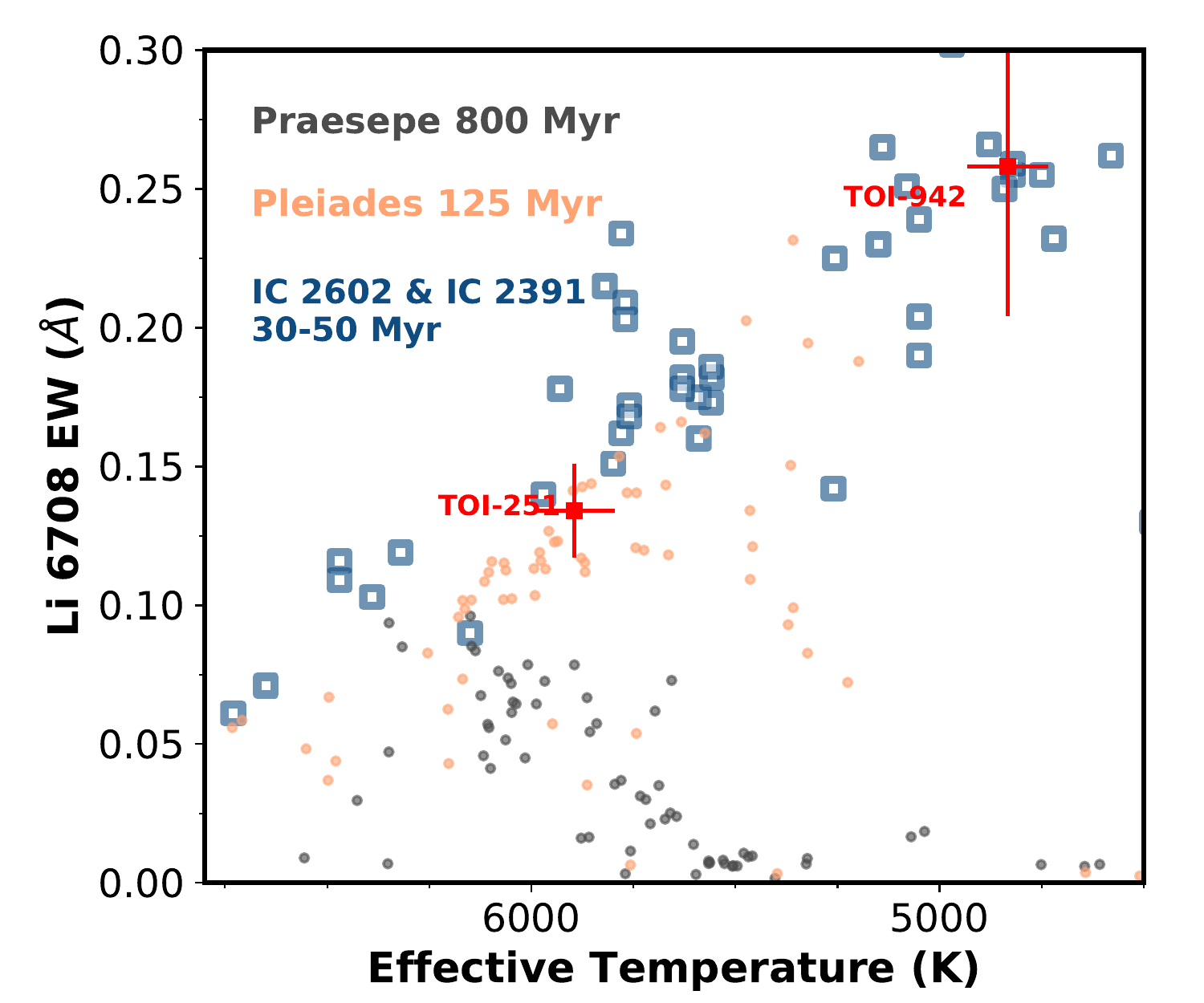} \\
    \caption{Both \starA{} and \starB{} exhibit strong lithium absorption at 6708\,\AA. We fit this absorption feature with a Gaussian doublet at $6707.76\,\AA$ and $6707.91\,\AA$, and a single Fe I line at $6707.43\,\AA$. The absorption features and best fit models of \starA{} are shown in the \textbf{left panel}, \starB{} on the \textbf{middle panel}. The \textbf{bottom panel} compares the equivalent widths of the lithium feature against stars in the IC 2602, IC 2391, Pleiades, and Praesepe clusters and associations. Measurements for Pleiades and Praesepe were obtained as part of this work using the methodology described above. Measurements for IC 2602 and IC 2391 are adopted from \citet{1997A&A...323...86R} and \citet{2001A&A...372..862R}.  }
    \label{fig:Li}
\end{figure*}

Figure~\ref{fig:Li} shows the Li 6708\,\AA{} equivalent widths of the target stars, in comparison with the same measurements for membership stars. The Li line strength of \starA{} agrees with that of Pleiades members, whilst \starB{} exhibits significantly stronger Li absorption than equivalent stars in Pleiades. 

\subsection{Activity-age relationships}

Both stars exhibit significant chromospheric emission in the near-ultraviolet calcium II H and K lines and the near-infrared calcium II triplet. Both stars are also detected in X-ray with the ROSAT all-sky survey, while \starB{} is detected in the GALEX NUV band. Chromospheric emission is a proxy for stellar rotation, being generated due to the stronger magnetic dynamo of the rapidly rotating young stars \citep[e.g.][]{1984ApJ...279..763N}. We make use of the Mount Wilson $S_{HK}$ index to compare the chromospheric emission of our target stars from the TRES spectra against literature activity-age relationships, as explained in the next section.

\subsubsection{Calcium II HK emission}
\begin{figure*}
    \centering
    \begin{tabular}{cc}
    \includegraphics[width=0.4\textwidth]{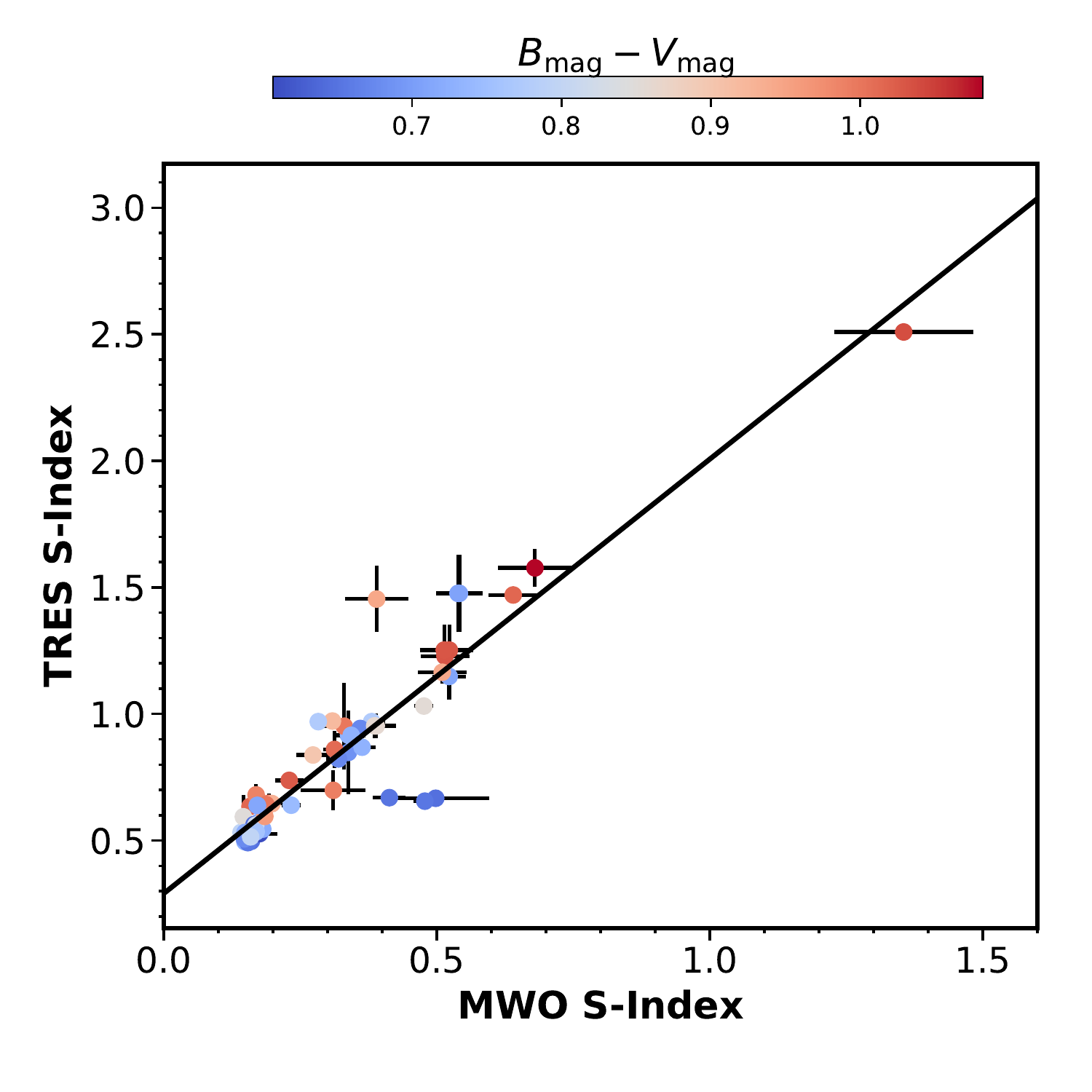}          & 
    \includegraphics[width=0.4\textwidth]{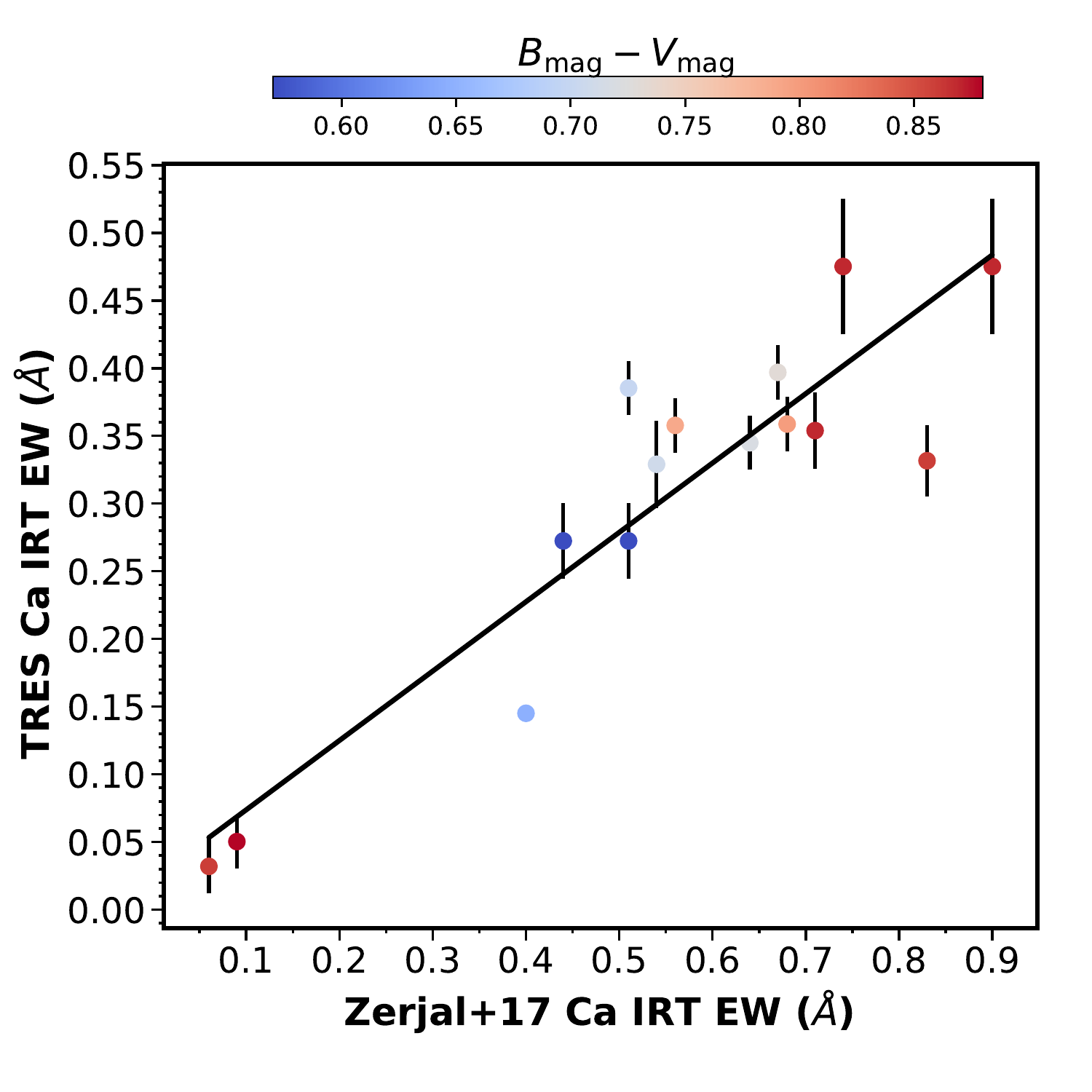}          \\
    \end{tabular}
    \caption{Calibrations for our measurements of the calcium II HK and infrared triplet activity indices. \textbf{Left panel} shows the $S_{HK}$ index of Mount Wilson stars as measured by TRES observations. We find agreement between our measurements and those from literature to within $\Delta S_{HK} = 0.078$. \textbf{Right panel} shows the equivalent width measurements of the calcium II infrared triplet core emission for stars that have both archival TRES observations and were surveyed in \citet{2017ApJ...835...61Z}. We find that a linear transformation is also sufficient to transform our measured values from TRES to those reported in literature, with a resulting scatter in the equivalent widths of $\Delta \mathrm{EW}_{\mathrm{IRT}} = 0.053\,\AA$.  }
    \label{fig:sindex-calibration}
\end{figure*}

Both calcium II H $(3969\,\AA)$ and K $(3934)$ lines are captured within TRES echelle orders. We measure calcium II HK emission core emissions in a band $1\,\AA$ wide centered about each line. The baseline flux is estimated from $5\,\AA$ wide bands over the continuum regions on either side of the line center. Figure~\ref{fig:sindex-calibration} shows our calcium line strength measurements for Sun-like stars $(5000 < T_\mathrm{eff} < 6000\,\mathrm{K})$ against that of catalog $S_{HK}$ values of the same stars from the Mount Wilson Observatory HK Project \citep{1978ApJ...226..379W,1978PASP...90..267V,1991ApJS...76..383D,1995ApJ...438..269B}. With the exception of a few active outlying stars, we find that our line emission flux estimates can be translated to the $S_{HK}$ index with a simple linear transformation, with an uncertainty of $\Delta S_{HK} = 0.078$. The calcium II HK line strength, as a function of the bolometric flux $(R'_{HK})$, is then calibrated using the relationship from \citet{1984ApJ...279..763N}, via the \emph{PyAstronomy} \citep{pya} \emph{SMW\_RHK} function. The $S_{HK}$ and $R'_{HK}$ values for the target stars are listed in Table~\ref{tab:stellar}. The uncertainties in these values are computed as the quadrature addition of uncertainty in the TRES emission flux to $S_{HK}$ calibration, and the scatter of the measurements between each observation.

The calcium II HK luminosity can be correlated with stellar age. We make use of the calibration from \citet{2008ApJ...687.1264M} (Equation 3) to calculate the calcium II HK ages of our target stars. \starA{} has an age of $27_{-13}^{+21}$\,Myr, with a $3\sigma$ regime ranging from 3 to 170\,Myr, consistent with the gyrochronology estimate. \starB{} is estimated to have an age of $10_{-6}^{+12}$ Myr, with a $3\sigma$ upper limit of 100\,Myr. The activity age estimate is significantly younger than that from gyrochronology. We note that with $\log R'_{HK} = -4.019 \pm 0.064$, the calcium II HK emission from \starB{} is near the limits of the calibrated range of the  \citet{2008ApJ...687.1264M}  relations of $-5.0 < \log R'_{HK} < -4.0$, corresponding to a lower age boundary of 8 Myr. Stars as young as \starB{} have saturated chromospheric emission features, making it difficult for us to derive precise ages from these spectroscopic indicators.

\subsubsection{Infrared calcium II triplet emission}

The infrared calcium II triplet at $8498\,\AA$, $8542\,\AA$ and $8662\,\AA$ also exhibit line core emission in active stars. Both \starA{} and \starB{} exhibit strong core emission in the infrared. The $8498\,\AA$ and $8542\,\AA$ lines are well placed within the TRES spectral orders. To measure the core emission equivalent widths of these lines, we first fit and remove a synthetic spectral template, and fit the residuals about the calcium lines with a Gaussian profile. The synthetic template is an ATLAS9 \citep{Castelli:2004} atmosphere model, convolved with the instrumental and rotational broadening kernel of the target star. The synthetic template is then subtracted from the continuum normalized observed spectrum. The resulting residuals about each calcium line are fitted with a Gaussian, with width corresponding to the line broadening of the spectrum, and with centroid fixed to the expected central wavelength of each line. 

Figure~\ref{fig:sindex-calibration} shows the equivalent widths as measured from TRES against the same stars that were characterized by \citet{2017ApJ...835...61Z}. For Sun-like stars, a linear transformation between our TRES observations and the literature values is sufficient, with a scatter in the resulting relationship of $\Delta \mathrm{EW}_{\mathrm{IRT}} = 0.053\,\AA$. 

We measure calcium II infrared triplet equivalent widths of $\mathrm{EW}_{\mathrm{IRT}} = 0.61 \pm 0.10\,\AA$ for \starA{} and $1.73 \pm 0.17\,\AA$ for \starB{}. \citet{2017ApJ...835...61Z} offers a qualitative age -- calcium triplet relationship from their calibration of cluster stars. The infrared triplet emission of \starA{} makes it compatible with similar stars in the 100-1000\,Myr range, whilst \starB{} falls in the $<100$\,Myr age range. 


\subsubsection{X-ray emission}

Young, rapidly rotating stars are also known to exhibit X-ray emission. \starA{} is an X-ray source in The Second ROSAT Source Catalog of Pointed Observations \citep{2000yCat.9030....0R}, while \starB{} is an X-ray source in the ROSAT all-sky faint source catalog \citep{2000yCat.9029....0V}. Their X-ray count rates, hardness ratios, and luminosities are provide in Table~\ref{tab:stellar}. We adopt the calibration provided in \citet{1995ApJ...450..401F} for the conversion between the X-ray count rate $(CTS)$ to an X-ray luminosity $(L_x)$, finding an X-ray luminosity of $\log(L_x/L_\mathrm{bol}) = -4.45 \pm 0.36$ for \starA{}, and $ -3.12 \pm 0.16$ for \starB{}. Using the age -- X-ray luminosity relationship from \citet{2008ApJ...687.1264M} (Equation A3), we derive an estimated age of $280_{-180}^{+550}$\,Myr for \starA{}, but the $3\sigma$ age upper limit is not constraining due to the scatter in the distribution. For \starB{}, we find an X-ray age of $7_{-4}^{+11}$ Myr, with a $3\sigma$ upper limit of 250\,Myr. Like our Ca II HK and infrared triplet estimates, the X-ray luminosity is saturated for this calibration above $\log(L_x/L_\mathrm{bol}) =4.0$ (corresponding to an 8 Myr lower limit). \starB{} is more active than the calibrated range of the \citet{2008ApJ...687.1264M} relationships, so our age estimate is at best qualitative. 

\subsubsection{NUV detection}

Ultraviolet emission from active chromospheres can also be an indicator of youth for Sun-like stars. \starB{} has a measurable flux in the NUV band of the GALEX all-sky catalogue \citep{2017ApJS..230...24B}. \citet{2011AJ....142...23F} calibrated a $NUV-J$ / $J-K$ colour-colour relationship using Hyades, Blanco 1, and moving group members, with a resulting scatter in the determined $\log$ ages of $0.39\,$dex. We adopt this relationship, and find an NUV age for \starB{} of $60_{-40}^{+130}$\,Myr. We note that there is significant scatter in the relationship between NUV flux and age, and as such the $3\sigma$ range of the NUV age estimate is only constraining to $<1\,$Gyr.

\subsection{Rotation Period and Gyrochronology}

\begin{figure*}
    \centering
    \begin{tabular}{cc}
        \includegraphics[width=0.4\linewidth]{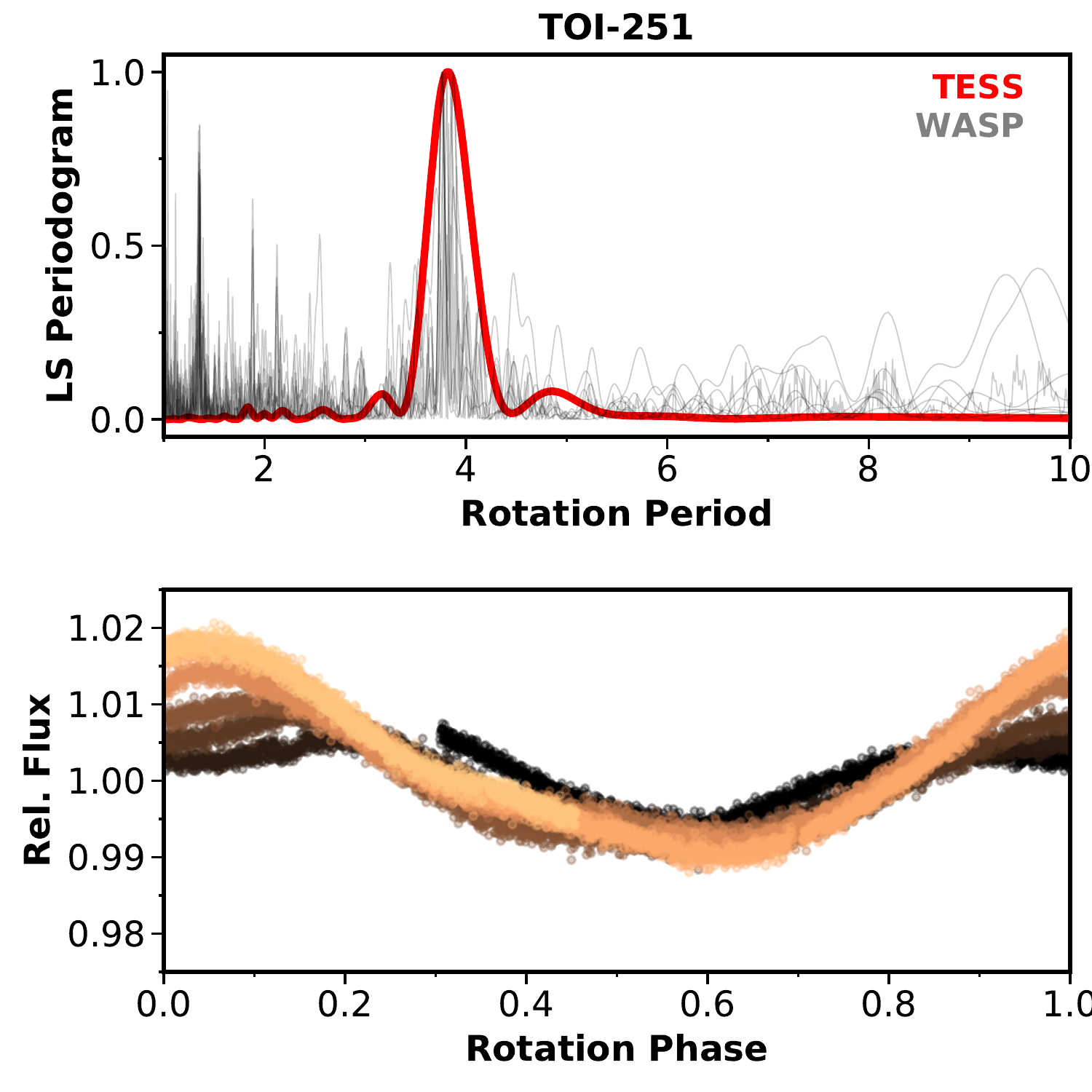} &
        \includegraphics[width=0.4\linewidth]{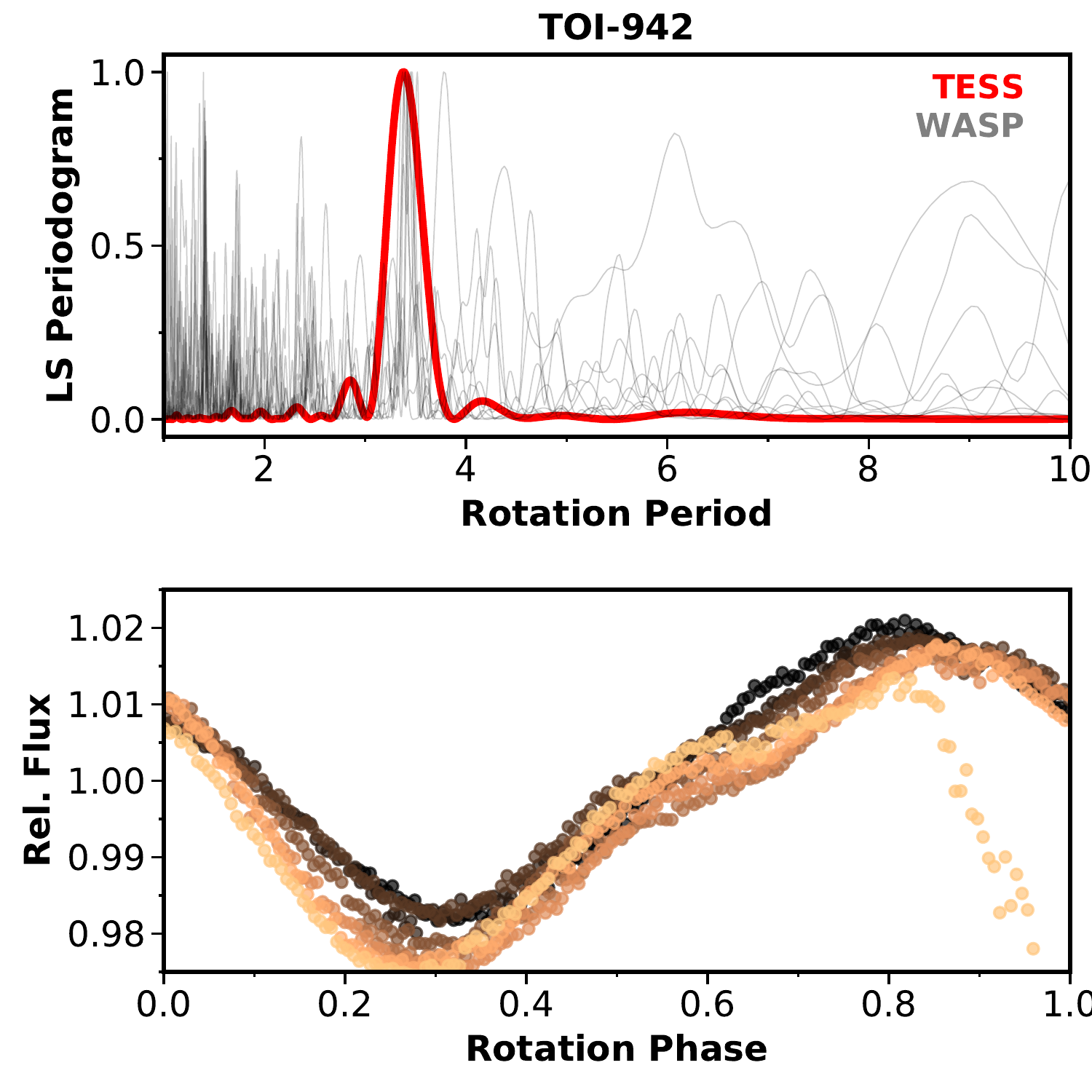} \\
    \end{tabular}
    \caption{Both \starA{} and \starB{} show significant spot modulation in their light curves. The top panels show the Lomb-Scargle periodogram of the \emph{TESS} light curves (red) and each campaign of the WASP observations (grey). The rotation period is consistently detected in the single sector of \emph{TESS} observations, and the eight years of monitoring from WASP. The bottom panels show the \emph{TESS} light curve folded to the rotation period. Each rotation period is over-plotted with a slightly different color gradient, such that it is easier to discern the spot variations over the course of the \emph{TESS} observations. }
    \label{fig:rot_lc}
\end{figure*}

\starA{} and \starB{} exhibit significant spot-induced rotational modulation in their \emph{TESS} light curves. These light curves, folded over the rotational period, are shown in Figure~\ref{fig:rot_lc}. The periodograms from a Lomb-Scargle analysis of the single sector of \emph{TESS} observations, as well as each observing campaign of the WASP observations, are shown. 

From the \emph{TESS} light curves, we measure a rotation period for \starA{} of $3.84 \pm 0.48$\,days, while \starB{} has a rotation period of $3.40\pm0.37$\,days. Similarly, the WASP observations yielded rotation periods of $3.799 \pm 0.047$\,days for \starA{}, and $3.41 \pm 0.49$ for \starB{}. The longevity of this activity signal from \emph{TESS} and WASP gives us confidence that the periodicity we quote is the rotation period of the host stars. 

We adopt a few gyrochronology relationships to estimate the ages of \starA{} and \starB{}. These relationships were calibrated by interpolating the slow rotating sequence in well characterized clusters. We note that by adopting these relationships, we are making the assumption that \starA{} and \starB{} follow the age spin-down trends seen in slow rotators amongst Sun-like stars. Many of these young clusters also show a spread in rotation at a given mass.  By Pleiades age, the rapid rotators are usually all binaries \citep[e.g.][]{2016ApJ...822...47D,2018AJ....156..275S}, but at the youngest ages, we cannot confirm that our stars lie on the main sequence. If \starA{} and \starB{} are rapid rotators for their age, then their ages will be difficult to estimate using gyrochronology. Given that both stars exhibit significant chromospheric activity and Lithium absorption, we think it is reasonable to assume these stars are relatively young. By assuming that they are also slow rotators for their age, we can apply gyrochronology relations to derive an additional age constraint.

Using the relationship in \citet{2007ApJ...669.1167B}, we find an age $T_\mathrm{gyro}$ of $110_{-60}^{+30}$ Myr for \starA{}, with a $3\sigma$ age range of 40-220\,Myr. \starB{} is considerably younger at $50 \pm 13$ Myr, with a $3\sigma$ age range of 20-90\,Myr. The uncertainties are the quadrature addition of the intrinsic uncertainties in the gyrochronology relationship, as prescribed in \citet{2007ApJ...669.1167B}, and the uncertainties resulting from the \emph{TESS} rotation period measurements. Similarly, applying the age relationship from \citet{2008ApJ...687.1264M}, we get an age range of $70-318$ Myr for \starA{}, and $41-155$ Myr for \starB{}. 

\begin{figure*}
    \centering
    \includegraphics[width=0.6\textwidth]{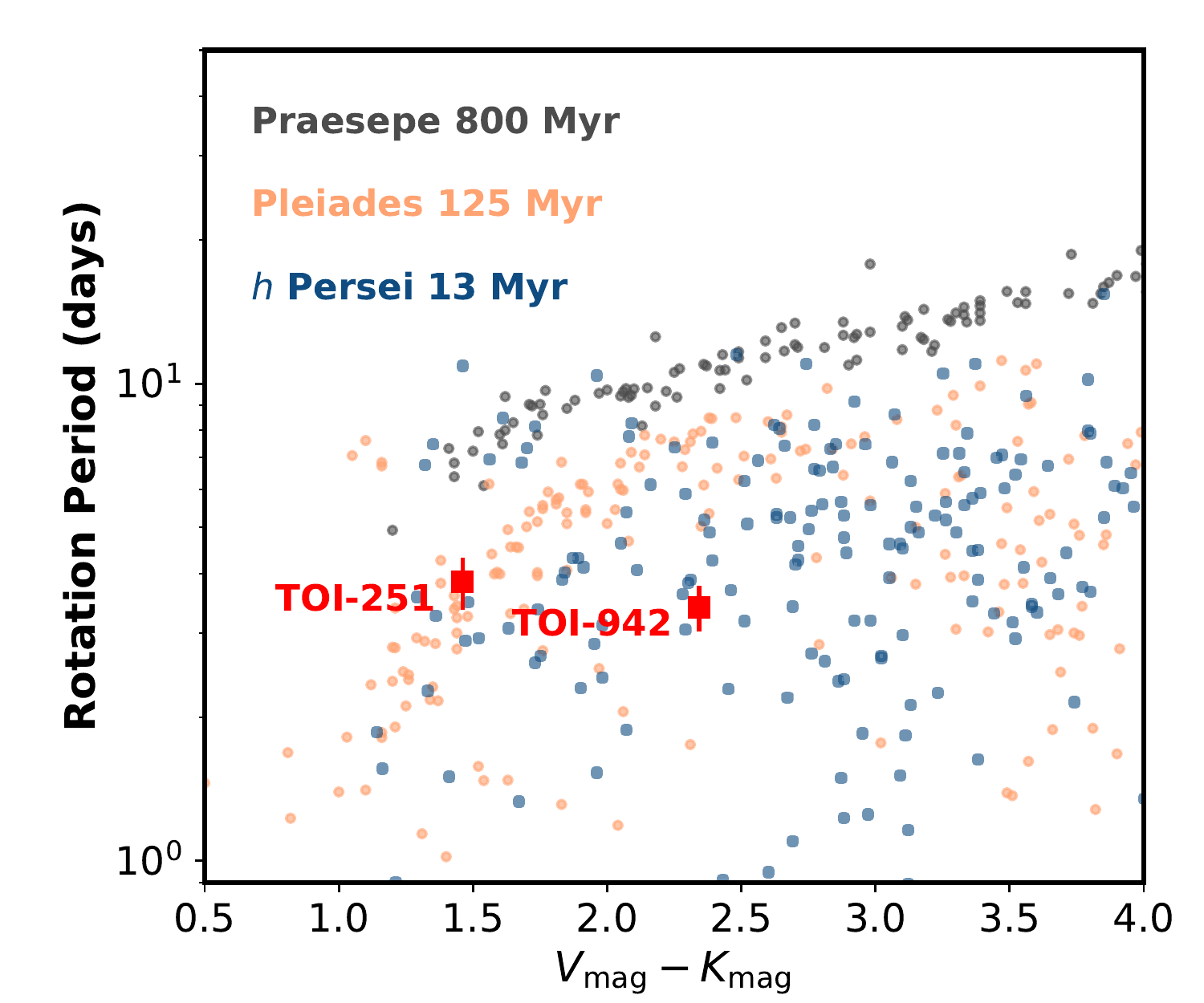} 
    \caption{A comparison between the rotation periods of \starA{} and \starB{} against stars from known clusters and associations. The blue squares mark stars from the 13 Myr old $h$ Persei cluster as measured by \citet{2013A&A...560A..13M}, with colors dereddened from 3D dust maps.  The orange points show the distribution of rotation periods for stars in the 120\,Myr Pleiades cluster from \citet{2016AJ....152..113R}. The grey points show stars in the $\sim 800$ Myr Praesepe cluster measured by \citet{2017ApJ...839...92R}. \starA{} has rotation periods similar to stars from the Pleiades. \starB{} appears younger than the majority of single stars from the Pleiades, but older than Sun-like stars near the zero-age main sequence in the $h$ Persei cluster.}
    \label{fig:Prot}
\end{figure*}

Qualitatively, we can compare the rotation periods of our target stars against those in known clusters and associations. Figure~\ref{fig:Prot} shows the rotation periods of the target stars against known members of the 120\,Myr old Pleiades measured by \citet{2016AJ....152..113R}, and 800\,Myr old Praesepe measured by \citet{2017ApJ...839...92R}. For comparison, we also plot the rotation of members of the $h$ Persei cluster, at an age of $13$\,Myr from \citet{2013A&A...560A..13M}, to illustrate that both \starA{} and \starB{} are older than some of the youngest cluster and associations. The $h$ Persei members are marked by the blue squares in Figure~\ref{fig:Prot}. The colors of stars in $h$ Persei have been de-reddened according to the 3D dust maps via \emph{dustmap} \citep{2018JOSS....3..695M} using maps from \citet{2019ApJ...887...93G}. In this qualitative comparison, the rotation period of \starA{} agrees well with the Pleiades population, supporting the gyrochronology estimate of $\sim 110$ Myr. \starB{} is rotating faster than an equivalent star within the Pleiades distribution, and as such supports our estimate of it being younger than 100 Myr. We note though that there is a possibility for our targets to be young, rapid rotators, as with stars in the significantly younger $h$ Persei cluster.


\subsubsection{Infrared excess}

The spectroscopic and gyrochronology age estimates described above have placed meaningful upper age limits. Lower limits from these measurements are more difficult for \starB{} given that many of its activity indicators are saturated. 

A qualitative argument for the lower limits of both \starA{} and \starB{} can be made due to their lack of any infrared excess in the WISE bands. The spectral energy distribution of \starA{} and \starB{} are shown in Figure~\ref{fig:sed}. Disks can be traced by $H\alpha$ emission and infrared excess, and typically dissipate by $\sim 5-10$ Myr \citep[See reviews by ][]{2009AIPC.1158....3M,2011ARA&A..49...67W}. As such, we adopt a lower limit of 10\,Myr for the age of \starB{}. 

\subsubsection{Color-magnitude diagram}

\begin{figure}
    \centering
    \includegraphics[width=0.5\textwidth]{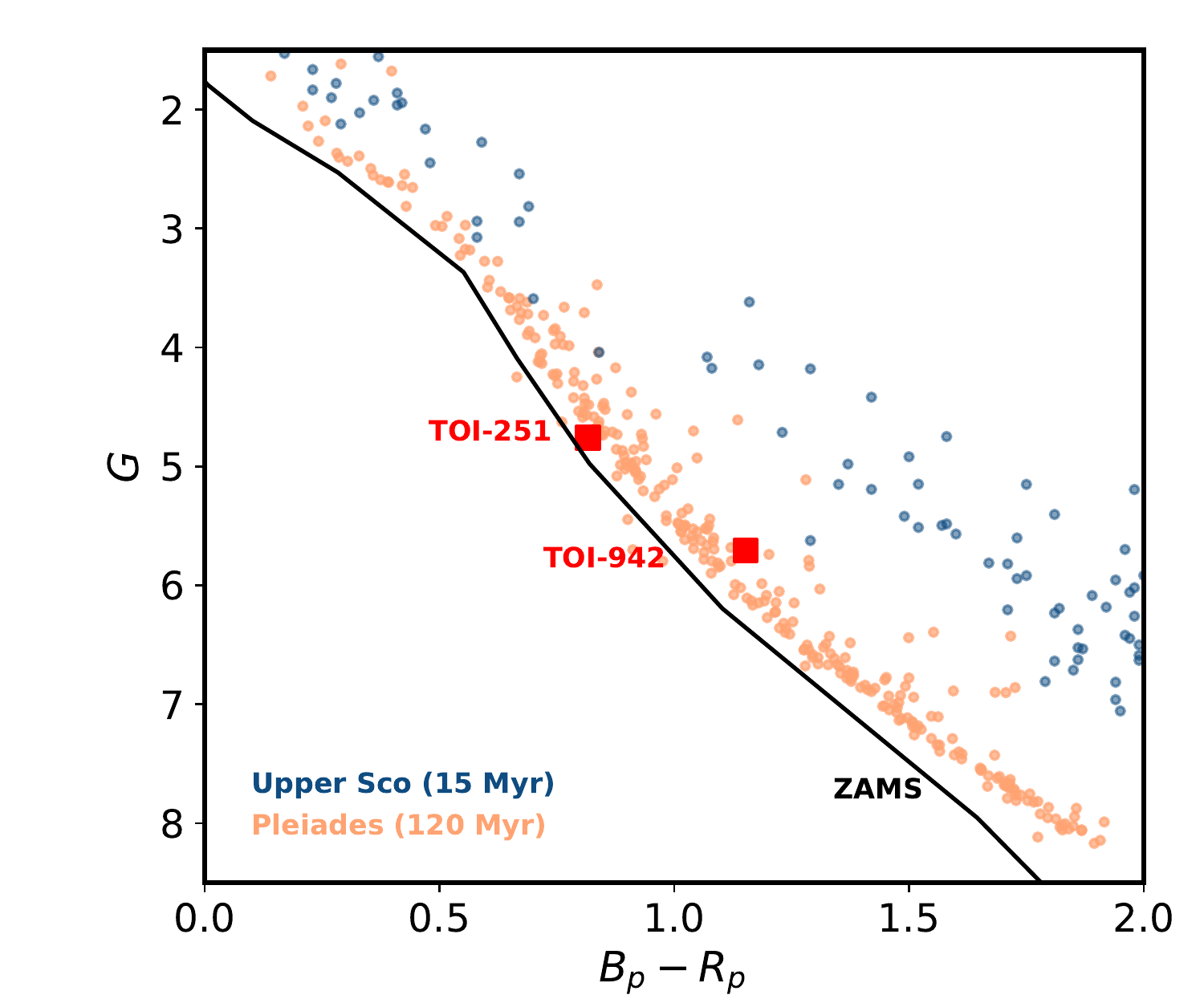}
    \caption{The Gaia color-magnitude diagram of \starA{} and \starB{}. For comparison, stars from the 120\,Myr old Pleiades cluster \citep{2019A&A...628A..66L}, and the 15 Myr old Upper Sco association are shown \citep{2019A&A...623A.112D}. \starA{} lies close to the distribution of stars in the Pleiades cluster. \starB{} lies marginally above stars in the Pleiades, but is clearly older than the pre-main sequence stars in Upper Sco. For referece, the zero-age main sequence from the MIST isochrones is marked by the black line \citep{2016ApJS..222....8D}. }
    \label{fig:CMD}
\end{figure}

Further lower bound age limits may be inferred by comparing the colors and magnitudes of the target stars against members of well known clusters. Figure~\ref{fig:CMD} shows the Gaia color-magnitude diagram for \starA{} and \starB{}. Stars from the 120\,Myr old Pleiades cluster \citep{2019A&A...628A..66L} and 15 Myr old Upper Sco association \citep{2019A&A...623A.112D} are shown for comparison. 

\starA{} is consistent with having an age similar to stars in the Pleiades cluster, as per our spectroscopic and gyrochronology estimates above. Having reached the main sequence, it is difficult to estimate its age from the spectral energy distribution, and the ages from isochrone models provide no further constraints to the age of the system. 

\starB{} lies marginally above the zero-age main sequence. The activity-based age estimates described above have trouble placing an lower bound on the age of \starB{}. From the color-magnitude diagram, \starB{} clearly sits below stars from the 15 Myr old Upper Sco association. As part of the global analysis, the isochrone-fitted age from Section~\ref{sec:modelling} provides a $3\sigma$ lower age limit of 23 Myr, consistent with that provided by gyrochronology. 

\subsection{Kinematics} 

To the best of our knowledge, neither TOI 251 nor TOI 942 is a member of any known
coeval stellar population.
To check, we searched the CDIPS target star
list (\citealt{bouma_cluster_2019} Table~1), which is a concatenation of stars from
across the literature reported to be in known moving groups and open clusters.
This concatenation included large surveys ($>10^5$ cluster stars) such as those of
\citet{kharchenko_global_2013},
\citet{dias_proper_2014},
\citet{oh_comoving_2017},
\citet{cantatgaudin_gaia_2018},
\citet{gaia_hr_2018},
\citet{zari_3d_2018},
\citet{cantatgaudin_expanding_2019}, and
\citet{kounkel_untangling_2019}.
It also included targetted surveys, for instance those of
\citet{roser_deep_2011},
\citet{rizzuto_multidimensional_2011},
\citet{kraus_tucanahor_2014},
\citet{bell_32ori_2017},
\citet{gagne_banyan_XII_2018},
\citet{gagne_banyan_XI_2018}, and
\citet{gagne_banyan_XIII_2018}.
We also verified that we could not place these targets into any known associations via the online BANYAN
$\Sigma$ tool \citep{gagne_banyan_XI_2018}.


\subsection{Summary of age estimates}

\begin{figure}
    \centering
    \includegraphics[width=1\linewidth]{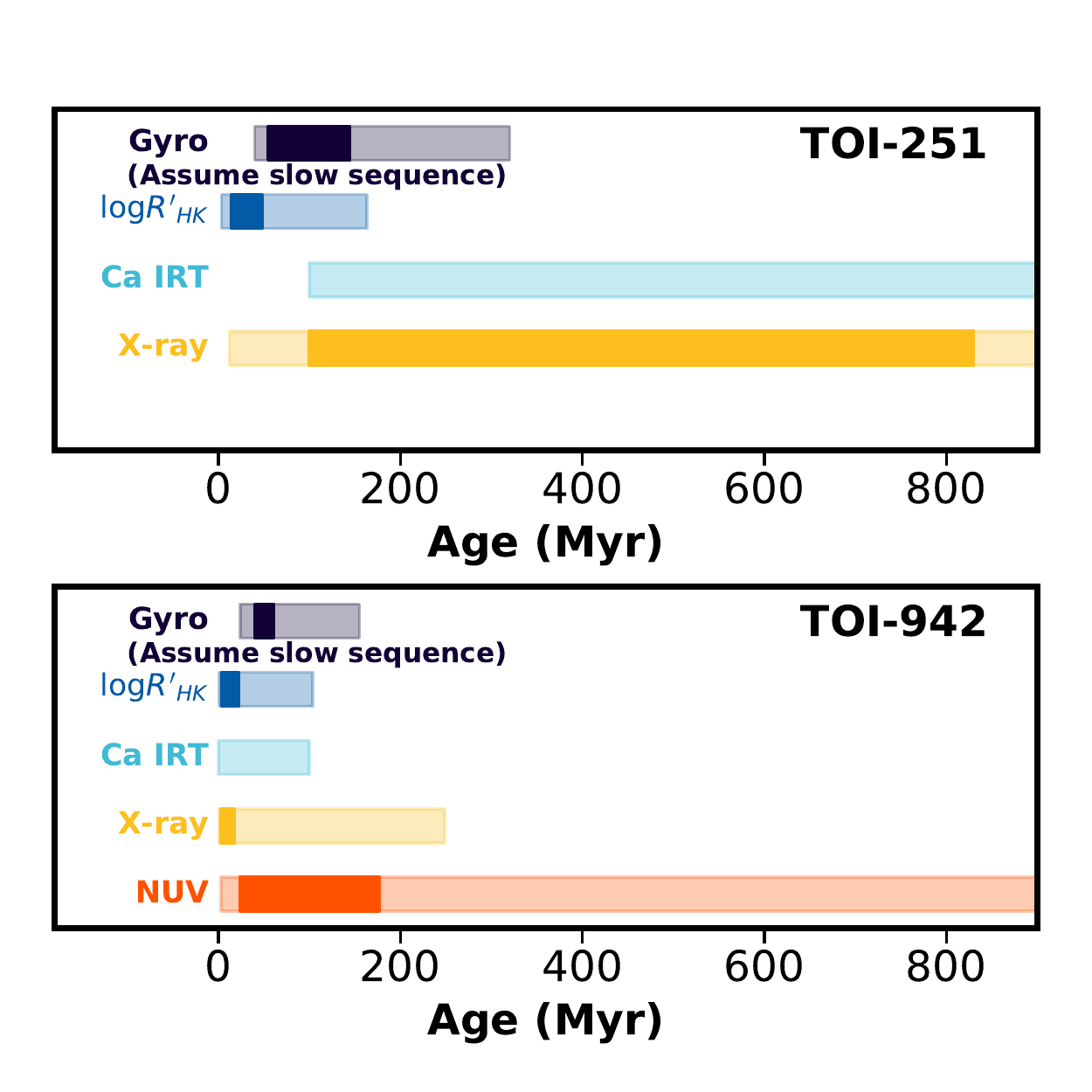}
    \caption{A summary of the age-activity indicators for \starA{} and \starB{}. The $1\sigma$ (solid) and $3\sigma$ (lightened) age ranges from gyrochronology, spectroscopic and photometric activity indicators are marked. We adopt a final age estimate for \starA{} of \starAage{}\,Myr, and for \starB{} of \starBage{}\,Myr.  }
    \label{fig:age}
\end{figure}

Figure~\ref{fig:age} summarizes the ages of \starA{} and \starB{} as estimated from the age indicators. For \starA{}, gyrochronology and chromospheric Ca II HK emissions provide constraining age estimates. We adopt an estimated age range of 40-320\,Myr for \starA{}, encompassing the $3\sigma$ upper range of both gyrochronology relationships we tested, and the upper limit of chromospheric Ca II HK emission estimates -- the two age indicators that yielded constraining estimates in our analysis. This age range also agrees with the estimates from the less constraining Ca II infrared triplet and X-ray emission estimates, which put the age of \starA{} below 1 Gyr. 

\starB{} is clearly more active than \starA{} and members of the Pleiades cluster. Constraining measurements for the age of \starA{} come from gyrochronology, Ca II HK emissions, and the Ca II infrared triplet emissions, placing a $3\sigma$ age range at $20-160$\,Myr. The age estimates from X-ray and NUV emissions are less constraining, but still consistent with a young age for \starB{}. The fact that the rotation of the star is significantly slower than similar stars in the 13 Myr $h$ Persei cluster at zero-age main sequence, and the lack of infrared excess for \starB{} allowed us to place an approximate lower limit of 20\,Myr. 

\section{System modeling}
\label{sec:modelling}

To derive accurate system parameters, we present a series of global models for each system incorporating the \emph{TESS} discovery light curve, ground-based follow-up light curves, radial velocities, spectroscopic and broadband atmospheric parameters, and stellar isochrone constraints. 

The light curves of young stars exhibit large variations due to spot modulation. The variability signal often dwarfs the planetary transits, and as such needs to be carefully considered so as to yield un-biased system parameters. The modeling of spot activity is simplified by the quasi-sinusoidal nature of the light curves, with well defined periods that can be easily modeled. We adopt the \emph{celerite} package \citep{celerite} to model the stellar variability via its simple harmonic oscillator kernel, with free parameters describing the frequency of the stellar variability $\log \omega_0 =\log  1 / P_\mathrm{rot}$, the dampening factor $\log Q_0$, and the power of the oscillator $\log S_0$. We impose a Gaussian prior on $\log \omega_0$ based on the rotation period of each star, with the specific limits shown in Table~\ref{tab:planetparam}. Linear uniform priors were imposed for $\log Q_0$ and $\log S_0$. The follow-up light curves are also included in the global analysis. These light curves are not modeled as part of the Gaussian process. To account for any environmental systematics in the light curves, we instead include a linear trend for each transit, a linear coefficient detrending against airmass, and an additional offset term for meridian flips. 

The transits are modeled via the \emph{batman} package \citep{2015PASP..127.1161K}, with free parameters defining the periods $P$, transit center ephemeris $T_c$, planet-star radius ratio $R_p/R_\star$, and line of sight inclination $i$ for each transiting planet. Limb darkening parameters are interpolated from \citet{Claret:2011} and \citet{2017A&A...600A..30C}, and fixed through the modeling process. The radial velocities are further modeled by circular orbits, with masses of each planet $m_p$, and a jitter term $\sigma_\mathrm{rv}$ to account for the radial velocity jitter. Given the large stellar-activity induced jitter for these young stars, we only provide upper limits on the masses of the planets reported here. 

In addition, we model the stellar parameters simultaneously with the transits and radial velocities. The stellar properties are modeled via the MESA Isochrones \& Stellar Tracks \citep[MIST][]{2016ApJS..222....8D}, interpolated using stellar evolution tracks with ages below 1\,Gyr and with initial masses $0.6 < M_\star < 1.6\,M_\odot$. Free parameters include the stellar mass $M_\star$, radius $R_\star$, and stellar metallicity [M/H]. These are constrained by the spectroscopic effective temperature $T_\mathrm{eff}$, surface gravity $\log g_\star$, and metallicity as measured by SPC from the TRES spectra. We also incorporate photometric magnitudes from visual HIPPARCOS \citep{1997AA...323L..49P}, \emph{Gaia} \citep{2018AA...616A...1G} and near infrared 2MASS bands \citep{2006AJ....131.1163S} in the modeling of the spectral energy distribution. We incorporate de-biased \emph{Gaia} DR2 distance estimates from \citet{2018AJ....156...58B} into the SED analysis. Though we have estimates of the stellar age, we apply no age prior to the stellar evolution modeling beyond limiting the ages to below 1\,Gyr. The resulting age posteriors are largely uninformative, allowing for all ages within the given range. 

\begin{figure}
    \centering
    \textbf{TOI-251}\\
    \includegraphics[width=0.8\linewidth]{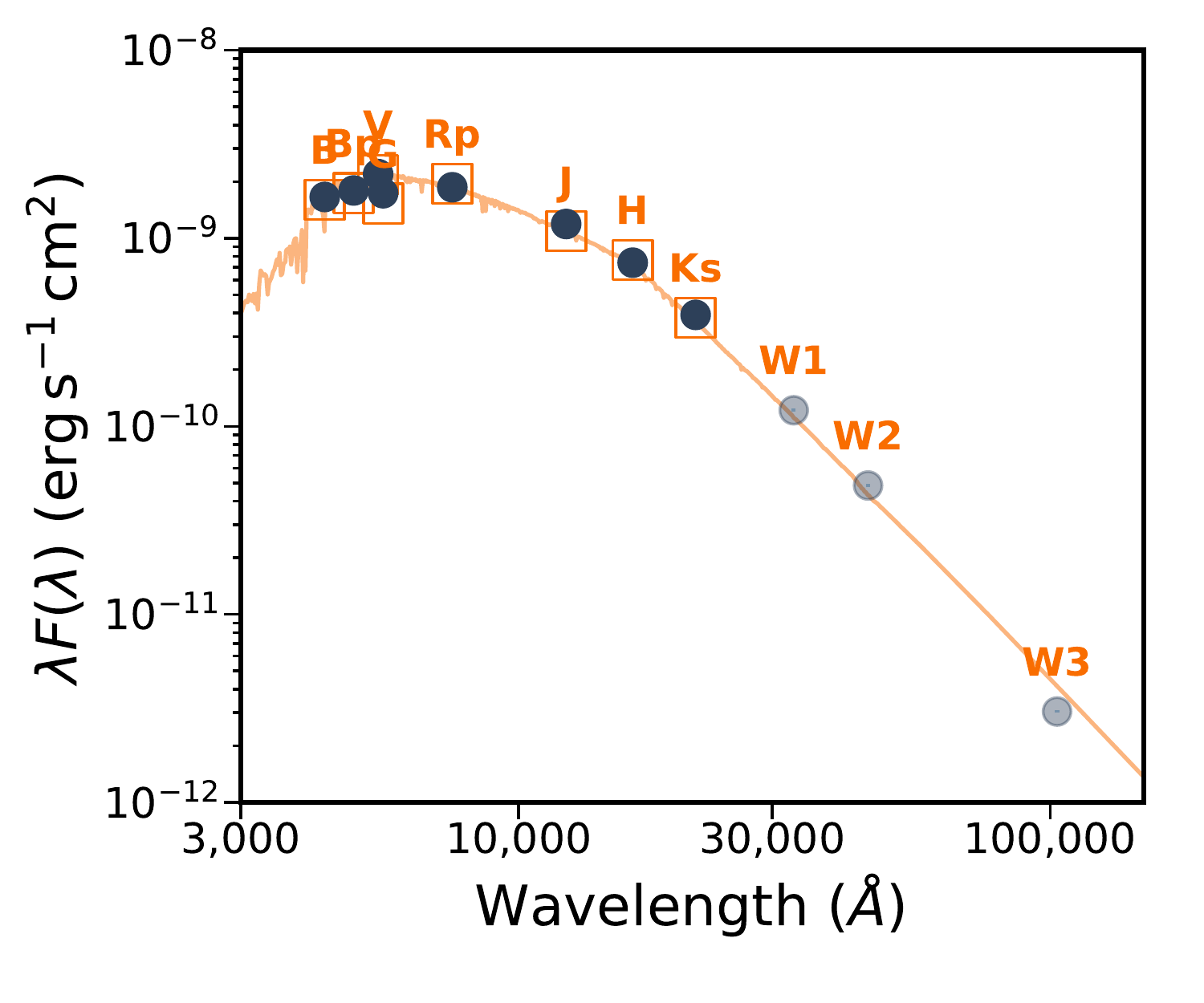}\\
    \textbf{TOI-942}\\
    \includegraphics[width=0.8\linewidth]{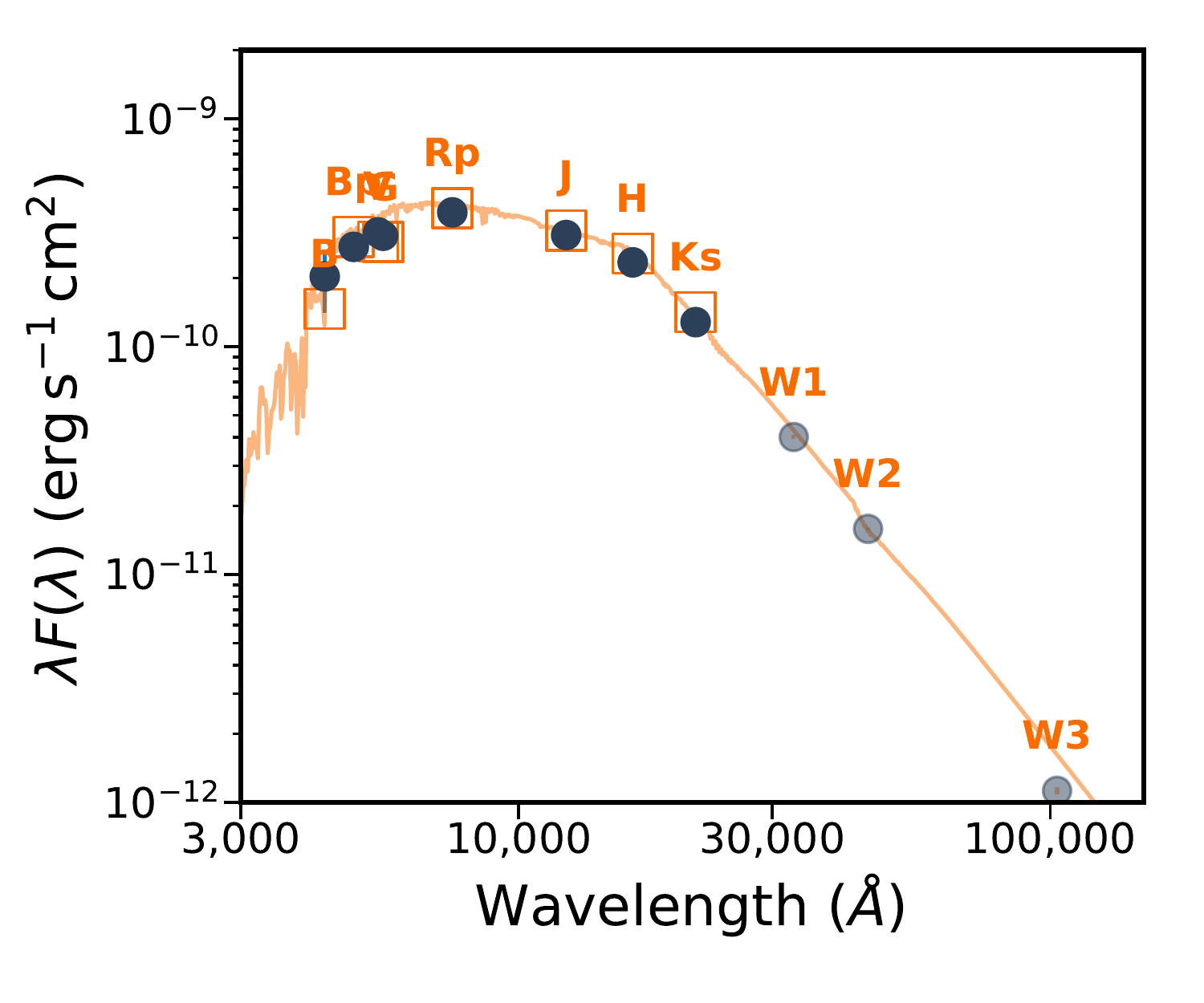}
    \caption{Spectral energy distributions (SED) of \starA{} and \starB{} over the HIPPARCOS \citep{1997AA...323L..49P}, \emph{Gaia} \citep{2018AA...616A...1G}, and 2MASS bands \citep{2006AJ....131.1163S} bands are plotted in navy. The MIST isochrone predicted fluxes in each band are marked by the open orange squares. The best fit ATLAS9 \citep{Castelli:2004} synthetic atmosphere model is overlaid in orange. The WISE $W_1$, $W_2$, and $W_3$ fluxes of the target stars are also shown, though they are not used to constrain the SED model. }
    \label{fig:sed}
\end{figure}

The final parameters for the host stars are presented in Table~\ref{tab:stellar}, and the planet properties are presented in Table~\ref{tab:planetparam}. The spectral energy distributions are shown in Figure~\ref{fig:sed}, with the best fit atmospheric model overlaid.

\begin{deluxetable*}{lrrr}
\tablewidth{0pc}
\tabletypesize{\scriptsize}
\tablecaption{
    Stellar parameters
    \label{tab:stellar}
}
\tablehead{ \\
    \multicolumn{1}{c}{~~~~~~~~Parameter~~~~~~~~}   &
    \multicolumn{1}{c}{\starA{}} &
    \multicolumn{1}{c}{\starB{}} &
    \multicolumn{1}{c}{Source} 
}
\startdata
\sidehead{Catalogue Information}
~~~~TIC \dotfill & 224225541 & 146520535 & \citet{2018AJ....156..102S}\\
~~~~Tycho-2 \dotfill & 7520-00369-1 & 5909-00319-1 & \citet{2000AA...355L..27H} \\
~~~~\emph{Gaia} DR2 Source ID \dotfill & 6539037542941988736 & 2974906868489280768 & \citet{2018AA...616A...1G}\\
~~~~\emph{Gaia} RA (2015.5) \dotfill & 23:32:14.9 & 05:06:35.91 & \citet{2018AA...616A...1G}\\
~~~~\emph{Gaia} DEC (2015.5) \dotfill& -37:15:21.11 & -20:14:44.21 & \citet{2018AA...616A...1G}\\
~~~~\emph{Gaia} $\mu_\alpha$ $(\mathrm{mas}\,\mathrm{yr}^{-1})$ \dotfill& $ 44.639 \pm 0.074$ & $15.382 \pm 0.034$ & \citet{2018AA...616A...1G}\\
~~~~\emph{Gaia} $\mu_\delta$ $(\mathrm{mas}\,\mathrm{yr}^{-1})$ \dotfill& $1.902 \pm 0.070$  & $-3.976 \pm 0.040$ & \citet{2018AA...616A...1G}\\
~~~~\emph{Gaia} DR2 Parallax $(\mathrm{mas})$ \dotfill & $10.019 \pm 0.044 $ & $6.524 \pm 0.029 $ & \citet{2018AA...616A...1G}\\
~~~~Systemic Radial Velocity (\kms) \tablenotemark{a}        
                          \dotfill    & \starAsystemic{} & \starBsystemic{} &  ...\\
~~~~U $(\kms)$ \dotfill & $-19.454 \pm 0.085$ & $-15.729 \pm 0.030$ &\\
~~~~V $(\kms)$ \dotfill & $-7.190 \pm 0.045$ & $-22.347 \pm 0.045$&\\
~~~~W $(\kms)$ \dotfill & $-4.550 \pm 0.039$ & $-5.252 \pm 0.040$&\\
\sidehead{Stellar atmospheric properties}
~~~~$\teffstar$ (K)  \dotfill       &  \starAteff{} & \starBteff{}  & \\
~~~~$\feh$ \tablenotemark{a} \dotfill                &  \starAfeh{} & \starBfeh{}  & \\
~~~~$\vsini$ (\kms)\dotfill        &  \starAvsini{} & \starBvsini{} &\\
~~~~$v_\mathrm{macro}$ (\kms)\dotfill        &  \starAvmac{} & \starBvmac{} &  \\
\sidehead{Stellar activity properties}
~~~~$P_\mathrm{rot}$ (d) \dotfill        &  $3.84 \pm 0.48$ & $3.40\pm0.37$ & \\
~~~~$S_{HK}\,(\AA)$ \dotfill        &  $0.616 0.083$ & $1.06 \pm 0.15$ &  \\
~~~~$\log R'_{HK}$ \dotfill        &  $-4.119 \pm 0.066$ & $-4.019 \pm 0.064$ & \\
~~~~Ca II IRT EW $(\AA)$ \dotfill        &  $0.61 \pm 0.10$ & $1.73 \pm 0.17$ & \\
~~~~ROSAT X-ray Counts $CTS$ \dotfill        & $0.01477$  & $0.0446 \pm 0.0130$ & \citet{2000yCat.9030....0R,2000yCat.9029....0V} \\
~~~~ROSAT X-ray Hardness Ratio $HR_1$ \dotfill        & $-0.39\pm0.12$  & $0.17 \pm 0.28$ &  \citet{2000yCat.9030....0R,2000yCat.9029....0V}  \\
~~~~X-ray luminosity $\log Lx/L_\mathrm{bol}$ \dotfill        & $-4.45 \pm 0.36$  & $-3.12 \pm 0.16$ & \\
~~~~Li 6708 EW $(\AA)$ \dotfill        & $0.134 \pm 0.017$  & $0.257 \pm 0.054$ & \\
\sidehead{Photometric properties}
~~~~GALEX NUV (mag)\dotfill               &  & $18.507 \pm 0.049$ & \citet{2017ApJS..230...24B}\\
~~~~TESS $T$ (mag)\dotfill               & $9.3258 \pm 0.0061$ & $11.0462 \pm 0.0066$ & \citet{2018AJ....156..102S}\\
~~~~\emph{Gaia} $G$ (mag)\dotfill               & $9.7541 \pm 0.0012$ & $11.6346 \pm 0.0014$ & \citet{2018AA...616A...1G}\\
~~~~\emph{Gaia} $Bp$ (mag)\dotfill               & $10.1070 \pm 0.0030$ & $12.1468 \pm 0.0034$ & \citet{2018AA...616A...1G} \\
~~~~\emph{Gaia} $Rp$ (mag)\dotfill               & $9.2910 \pm 0.0026$ & $10.9950 \pm 0.0029$ & \citet{2018AA...616A...1G}\\
~~~~TYCHO $B$ (mag)\dotfill               &  $10.528 \pm 0.068$ & $12.783 \pm 0.364$ & \citet{1997AA...323L..49P}\\
~~~~TYCHO $V$ (mag)\dotfill               &  $9.8870 \pm 0.0050$ & $11.982 \pm 0.026$ & \citet{1997AA...323L..49P}\\
~~~~2MASS $J$ (mag)\dotfill               & $8.766 \pm 0.020$ & $10.231 \pm 0.022$ & \citet{2006AJ....131.1163S} \\
~~~~2MASS $H$ (mag)\dotfill               & $8.498 \pm 0.044$ & $9.747 \pm 0.023$ & \citet{2006AJ....131.1163S}\\
~~~~2MASS $K_s$ (mag)\dotfill             & $8.426 \pm 0.027$ & $9.639 \pm 0.023$ & \citet{2006AJ....131.1163S}\\
~~~~WISE $W_1$ (mag)\dotfill             & $8.372 \pm 0.024$ & $9.576 \pm 0.024$ & \citet{2010AJ....140.1868W,2013yCat.2328....0C}\\
~~~~WISE $W_2$ (mag)\dotfill             & $8.399 \pm 0.020$ & $9.609\pm0.020$ & \citet{2010AJ....140.1868W,2013yCat.2328....0C}\\
~~~~WISE $W_3$ (mag)\dotfill             & $8.368 \pm 0.022$ & $9.453 \pm 0.039$ & \citet{2010AJ....140.1868W,2013yCat.2328....0C}\\
\sidehead{Stellar properties}
~~~~$\mstar$ ($\msun$)\dotfill      & \starAmass{} &  \starBmass{} &\\
~~~~$\rstar$ ($\rsun$)\dotfill      & \starAradius{} &  \starBradius{} &\\
~~~~$\loggstar$ (cgs)\dotfill       & \starAlogg{} & \starBlogg{} &\\
~~~~$\lstar$ ($\lsun$)\dotfill      & \starAlum{} & \starBlum{}  &\\
~~~~Line of sight inclination $I_*\,(^\circ)$\dotfill        & \starAirot{} & \starBirot{} &\\
~~~~Age (Myr)\dotfill           &    \starAage{}&  \starBage{}  &\\
~~~~Distance (pc) \dotfill      &    \starAdist{} & \starBdist{}   &\\
\enddata
\tablenotetext{a}{
  Derived from the global modeling described in Section~\ref{sec:modelling}, co-constrained by spectroscopic stellar parameters and the \emph{Gaia} DR2 parallax.\\
}

\end{deluxetable*}



\begin{deluxetable*}{lrrr}
\tablewidth{0pc}
\tabletypesize{\scriptsize}
\tablecaption{
    Orbital and planetary parameters 
    \label{tab:planetparam}
}
\tablehead{ \\
    \multicolumn{1}{c}{~~~~~~~~Parameter~~~~~~~~}   &
    \multicolumn{1}{c}{\starA{}b} &
    \multicolumn{1}{c}{\starB{}b} &
    \multicolumn{1}{c}{\starB{}c} 
}
\startdata
\sidehead{\Lc{} parameters}
~~~$P$ (days)             \dotfill    & \starAperiod{} & \starBperioda{} & \starBperiodb{} \\
~~~$T_c$ (${\rm BJD-TDB}$)    
      \tablenotemark{a}   \dotfill    & \starATc{} & \starBTca{} & \starBTcb{} \\
~~~$T_{14}$ (days)
      \tablenotemark{a}   \dotfill    & \starATdur{} & \starBTdura{} & \starBTdurb{} \\
~~~$\arstar$              \dotfill    & \starAars{} & \starBarsa{} & \starBarsb{} \\
~~~$\rpl/\rstar$          \dotfill    & \starArprs{} & \starBrprsa{} & \starBrprsb{}  \\
~~~$b \equiv a \cos i/\rstar$
                          \dotfill    & \starAb{} & \starBba{} & \starBbb{} \\
~~~$i$ (deg)              \dotfill    & \starAinc{} & \starBinca{} & \starBincb{} \\
\sidehead{Limb-darkening and gravity darkening coefficients \tablenotemark{b}}
~~~$a_i'$                 \dotfill  & 0.3817  & ... & ... \\
~~~$b_i'$                 \dotfill  & 0.3393   &... & ... \\
~~~$a_Y$                \dotfill   & 0.1428  & ... & ...    \\
~~~$b_Y$                \dotfill  & 0.3642   & ... & ...  \\
~~~$a_\mathrm{Mearth}$                \dotfill   & 0.1925  &  0.3201 &  ... \\
~~~$b_\mathrm{Mearth}$                \dotfill  & 0.3552  & 0.2797 & ... \\
~~~$a_\mathrm{TESS}$      \dotfill  & 0.2831 & 0.4006 & ... \\
~~~$b_\mathrm{TESS}$      \dotfill    & 0.2873 & 0.2243 & ... \\
\sidehead{RV parameters}
~~~$K$ (\ms)              \dotfill    & \starAKrv{} & \starBKrva{} & \starBKrvb{}  \\
~~~$e$                    \dotfill    & 0 (fixed) & 0 (fixed) & 0 (fixed) \\
~~~RV jitter (\ms)        
                          \dotfill    & \starArvjitter{} & \starBrvjitter{} & ... \\
\sidehead{Gaussian process hyperparameters}    
~~~$\log \omega_0$ SHOT Frequency  \dotfill    & \starAGPw{}  & \starBGPw{}  & ...  \\
& (Gaussian prior $\mu = -1.35, \sigma = 0.13$) & (Gaussian prior $\mu= -1.22, \sigma= 0.11$) & \\
~~~$\log Q_0$ SHOT Quality factor \dotfill    & \starAGPQ{} & \starBGPQ{} & ... \\
~~~$\log S_0$ SHOT S  \dotfill    & \starAGPS{} & \starBGPS{} & ...\\
\sidehead{Planetary parameters}
~~~$\rpl$ ($R_\oplus$)       \dotfill    & \starAplrad{} &  \starBplrada{} & \starBplradb{}  \\
~~~$\mpl$ ($M_\mathrm{Jup}$)       \dotfill    & \starAplmass{} & \starBplmassa{} & \starBplmassb{} \\
~~~$a$ (AU)               \dotfill    & \starAa{} & \starBaa{} & \starBab{} \\
\enddata
\tablenotetext{a}{
    \ensuremath{T_c}: Reference epoch of mid transit that minimizes the
    correlation with the orbital period.
    \ensuremath{T_{14}}: total transit duration, time between first to
    last contact;
}
\tablenotetext{b}{
        Values for a quadratic law given separately for each of the filters with which photometric observations were obtained.  These values were adopted from the
        tabulations by \citet{Claret:2011} according to the
        spectroscopic an initial estimate of the stellar parameters. 
}
\end{deluxetable*}

\section{Discussion}
\label{sec:discussion}

We report the discovery and validation of planets around \starA{} and \starB{}. The two systems of small planets orbiting relatively young field stars, with \starA{} estimated to be \starAage{}\,Myr old, and \starB{} at \starBage{}\,Myr old. The mini-Neptune \starA{}\,b has an orbital period of \starAperiodshort{}\,days, and a radius of \starAplrad{}\,$R_\oplus$. Two inflated Neptunes orbit \starB{}, with \starB{}b in a \starBperiodashort{}\,day orbit with a radius of \starBplrada{}\,$R_\oplus$, and \starB{}c in a \starBperiodbshort{}\,day orbit with a radius of \starBplradb{}\,$R_\oplus$. 

\subsection{The radius-period diagram for young planets, and the detectability of smaller planets}

\begin{figure}
    \centering
    \includegraphics[width=1.\linewidth]{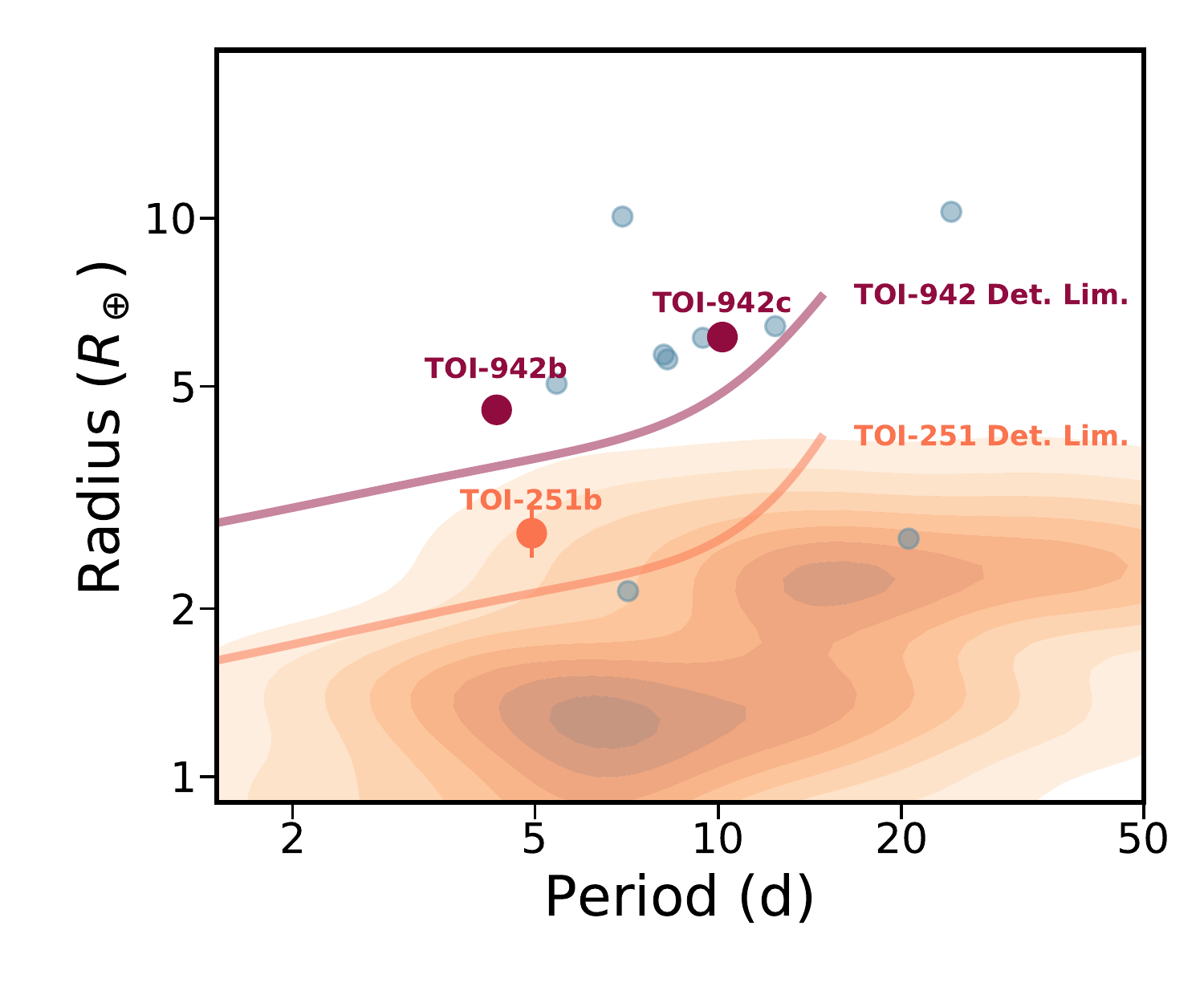}
    \caption{The period-radius distribution of close-in small planets. The contours mark the distribution of planets from the \emph{Kepler} survey characterized in \citet{2017AJ....154..109F}, with a gap between rocky and gaseous enveloped planets clearly seen. Planets around \starA{} and \starB{} are marked. The detection limits of planets around each of these stars are marked by horizontal lines. Planets around stars younger than 500\,Myr old are shown in cyan-colored points, including the planets around V1298 \citep{2019AJ....158...79D,2019ApJ...885L..12D}, K2-33b \citep{2016AJ....152...61M,2016Natur.534..658D}, DS Tuc Ab \citep{2019ApJ...880L..17N}, Kepler-63b \citep{2013ApJ...775...54S}, HIP67522\,b \citep{2020arXiv200500013R}, and TOI1726\,b and c  \citep{2020arXiv200500047M}. }
    \label{fig:period_radius}
\end{figure}

Discoveries of super-Earths and Neptunes around young stars give us a unique opportunity to explore the evolution of small planets in their early stages of evolution. Figure~\ref{fig:period_radius} shows the period-radius distribution of close-in small planets, comparing the planets characterized by the \emph{Kepler} survey and those subsequently discovered around stars younger than $500$\,Myr. Photoevaporation is key in sculpting this distribution, creating a sub-Saturn desert devoid of close-in Neptunes and super-Earths \citep[e.g.][]{2012ApJ...761...59L,2013ApJ...775..105O,2018MNRAS.479.5012O} and carving the radius valley between rocky and gaseous planets further out \citep[e.g.][]{2017AJ....154..109F,2017ApJ...847...29O}.  

\starA{}b straddles along the evaporation gap between solid cores and planets still retaining their gaseous envelopes. It is amongst the smallest planets known around stars younger than 500 Myr; only the planets around the Ursa Major moving group member TOI-1726 \citep{2020arXiv200500047M} are smaller. The planets around \starB{} lie in the sub-Saturn radius regime, similar in radii and periods to other young planetary systems at the $\sim 20-50$\,Myr age regime, such as DS Tuc Ab \citep{2019ApJ...880L..17N}, and V1298 Tau c and d \citep{2019ApJ...885L..12D}. 

The planets around \starA{} and \starB{}, as well as most other planets found around young stars, have larger radii than most in the \emph{Kepler} sample. However, it may be difficult to compare this sample of planets around young, active stars against those identified by the \emph{Kepler} sample around quiet stars with higher precision light curves.

To see if smaller planets could be recovered from similar light curves, we performed a signal injection and recovery exercise on the \emph{TESS} observations of \starA{} and \starB{}. Figure~\ref{fig:period_radius} shows the detection thresholds derived from 1000 injected planets across the radius-period space. Each injection is drawn from a uniform distribution in period between 1 and 15 days, radius from 0.5 to 15 $R_\oplus$, and impact parameter from 0 to 1. A circular orbit is assumed for every injected planet. The light curves are then detrended via a cosine-filtering \citep{2010A&A...521L..59M} algorithm, and the transits recovered via a BLS \citep{2002A&A...391..369K} search. The injection and recovery exercise shows that the planets we discovered are nearly the smallest detectable planets. The detection thresholds worsen significantly for planets in orbits longer than 10 days. Importantly, the predominant population of close-in small planets with $R_p < 2 R_\oplus $, and those with $R_P < 3 R_\oplus$ and orbital periods $>10$\,days, would not be detectable around either star. Similar injection and recovery exercises of previous analyses \citep[e.g.][]{2019ApJ...880L..17N,2020arXiv200500013R} show similar results, that our detectability for planets smaller than $\sim 2-3\, R_\oplus$ is only complete at periods shorter than 10 days.

\subsection{Orbital eccentricity and prospects for obliquity measurements}

One main tracer for the origins of close-in planets is the properties of the orbits they currently inhabit. High eccentricity or highly oblique orbits can be indicative of strong dynamical interaction in the early history of these planets. 

We modified our analysis (Section~\ref{sec:modelling}) to allow eccentricity, modeled as $\sqrt e \cos \omega$ and $\sqrt e \sin \omega$, to be free during the global modeling. To prevent the large jitter of radial velocities and systematic effects from ground base transit observations from influencing our eccentricity estimates, we make use of only the \emph{TESS} light curves and the SED in this analysis. 

\begin{figure}
    \centering
    \includegraphics[width=0.9\linewidth]{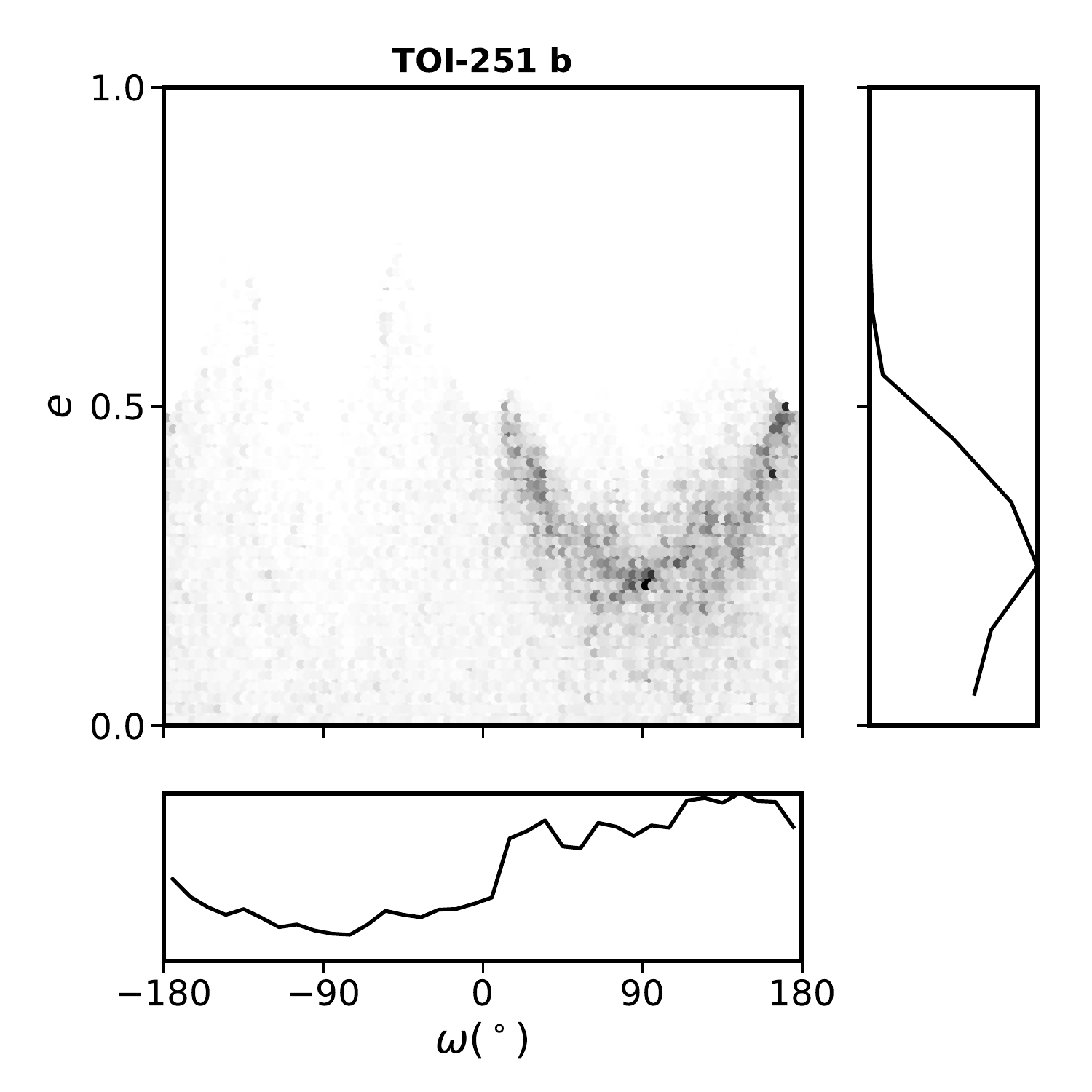}
    \caption{Eccentricity posterior for \starA{}b, as constrained by the \emph{TESS} transit duration and inferred stellar parameters. We find that the eccentricity of \starA{}b can only be loosely constrained to $e < 0.64$ at $3\sigma$ significance.  }
    \label{fig:ecc_toi251}
\end{figure}

\begin{figure}
    \centering
    \includegraphics[width=0.9\linewidth]{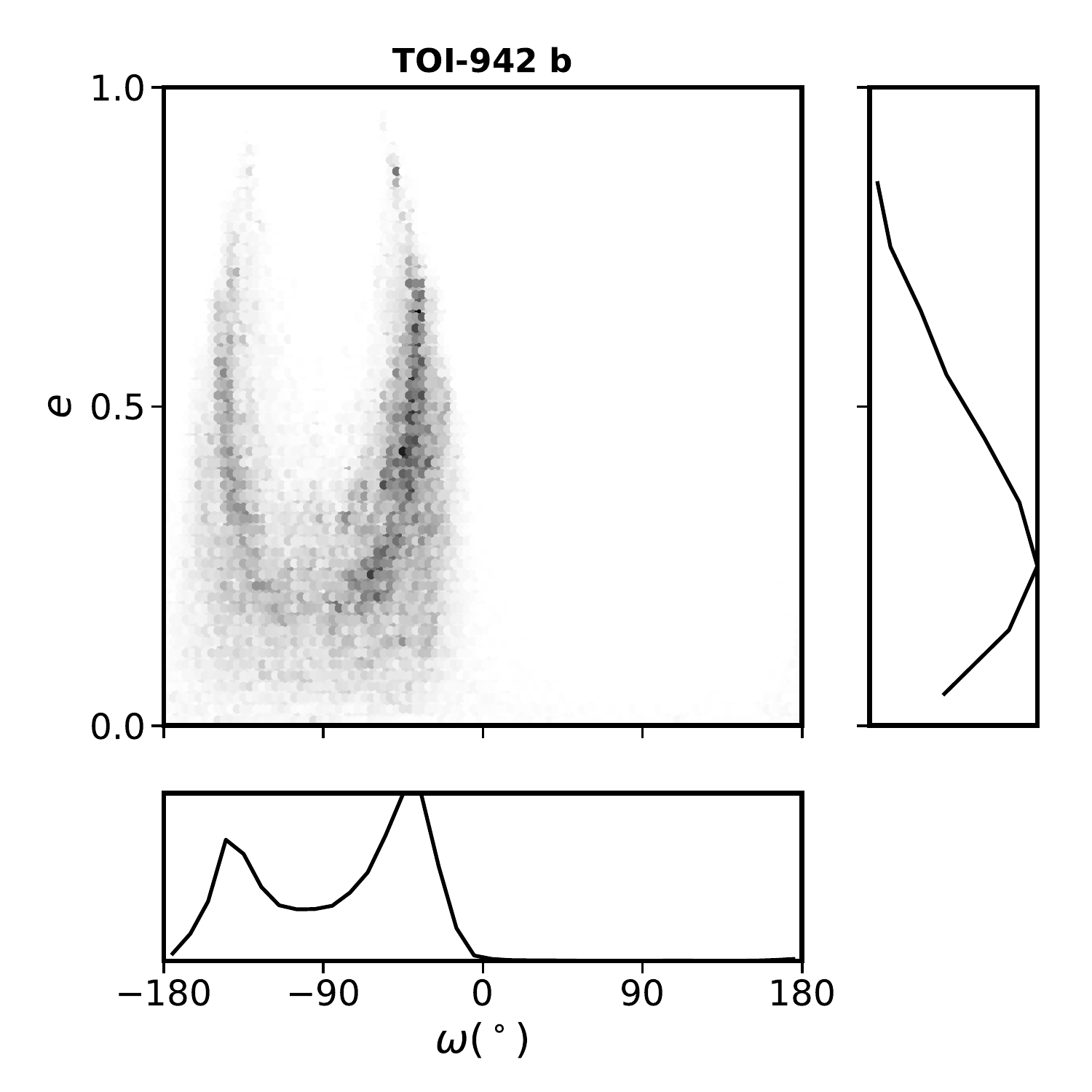}\\
    \includegraphics[width=0.9\linewidth]{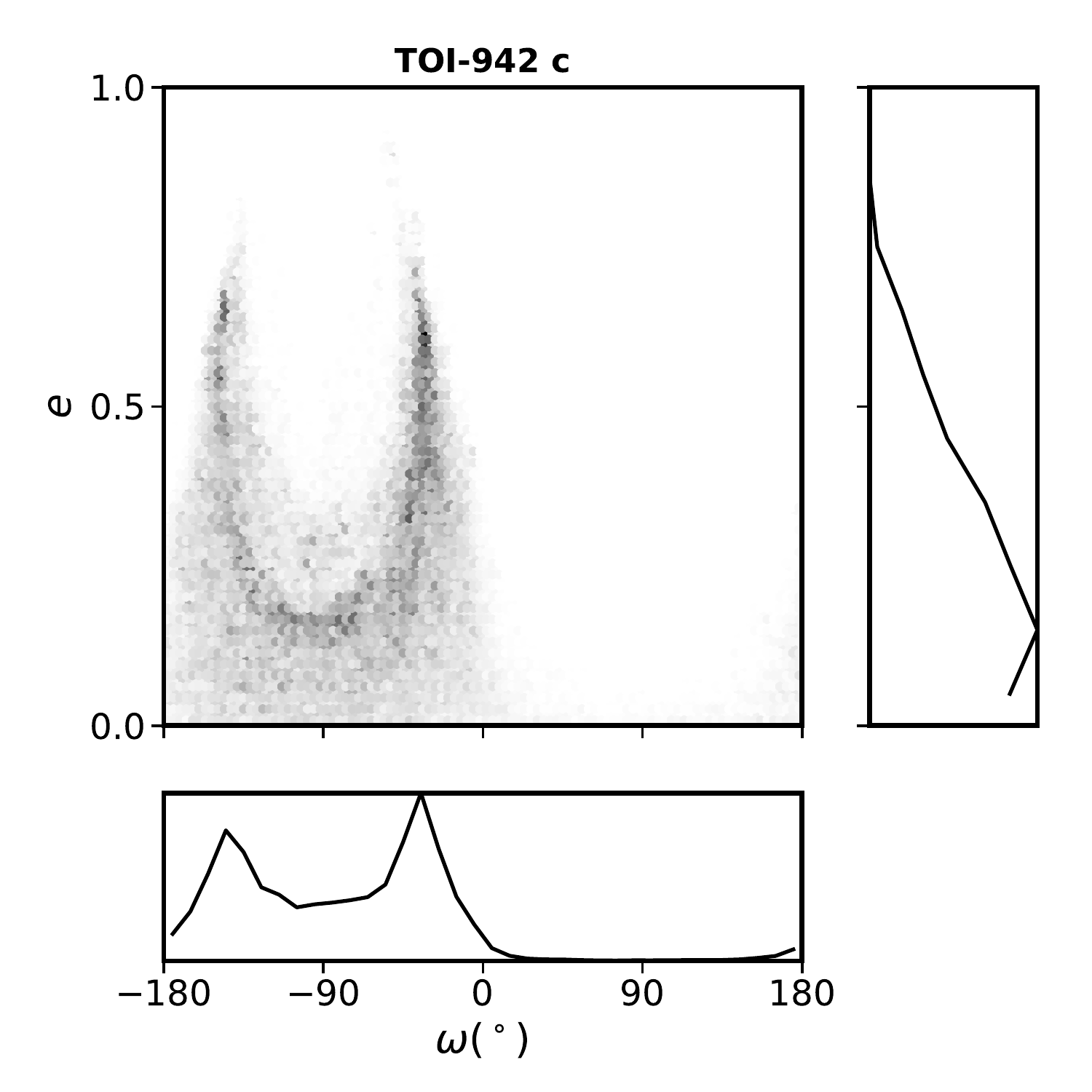}\\
        \caption{Eccentricity posteriors for \starB{}b and c based on their \emph{TESS} transit durations and inferred host star parameters. The eccentricities are poorly constrained with existing observations, with only highly eccentric orbits ruled out. The eccentricity of \starB{}b is constrained to be $e<0.87$, and $e<0.80$ for \starB{}c at $3\sigma$ significance.}
    \label{fig:ecc_toi942}
\end{figure}

Figure~\ref{fig:ecc_toi251} and \ref{fig:ecc_toi942} show the resulting posteriors in eccentricity $e$ and longitude of periastron $\omega$. For both systems, the eccentricity can only be loosely constrained with existing datasets. We find $3\sigma$ upper limits $e < 0.64$ for \starA{}b, $<0.87$ for \starB{}b, and $< 0.80$ for \starB{}c. 

Given the difficulty in measuring precise radial velocity orbits for these planets around young, active stars, a potentially easier observational constraint on their dynamical histories is their projected orbital obliquity angles. The obliquity angle is the relative angle between the orbital planes of the planets and the rotation axis of the host star. Planets that experienced significant dynamical interactions in their past can often be found in highly oblique orbits \citep[e.g.][]{2017haex.bookE...2T}. 

The line of sight inclination of the host stars can be used to approximate the orbital obliquity for the planets. Following \citet{2020AJ....159...81M}, we compute the stellar inclination angle $I_*$ via the isochrone-derived stellar radius $R_\star$, the spectroscopic rotational broadening velocity $\vsini$, and the photometric rotation period $P_\mathrm{rot}$ of each system. Though tight constraints cannot be placed with our existing observations, \starA{} and \starB{} appear both to be well aligned, with $3\sigma$ lower limits of $I_\star > 45^\circ$ for both stars (see Table~\ref{tab:stellar} for their precisely derived values). 

A much more precise technique of measuring the orbital obliquities of these transiting systems is to observe their spectroscopic transits via the Rossiter-McLaughlin \citep{1924ApJ....60...15R,1924ApJ....60...22M} effect, or via the Doppler tomographic transit shadow detection \citep{1997MNRAS.291..658D,2010MNRAS.407..507C}. Based on our measured rotational broadening velocity, \starA{}b should exhibit an in-transit velocity anomaly of $\sim 4\,\ms$, while \starB{}b and \starB{}c should exhibit significant velocity variations at the levels of $20-30\,\ms$. Planets around bright, young, rapidly rotating stars are good targets for mapping the orbital obliquity -- age relationship, and determining the origins of the abundance of small planets around Sun-like stars. 

\section{Conclusions}

\starA{} and \starB{} are two field stars exhibiting photometric and spectroscopic signatures of youth, hosting transiting Neptunes that were identified by \emph{TESS} observations.

\starA{}b is a $\starAplrad{}\,R_\oplus$ mini-Neptune in a $\starAperiodshort{}$\,day period around a G-type star. The period and transit ephemeris of the system were refined by three ground-based follow-up observations that successfully detected the 1\,mmag transit events on the target star. We were able to eliminate false positive scenarios with extensive spectroscopic follow-up and speckle imaging of the host star, validating the planetary nature of the transits. 

We estimated the age of \starA{} to be \starAage{}\,Myr, based on its photometric rotation period, spectroscopic calcium II core emission in the HK and infrared triplet lines, and the presence of X-ray emission from the ROSAT all-sky survey. The rotation period and 6708\,\AA{} lithium absorption strength are comparable to that of stars in the Pleiades cluster, agreeing with our age estimates for the system. 

\starB{} hosts two Neptune sized planets around a K-type star. \starB{}b is a $\starBplrada{}\,R_\oplus$ planet in a $\starBperiodashort{}$\,day period orbit, while \starB{}c has a radius of $\starBplradb{}\,R_\oplus$ and an orbital period of $\starBperiodbshort{}$\,days. The transits of \starB{}b and c were both successfully recovered by an extensive ground-based follow-up campaign with the MEarth telescope array. The planetary nature of the system were further validated by diffraction limited imaging and spectroscopic analyses. 

The age of \starB{} is significantly younger than \starA{}, estimated at \starBage{}\,Myr. It exhibits strong X-ray and calcium emission that are stronger and beyond the range of calibrated literature age-activity relationships. The rotation period and lithium absorption strength of \starB{} suggest that it has an age younger than the Pleiades cluster, and more in agreement with younger stellar associations.  

\starA{} and \starB{} are examples of young planet hosting field stars that can contribute significantly to characterizing the relationship between planet properties and their ages. \emph{TESS} is likely to yield numerous systems like \starA{} and \starB{}, amenable to extensive follow-up observations that can characterize the orbital and atmospheric properties of planets at early stages of their evolutions.

\acknowledgements  
The authors became aware of a parallel effort on the characterization of TOI-942 by Carleo et al. in the late stages of the manuscript preparations. The submissions are coordinated, and no analyses or results were shared prior to submission. 

Work by G.Z. is supported by NASA through Hubble Fellowship grant HST-HF2-51402.001-A awarded by the Space Telescope Science Institute, which is operated by the Association of Universities for Research in Astronomy, Inc., for NASA, under contract NAS 5-26555.
The MEarth Team gratefully acknowledges funding from the David and Lucile Packard Fellowship for Science and Engineering (awarded to D.C.). This material is based upon work supported by the National Science Foundation under grants AST-0807690, AST-1109468, AST-1004488 (Alan T. Waterman Award), and AST- 1616624, and upon work supported by the National Aeronautics and Space Administration under Grant No. 80NSSC18K0476 issued through the XRP Program. This work is made possible by a grant from the John Templeton Foundation. The opinions expressed in this publication are those of the authors and do not necessarily reflect the views of the John Templeton Foundation.
This research has made use of the NASA Exoplanet
Archive, which is operated by the California Institute of Technology,
under contract with the National Aeronautics and Space Administration
under the Exoplanet Exploration Program. 
Funding for the TESS mission is provided by NASA's Science Mission directorate. We acknowledge the use of public TESS Alert data from pipelines at the TESS Science Office and at the TESS Science Processing Operations Center. This research has made use of the Exoplanet Follow-up Observation Program website, which is operated by the California Institute of Technology, under contract with the National Aeronautics and Space Administration under the Exoplanet Exploration Program. This paper includes data collected by the TESS mission, which are publicly available from the Mikulski Archive for Space Telescopes (MAST).
Resources supporting this work were provided by the NASA High-End Computing (HEC) Program through the NASA Advanced Supercomputing (NAS) Division at Ames Research Center for the production of the SPOC data products.
This work makes use of data from the Mount Wilson HK-Project. The HK\_Project\_v1995\_NSO data derive from the Mount Wilson Observatory HK Project, which was supported by both public and private funds through the Carnegie Observatories, the Mount Wilson Institute, and the Harvard-Smithsonian Center for Astrophysics starting in 1966 and continuing for over 36 years.  These data are the result of the dedicated work of O. Wilson, A. Vaughan, G. Preston, D. Duncan, S. Baliunas, and many others. This work makes use of observations from the LCOGT network.
Some of the observations in the paper made use of the High-Resolution Imaging instrument(s) ‘Alopeke (and/or Zorro). ‘Alopeke (and/or Zorro) was funded by the NASA Exoplanet Exploration Program and built at the NASA Ames Research Center by Steve B. Howell, Nic Scott, Elliott P. Horch, and Emmett Quigley. ‘Alopeke (and/or Zorro) was mounted on the Gemini North (and/or South) telescope of the international Gemini Observatory, a program of NOIRLab, which is managed by the Association of Universities for Research in Astronomy (AURA) under a cooperative agreement with the National Science Foundation. on behalf of the Gemini partnership: the National Science Foundation (United States), National Research Council (Canada), Agencia Nacional de Investigación y Desarrollo (Chile), Ministerio de Ciencia, Tecnología e Innovación (Argentina), Ministério da Ciência, Tecnologia, Inovações e Comunicações (Brazil), and Korea Astronomy and Space Science Institute (Republic of Korea).

\facility{FLWO 1.5\,m, CHIRON, TESS, LCOGT, MEarth, Gemini, Subaru}
\software{lightkurve \citep{lightkurve}, emcee \citep{2013PASP..125..306F}, Astropy \citep{2013A&A...558A..33A,2018AJ....156..123A},\texttt{AstroImageJ}\citep{Collins:2017}, PyAstronomy \citep{pya}, TESSCut \citep{2019ascl.soft05007B}}

\bibliographystyle{apj}
\bibliography{refs}

\end{document}